\documentclass[preprint,12pt]{aastex}
\newcommand{\eten}[1]{\mbox{$10^{#1}$}}

\newcommand{\degree}{\mbox{$^{\circ}$}}
\newcommand{\am}{\mbox{\arcmin}}
\newcommand{\as}{\mbox{\arcsec}}

\newcommand{\kms}{\mbox{km s$^{-1}$}}

\newcommand{\um}{$\mu$m}
\newcommand{\yr}{\mathrm{yr}}
\newcommand{\erg}{\mathrm{erg}}

\def\lsim {$\rlap{\raise.4ex\hbox{$<$}}\lower.55ex\hbox{$\sim$}\,$}

\newcommand{\lsun}{\mbox{L$_\odot$}}
\newcommand{\msun}{\mbox{M$_\odot$}}

\newcommand{\lbol}{\mbox{$L_{bol}$}} 

\newcommand{\rsun}{\mbox{$R_{\odot}$}}

\newcommand{\mean}[1]{\mbox{$\langle#1\rangle$}} 
 
\newcommand{\co}{$^{12}$CO}
\newcommand{\coo}{$^{13}$CO}

\input{epsf}

\begin{document}

               
\title {\bf A Systematic Search for Molecular Outflows Toward Candidate Low-Luminosity Protostars and Very Low Luminosity Objects}
\author{Kamber R. Schwarz\altaffilmark{1,2}, Yancy L. Shirley\altaffilmark{1,3},
        and Michael M. Dunham\altaffilmark{4}
	}
\altaffiltext{1}{Steward Observatory, 933 N Cherry Ave., Tucson, AZ 85721}
\altaffiltext{2}{2009-2010 NASA Space Grant Intern.  Senior Honors Thesis for the Honors College at The University of Arizona.}
\altaffiltext{3}{Adjunct Astronomer, The National Radio Astronomy Observatory}
\altaffiltext{4}{Department of Astronomy, Yale University, P.O. Box 208101, New Haven, CT 06520}
 
\begin{abstract}

We present a systematic single-dish search for molecular outflows toward a sample
of 9 candidate low-luminosity protostars and 30 candidate Very Low Luminosity Objects (VeLLOs; 
$L_{int} \leq 0.1$ \lsun ).  The sources are identified using data from the 
\textit{Spitzer Space Telescope} catalogued by Dunham et al. toward nearby
($D < 400$ pc) star forming regions.
Each object was observed in \co\ and \coo\ J =
$2\,\rightarrow\,1$ simultaneously using the sideband 
separating ALMA Band-6 prototype
receiver on the Heinrich Hertz Telescope at $30$\as\ resolution.  
Using 5-point grid maps we identify five new potential outflow candidates and make 
on-the-fly maps of the regions surrounding sources in the dense cores
B59, L1148, L1228, and L1165.  Of these new outflow candidates, only the map of 
B59 shows a candidate blue outflow lobe associated with a source in our
survey. 
We also present larger and more sensitive maps of the 
previously detected L673-7 and the L1251-A IRS4 outflows and 
analyze their properties in comparison to other outflows from
VeLLOs. 
The accretion luminosities derived from the outflow properties of the VeLLOs with 
detected CO outflows are higher than the observed internal luminosity of the protostars, 
indicating that these sources likely had higher accretion rates in the past.
The known L1251-A IRS3 outflow is detected but not remapped.
We do not detect clear, unconfused signatures of red and blue 
molecular wings toward the other 31 sources in the survey indicating
that large-scale, distinct outflows 
are rare toward this sample of candidate protostars.
Several potential outflows are confused with kinematic
structure in the surrounding core and cloud.  
Interferometric imaging is needed to disentangle large-scale molecular cloud 
kinematics from these potentially weak protostellar outflows.

\end{abstract}



\section{Introduction} \label{intro}

The incredible sensitivity of the \textit{Spitzer Space Telescope} has
resulted in the study of a new class of very low luminosity objects (VeLLOs).
Formally, a VeLLO is defined as a protostar with an internal
luminosity of $< 0.1$ \lsun .  The initial \textit{Spitzer}
testing and validation observations for the Cores to Disk Legacy team (c2d) 
detected a new low-luminosity protostar in the previously classified
starless core L1014 (Young et al. 2004).  
Follow-up observations revealed that L1014 was indeed a VeLLO
with a weak molecular outflow (Bourke et al. 2005) and centimeter continuum
emission consistent with a thermal jet (Shirley et al. 2007).  
Since the discovery of L1014, the c2d project has completed a 
census of the embedded protostellar population in the nearby ($D < 400$ pc)
low-mass star-forming molecular clouds (see Evans et al. 2009 for
a summary).  The c2d survey is sensitive to bolometric luminosities of
$L_{bol} \sim 0.004 \left(\frac{D}{140 pc}\right)^2$ \lsun\ and
can identify essentially all low-mass protostars with accretion
rates down to 
$\frac{dM_{acc}}{dt} = 1.3 \times 10^{-9} \frac{R_{proto}}{1 \rsun } \frac{0.1 \msun }{M_{proto}} 
\left( \frac{D}{140 pc}\right)^2$ \msun /yr (see Table 1 for the adopted
distances to regions observed).  
The subset of embedded low luminosity protostars ($L_{bol} \leq 1$ \lsun ) 
including VeLLOs were catalogued by Dunham et al. (2008) 
and provide the most complete collection that may be systematically studied.

The true nature of VeLLOs are a subject of much current research.  
Some VeLLOs may be protostars in the incipient stages
of accretion, the observationally 
elusive first hydrostatic core phase of their evolution (Larson 1969; see Machida et al. 2008
for a recent theoretical simulation).
Alternatively, some VeLLOs may be more evolved protostars which happen to be
observed in a low accretion state.  Observations of the molecular outflow,
a byproduct of magnetized accretion onto a protostar, provide a potential
diagnostic to constrain these two possibilities.  Molecular outflows may provide
a fossil record of the accretion history of the source.  


The source IRAM04191+1522 was one of the first discoveries of a
low-luminosity source ($L_{int} = 0.08$ \lsun ; Dunham et al. 2006) 
driving a large-scale molecular
outflow (Andr\'e et al. 1999, Belloche et al. 2002).  Since the identification
of the outflow around the VeLLO L1014, five more VeLLOs have had
outflows detected: L1521F (Bourke et al. 2006), MMS126 (Stanke et al. 2006),
Per 073 (Hatchell \& Dunham 2009), L673-7 (Dunham et al. 2010a), 
and L1148 (Kaufmann et al. 2011).
Four out of the seven VeLLOs drive large-scale outflows that
were detected with single-dish telescopes (IRAM04191, L673-7, Per 073, and MMS126), 
two of the sources (L1014 and L1148) have outflows that are detected only with interferometers
(not detected with single-dish),
and one source (L1521F) has a near-infrared scattered light nebula indicative
of outflow activity but no direct detection of the outflow.  
In addition, there are now a few objects with interferometrically 
detected outflows, but no detected
\textit{Spitzer} infrared driving source: L1448 IRS2E1 (Chen et al. 2010), 
Per-Bolo 58 (Dunham et al. 2011),
L1451-mm (Pineda et al. 2011), and CB17MMS (Chen et al. 2012).
These objects are candidates for first hydrostatic cores although follow-up
observations are still needed to confirm their status.

From maps of the emission in the molecular outflow line wings, 
the dynamical age, mass, momentum, and total outflow force may be
estimated (see Cabrit \& Bertout 1992). Resulting estimates 
of the average accretion luminosity are 
typically larger than the observed source luminosity implying 
that the accretion rate
must have been larger in the past (e.g. Dunham et al. 2010a).  
This is another version of the
famous ``luminosity problem" for low-mass protostars, 
elucidated by Kenyon et al. (1990), where protostars are not luminous
enough to account for the expected accretion luminosity based on
estimates of the protostellar lifetime.  A possible solution 
to this problem is that the accretion rate is variable with time. 
Modeling of source counts of protostars in different evolutionary
stages from the c2d survey 
indicates that low luminosity protostars may accrete most of their mass over a period 
of a few $10^5$ years through episodic events (see Dunham et al. 2010b).  
In order to test this hypothesis, 
comprehensive surveys for molecular outflows around low luminosity
protostas and VeLLOs are needed.

The published searches for molecular outflows associated with VeLLOs 
have been heterogenous to date toward a very small fraction of the
VeLLO population in nearby clouds and cores.  
In this paper, we present the first systematic survey for 
large-scale molecular outflows toward the sample of 39 isolated,
candidate low-luminosity protostars
(including 30 VeLLOs) in the J $= 2 \rightarrow 1$ 
lines of \co\ and \coo\ (see Table 1) .  Our main goal is to characterize the
frequency of kinematically distinct large-scale outflows toward VeLLOs and 
to determine the properties (mass, age, momentum, force, etc.) of any detected
outflows.  In \S2 we describe the observing techniques and our 
sample selection.
In \S3 we analyze the properties of 3 detected outflows in the sample as
well as describe the properties of the numerous non-detections.  In \S4 we  
discuss the nature of the VeLLO outflows.

\section{Observations} \label{obs}

Observations were performed on 9 days 
between December 2009 and March 2011 (see Table 2)
using the sideband separating ALMA Band-6 prototype receiver on 
the 10-m Heinrich Hertz Telescope.  This instrument permits
simultaneous dual polarization observations of 
\co\ J =$ 2 \rightarrow 1$ and \coo\  J = $2 \rightarrow 1$ in the 
LSB and USB respectively with a 5 GHz IF. Observations were
performed in 4-IF mode (dual polarization and sideband separating)
using filter bank backends with 128 channels at 250 kHz resolution
(0.34 km/s and 0.32 km/s at 220 GHz and 230 GHz respectively).

Pointing was carried out using the planets Jupiter, Mars, and Saturn
with a typical rms pointing uncertainty of $< 5$\as . 
Position switching observations with a $5^{\prime}$ OFF position
in azimuth were used to calculate the main beam efficiency
($T_{mb} = T_A^* / \eta_{pol}$; Mangum 1993).
The calculated efficiencies for each observing session are given in Table 2. 
Sideband rejections were measured to be $> 13$ dB and were not corrected for.  
Variation in $\eta_{pol}$ likely reflect calibration variations and changes 
in the coupling between the receiver optics and the telescope between observing
seasons.

\subsection{Observing the Sample}

Our target objects were selected from the catalogue of low luminosity
protostars in the Cores to Disk Legacy sample compiled in 
Dunham et al. (2008) (their Tables 3 - 8). Throughout this paper,
we use the region name followed by the source number in
parenthesis listed in the tables of Dunham et al. (2008)
(i.e. B35A (007)).
The Dunham catalog contains 199 sources.
All isolated sources ($> 30$\as\ from a IRAS source) with suitable declinations
observable from the Heinrich Hertz Telescope ($\delta > -30$\degree ) 
from Dunham Table 3 (Cores) and Dunham Table 7 (Ophiuchus),
which had not been observed previously,
were included in our final target list of 39 sources.  
We excluded candidates within a beam width, 30\as , of a known IRAS source,
since it is unlikely that we will be able
to disentangle the outflow from the low luminosity protostar
from the stronger IRAS source. 
We also exclude some sources and regions where known single-dish searches for molecular outflows have already 
occurred (e.g. IRAM04191 Andr\'e et al. 1999, L1521F Bourke et al. 2006, L328  Lee et al. 2009, 
L1014 Bourke et al. 2007, CB130 Kim et al. 2011, and LFAM 26 Bussmann et al. 2007, 
Nakamura et al. 2011). 
We exclude sources in Perseus (Dunham Table 4) and
Serpens (Dunham Table 8) because those regions have been recently mapped 
at $\leq 30$\as\ resolution in transitions of \co\ (Hatchell et al. 2007, Curtis et al. 2010;
Burleigh \& Bieging in preparation). 
We did not observe the Lupus and Chamaeleon II regions as they
are too far south and those regions are the the subject of a
southern hemisphere companion survey (Dunham et al., in preparation).
The final observed sample consists of 39 candidate protostars, 
listed in Table 1.  Of this sample, 30 sources are classified as 
VeLLOs $L_{IR} < 0.1$ \lsun\ and all sources
have $L_{IR} < 0.45$ \lsun .

For each source in Table 1, 
a 5-point map was obtained with 30$\as$ spacing and 
a minimum of 3 minutes absolute reference position switching. 
Clean OFF positions were found by manually checking for emission 
free regions using data from the Dame et al. (2001) survey included in SkyView CO 
maps\footnote{The SkyView Virtual Observatory http://skyview.gsfc.nasa.gov}. 
Before using an OFF position we integrated for three to 
four minutes using a second reference OFF to insure that it was truly free of CO emission. A linear baseline was subtracted from all maps and all data were reduced using custom routines in the GILDAS (CLASS and GREG) software package\footnote{http://iram.fr/IRAMFR/GILDAS/}.

\subsection{OTF Mapping} \label{otf}

We mapped the sources L673-7 (031), L1251-A (045), B59 (022), B59 (023), 
L1148 (033), L1228 (036), and
L1165 (040).  The sources were chosen as either known outflow candidates where the outflow properties
were uncertain and could be improved or as new 
outflow candidates based on the existence of line wings in the observed 5-point spectra (\S3.1). 
For each of these sources we created an on-the-fly (OTF) map 
with dimensions 5$\am$ by 5$\am$, 4$\am$ by 4$\am$, or 3$\am$ by 3$\am$.  
For the L673-7 (031) outflow two regions of size 
3$\am$ by 3$\am$ offset by $\pm$120$\as$ in the directions 
of the red and blue wings were also mapped. 
All maps had 10$\as$ row spacing with a scan rate 
of 10$\as$ per second and were scanned boustrophedonically 
in the RA and DEC directions multiple times to improve the signal-to-noise
in the line wings.

On-the-fly maps were reduced and analyzed using custom routines in the GILDAS software
package.
A linear baseline was removed from all observations before they were scaled using the calculated efficiency.
 These scaled V$_{\mathrm{pol}}$ and H$_{\mathrm{pol}}$ observations were combined and then convolved and re-gridded using a Gaussian-tapered Bessel function (Mangum et al. 2008). The resulting beam sizes where 35.5\as\ for \co\ and  36.9\as\ for \coo .
Integrated intensity maps were created using the convolved and re-gridded spectra. The red and blue wings were partitioned into multiple velocity channels with the minimum channel size being the spectral resolution ($\delta V_{chan}$). Using these files we generated channel maps for each source with
the lowest contour at the two to three sigma level of integrated intensity ($\mathrm{\sigma_{I} = \mean{\sigma_{T}}\sqrt{\delta V_{chan}(v_{2}-v_{1})}}$ where $\mathrm{\mean{\sigma_{T}}}$ was found by averaging the baseline rms at several points in the map). These channel maps allowed us to discern any large scale outflow structure for
the outflow candidates.

\section{Results} \label{results}

\subsection{Detections \& Non-detections} \label{detections}

The line properties for the center spectrum of each source 
are tabulated in Table 3. 
The ratio of integrated intensities for \co\ to \coo\ ranges from 1.06 to 23.3 
though the median is just 3.51.
This ratio is much lower than the expected value of approximately 50 for
optically thin emission (Wilson \& Rood 1994);  therefore we conclude that the \co\ 
emission in these regions is typically very optically thick.
This agrees with the general shape of line profiles seen in Figures 1 - 10.
Some sources display self-absorbed \co\ line profiles with \coo\ peaking
near the \co\ absorption dip.  The \coo\ line profiles are typically
well fit by Gaussian line profiles with an average $\Delta v = 2.12 \pm 0.98$ km/s. 
This linewidth is much larger than the sound speed in 10 K molecular gas
($0.077$ km/s for CO) indicating that the motions are likely characterized by turbulence.
Unfortunately, at the velocities of \co\ line wings, \coo\ is typically not detected; therefore, we are
unable to make a direct optical depth correction to the \co\ line wing emission.

The 5-point maps for each source (Fig. 1-10) 
were visually classified 
according to their spectral features (Table 3).  Sources consisting of a single 
component were flagged with an ``s" while sources consisting of 
multiple components were are denoted by ``mc". 
Just over half of the sources have single component \co\
spectra ($53.3$\%) while one third of the \co\ spectra
appear to have multiple velocity components ($33.3$\%).
Self absorbed sources are distinguished from sources with 
multiple components by comparing to the \coo\ line:
sources with \coo\ which peaks near the \co\ absorption dip 
are considered to be self absorbed and are denoted by the
flag ``sa".  The remaining $13.4$\% of sources show signatures of
self-absorption in \co .

Sources displaying evidence of blue or red wings in \co\ are flagged 
with ``bw" and ``rw" respectively. 
Seven sources ($23$\%) 
have either a prominent blue wing or a prominent red wing.  These are
the candidate outflow sources that we explore with OTF
observations.  The known outflow sources L673-7 (031),
L1251 (044), and L1251 (045) are successfully recovered as outflow candidates
in our visual flagging scheme.  The visual search for outflow
signatures is challenging because several sources 
have confused spectra with multiple velocity components.  
Nearby IRAS sources, despite being more than one beam ($30$\as )
away, also contribute to the confusion in some spectra. 
The sources L673 (027), (028), and (030) are examples of complicated,
multi-peaked spectra which appear to show evidence for an outflow, 
but all three sources are $< 50$\as\ from IRAS19180+1116 whose
outflow likely dominates the spectra shapes in this region.
From the spectra in Figures 1 - 10, we visually selected
five additional candidates, B59 (022), B59 (023), L1148 (033), 
L1228 (036), and L1165 (040), to follow up with OTF mapping that 
show clear evidence of a blue or red wing.

Each candidate protostar in the Dunham catalog is assigned a group number
(1 through 6) which indicates the likelihood that the object
is an embedded protostar (see Table 1 for the
group numbers).  Sources in group 1 
are verified protostars with the certainty decreasing
with increasing group number. The nature of sources in group 6
is very uncertain due to a lack of supporting observations 
(i.e. no evidence that the source is embedded within a core, etc.).
All three of the known outflow sources that we mapped belong to group 1
(L673-7 (031), L1251 (044), and L1251 (045)).
There are only two group 1 or group 2 objects in our sample
which were not selected as good outflow targets based solely on the
appearance of their \co\ spectra: 
L673 (027) and L1221 (041).  Unfortunately both of these sources
are nearby ($42$\as ) strong IRAS outflow sources.  As explained above,
the kinematics of the  L673 (027) region is severely complicate
by the IRAS19180+1116 outflow source.  L1221 (041) appears
to have extended $4.6 \mu$m emission in the NW-SE direction that is indicative of an
outflow, but unfortunately, the large outflow source IRAS22266+6845
to the northwest appears to overlap in orientation.  The effect of
the IRAS source outflow can be seen in the northernmost 5-point spectra
for L1221 (041) although no outflow signature is apparent at the position
of the VeLLO.  This region has been observed with the BIMA interferometer
(Lee et al. 2005) and a very tenuous detection of the VeLLO outflow is
indicated.  Deeper interferometric imaging is needed of L1221 (041) to confirm
this result.

Of the remaining 5 candidate outflow sources that we
mapped, all of them belong
to group 6 except B59 (023) which is in group 3.  Interestingly, B59 (023) 
is the only of those candidates
that shows evidence of an outflow being associated directly with the
source (\S3.4).  
Several of the Ophiuchus sources, (191) through (199) excluding
(197), were not detected in \co ; all of these non-detections
are in group 5.  This result indicates that those protostellar candidates
are not embedded objects and are not associated with molecular gas.


In summary, all of the confirmed outflow candidates reported in the literature and
in this survey belong to groups 1, 2, or 3.
In the following sub-sections we analyze in detail the properties
of the detected and non-detected outflows observed in the OTF maps.

\subsection{The Fully Mapped L673-7 (031) Outflow} \label{L673}

Dunham et al. (2010a) discovered a large outflow surrounding L673-7 (031) (L673-7-IRS), 
but were not sure that the full extent of the outflow was mapped. 
Here, we present a new map which extends the region covered 
by Dunham and fully map the outflow associated with this VeLLO
with higher signal-to-noise.
We follow the methodology described in Dunham et al. (2010a) to derive
the properties of the outflow.  We explicitly show the functional
dependence on the inclination in the calculations.  Also,
when appropriate, we always chose the lower limit in a calculation
such that the final outflow properties represent lower limits
to the actual values.


The outflow properties are first calculated per velocity channel per pixel, then summed over
those pixels covered by the outflow and all outflow velocity channels 
in the intervals $[-0.7, 5.3]$, and $[9.3,16.3]$ km/s (see Figure 11). 
Note that this velocity interval is more restrictive than the
velocity interval used in Dunham et al. (2010a) and is determined by the
velocity interval over which we statistically ($> 2\sigma_I$) 
detect the outflow signature in the map.  For each calculated quantity (mass,
momentum, etc.), we propagate
the statistical errors on a channel-by-channel basis.  The dominant source of statistical 
uncertainty is in the integrated intensity in the map.  We do not propagate the distance 
uncertainty (a systematic error which itself is poorly constrained) into these calculations.

A lower limit of the dynamical time, t$_{d}$, was found by dividing the extent of the outflow by the center 
velocity of the outflow wing relative to the v$_{LSR}$ of the protostar. 
We measure the extend of the outflow to be 290$\as$ compared to the 150$\as$ used by Dunham et al. (2010a).
We calculate $\mathrm{t_{d}=4.44\pm0.23\times\eten{4} \, \frac{\cos i}{\sin i}\ \yr}$ assuming an
angular uncertainty of half the beam size (15\arcsec ).
Using this value, we find the time averaged mass accretion rate is
$\mathrm{\langle\dot{M}_{acc}\rangle=2.78\pm0.14\times\eten{-7} \, \frac{\sin i}{\cos^{2} i}\ \msun\ \yr^{-1}}$ resulting
in a total protostellar mass accreted of $\mathrm{M_{acc}= \dot{M}_{acc}t_{d}} = \mathrm{1.23\pm0.09\times\eten{-2} \, \frac{1}{\cos i}\ \msun}$.
The resulting lower limit on the time averaged accretion luminosity ($L_{acc} = GM_{acc}\dot{M}_{acc}/R$ 
with $R = 3$\rsun ) is
$\mathrm{L_{acc}=0.036 \pm 0.003 \, \frac{\sin i}{\cos^{3} i}\ \lsun}$.  
The ratio of the outflow-derived accretion luminosity to the
internal luminosity is $2.1 \pm 0.7 \, \frac{\sin i}{\cos^{3} i}$
where we have assumed a 30\%\ uncertainty on $L_{IR}$.

While our calculated numbers are smaller than the values found in Dunham et al. (2010a), 
we are now confident that with this new map, the full extent of the L673-7 outflow has been mapped. 
This independent analysis still
supports the hypothesis that L673-7 has undergone episodic accretion; 
however, the discrepancy between the derived average accretion luminosity and the observed internal 
luminosity has decreased
significantly with our new analysis.  It must be kept in mind, though, that we have
selected lower limits to all derived outflow quantities except for the
inclination.  The unknown inclination of
the outflow is the largest source of systematic uncertainty.
Figure 12 shows that the
functional dependence of the outflow dynamical time, force, and time-averaged
accretion rate with inclination is monotonic for each quantity.  
For an inclination of $i_o$ = 34.3{\degree}, the correction factor 
$\frac{\sin i}{\cos^{3} i} = 1$.  
The probability that the real inclination of the outflow is
greater than an inclination $i_o$ is given by 
$P(i > i_o) = cos(i_o)$ for randomly oriented outflows.
The inclination for which $\frac{\sin i}{\cos^{3} i} \approx 1$  corresponds to an 
83\%\ confidence interval that the real outflow inclination is larger than $i_o$ = 34.3{\degree}. 
Thus, typical inclination angles will be larger than this value and the
derived quantities can easily be factors of a few times larger to over an order
of magnitude larger; therefore, any discrepancy
between the time-averaged accretion luminosity and the internal luminosity will 
also increase by the same factor (see \S4.1).

\subsection{The L1251 (045) Outflow} \label{IRS4}

L1251 (045) was first mapped by Lee et al. (2010) ($L_{IR} = 0.077$); 
however, only the properties of the larger, more distinct outflow associated with L1251 (044) 
were calculated ($L_{IR} = 0.156$).   L1251 (044) and (045) are labeled as IRS4 and IRS3
respectively in the Lee et al. (2010) paper.
In order to determine the properties of this adjacent outflow we 
re-mapped the L1251 (045) outflow using higher 
spacial resolution ($30$\as\ versus 
$48$\as\ in Lee et al.) and with higher signal-to-noise. 
We use the same method as for L673-7 to derive the outflow properties 
(\S3.2).  The velocity intervals for
the blue and red wings span from $[-10,-5]$ km/s 
and $[-1,0.5]$ km/s. 
The smaller velocity difference for the red wing is used to 
avoid integrating over a second velocity component which peaks $+3.8$ km/s 
from the $v_{LSR}$ of the primary component (see Figure 13).

The calculated outflow mass, energy and momentum are slightly larger
than for L673-7 despite this outflow being less distinct. 
Assuming a distance of 300 $\pm$ 50 pc (see Dunham et al. 2008)
we calculate an outflow dynamical time of 
$\mathrm{t_{d}=2.69\pm0.23\times\eten{4} \, \frac{\cos i}{\sin i}\ \yr}$.
This outflow has been powered for just more than half as
long (at the same
inclination) as the L673-7 outflow; therefore, the derived outflow
luminosity and force are higher for L1251-A (045) than for L673-7 (031).
We calculate  $\mathrm{\langle\dot{M}_{acc}\rangle=6.3\pm1.5\times\eten{-7} \, \frac{\sin i}{\cos^{2} i}\ \msun\ \yr^{-1}}$. 
The lower limit on the protostellar mass accreted is 
$\mathrm{1.7\pm0.6\times\eten{-2} \, \frac{1}{\cos i}\ \msun}$, 
making the lower limit for the accretion luminosity
$\mathrm{L_{acc} =  0.11 \pm 0.04 \, \frac{\sin i}{\cos^{3} i}  \, \lsun}$.  
The average (minimum) accretion luminosity is a factor of $1.5 \pm 0.7 \, \frac{\sin i}{\cos^{3} i}$ 
times higher than the observed internal luminosity.  For inclinations greater than
34.3\degree , the discrepancy will become larger (Figure 12).

\subsection{A New Potential Outflow Around B59 (023)}

We have identified a new potential outflow around the source B59 (023).
This source displays line wings from a blue lobe 
to the southwest direction from the source (Figure 14). 
This feature has also been confirmed in CO J = $3 \rightarrow 2$ 
observations with the ASTE 12m telescope (Dunham et al., in preparation). Unfortunately, 
the red wing is confused by a powerful 
outflow associated with 2MASS source J17112317-2724315 to the northeast (see Brooke et al. 2007).
It is not clear whether the blue shifted emission actually 
emanates from B59 (023), 
or whether we have mapped its full extent. 

Since we are only able to discern one wing of this purported outflow, 
we report the outflow properties found for the blue wing multiplied by two.
B59 (023) is assumed to be at the same distance ($D = 125$ pc) adopted
by the c2d team for the Pipe Nebula and Ophiuchus complex.
The derived outflow luminosity and force are significantly higher
than for L673-7 (031) and L1251 (045), in part due to the apparent 
small dynamical time.  B59 (023) is one of the nine sources in 
our survey that is technically not a VeLLO
($L_{IR} = 0.343$ \lsun ).  The (minimum) average accretion luminosity is
$1.5 \pm 0.5 \frac{\sin i}{\cos^{3} i}$ times higher than the observed bolometric luminosity;
however, we caution that the derived parameters may be overestimated
if the outflow indeed has a longer dynamical time.
Given that the region is too kinematically confusing for trustworthy analysis
using only a single dish, we believe this source to be a prime candidate 
for follow-up interferometric observations.

\subsection{OTF Map Non-detections}

All of the remaining OTF maps were kinematically confused and we were unable 
to  discern any large scale, distinct bipolar outflow structure
originating from our sources. The outflow lobes are not
obvious from analysis of the channel maps or individual spectra
in the HHT maps (see Figures 15-18). Thus, 
the remaining four sources cannot be classified as outflow candidates based
on their $^{12}$CO channel maps.  B59 (022) is west of (023) and its
spectra are also severely confused by the outflow from IRAS17081-2721
(visible across the northern edge of the channel maps).  There does appear
to be a red lobe to southwest of B59 (022), but the corresponding blue lobe
has strong red wing emission overlapping in the same position
(see the top two rows of the channel map in Figure 15).  The channel
maps for L1148 (033) show no distinct structure in the CO wing
emission (Figure 16). 
Outflow sources
can been seen in the maps of L1228 (036) and L1165 (040) but the emission is
clearly offset and not associated with the candidate protostars in
either case (Figures 17 and 18).  We calculate the mass sensitivity of our survey 
using the same technique as for the detected outflows 
but over an empty portion of the map. Using the map L1148 (033) and assuming a
conservative outflow
extent which covers half the map, the
$3\sigma$ limit for a mass detection is $1.38\times \eten{-3} \msun$ which 
is consistent with the uncertainties for the detected outflows. We display the 
maps of these non-detections for completeness.

\section{Analysis \& Caveats}

Of the 39 sources observed in this survey, three have previously 
known large-scale outflows.  We initially identified five additional protostars as 
outflow candidates; but of those five new candidates, only B59 (023) 
showed evidence of a newly detected candidate outflow although confusion with the 
strong nearby IRAS17081-2721 outflow makes estimation of its outflow 
properties difficult.  The VeLLO single-dish outflow detection
rate is only 2 out of 30 sources or $7$\%\ in this survey.  
Recently, Hatchell \& Dunham (2009) and Curtis et al. (2010) separately 
mapped and analyzed the outflow properties of protostars in the Perseus molecular cloud.
Outflow signatures from only one VeLLO was uniquely identified in their maps
(Per 073 in Hatchell \& Dunham; $L_{IR} = 0.018$ \lsun ) although
many sources with luminosities approaching $0.1$ \lsun\ were detected in
both surveys (e.g., Per 071 in Dunham Table 4 which is source 65 in 
Curtis et al. 2010; $L_{IR} = 0.105$ \lsun ).  If we
include the other previous VeLLO single-dish detections from the literature
(IRAM014191, Per 073, and MMS126), then the single-dish detection rate increases to 
$13$\% for 38 VeLLOs observed in cores.  
These detection rates are low and suggest that large scale outflows from VeLLOs 
in the Dunham catalog are rarely a dominant kinematic feature in the surrounding 
molecular cloud.

We compare the properties of the detected 
VeLLO outflows in this survey with those of other known 
VeLLOs and low-luminosity objects observed from the
Dunham catalogue.  The outflow properties are
taken from the individual paper references for each source
(IRAM01491 Andr\'e et al. 1999, 
MMS 126 Stanke et al. 2006, L1014 Shirley et al. 2007, L1448-IRS2E1 Chen et al. 2010,  
Per 071 073 105 106 Curtis et al. 2010, L1251A-IRS3 Lee et al. 2010,
L1148-IRS Kauffmann et al. 2011, Per-Bolo-58 Dunham et al. 2011, 
L1451-MM Pineda et al. 2011, LFAM 26 Nakamura et al. 2011, CB17 MMS Chen et al. 2012).  
Not all quantities plotted in Figure 19 were reported
directly in each paper; therefore, in those cases the quantities were
derived using the same equations as in this paper.  If a range of values
were given, the average value is plotted.  Errobars were rarely 
reported in the original references and are not shown in the figure.
The plotted values in Figure 19
do not make optical depth corrections or inclination corrections.
We used the literature values with the exception of
MMS 126 from Stanke et al. (2006) whom made an average optical
depth correction (factor of 3.5) and assumed an inclination
of 57.3\degree\ (we have reversed those corrections for comparison).

There is strong evidence that the outflow properties change as
protostars evolve from the Class 0 to the Class I phase; namely,
Class 0 sources drive faster, more massive molecular 
outflows than Class I sources (Arce \& Sargent 2006).  The outflow mass, 
momentum, and force decrease from the Class 0 to the Class I phase (see Curtis et al. 
2010 for a recent summary).  The survey of Bontemps
et al. (1996) studied low-mass protostars in the
same \co\ J = $2 \rightarrow 1$ transition that we used in this survey
toward a sample that spanned bolometric luminosities from $0.2$
to $41$ \lbol\ among sources from a heterogenous population of cores and clouds.  
Bontemps et al. found a correlation between the
outflow force and bolometric luminosity with Class 0 sources
having larger outflow force than Class I sources. 
Recently, Curtis et al. (2010) performed a similar analysis on the protostellar
population within a single cloud (Perseus) with sources extending to
lower luminosities and found similar results
for Class 0 and I sources.  The slopes and offsets of the Curtis correlations differ
from Bontemps correlations, but the trend of the Class 0 outflow force correlation
lying above the Class I correlation holds (see Figure 19).

We have plotted the outflow force of the protostars observed in this
paper along with VeLLOs and low-luminosity protostars from
the Dunham catalogue that were observed in the literature (Figure 19).
The Class 0 (solid line) and Class I (dashed line) 
correlations observed by Curtis et al. (2010) are over-plotted.
The three single-dish detections in this paper are generally more consistent 
with the Class 0 correlation than the Class I correlation.  
In general, however, when VeLLO outflows from the literature
are included, single-dish detections span both the Class 0 and Class I correlations.
The VeLLOs that are only detected
interferometrically (blue squares are L1014-IRS and L1148-IRS) 
are more consistent with the Class I correlation.  This is a fair
comparison since outflows from both VeLLOs were searched for and
not detected
with single-dish telescopes (Crapsi et al. 2005;
Kauffmann et al. 2011, this work - L1148 (032) ).  
This result is consistent with the conclusions drawn by 
Shirley et al. (2007) that L1014-IRS is a more evolved protostar
which was detected with \textit{Spitzer} while in a low
accretion state.

The time average accretion rate, time-averaged accretion luminosity,
total outflow mass, and dynamical time of the 
outflow are also plotted in Figure 19.  
All of the detected VeLLO outflows are less than $10^5$ years old.
There is a strong positive correlation between the observed outflow mass
and the observed dynamical time of the outflow.   
More luminous sources also tend to have higher mass outflow rates 
(and therefore higher accretion rates) although the correlation is less well
defined.  This is not surprising since the accretion luminosity
is linearly proportional to $dM/dt$.

The time-averaged accretion luminosity onto
the protostar is larger than the observed internal luminosity
for every single-dish detected protostar except for one 
source (MMS 126).  
The median ratio of the outflow-derived, time-averaged 
accretion luminosity and the observed internal luminosity of the protostar 
at the current epoch (current meaning when observed by \textit{Spitzer})
for all VeLLO outflows detected with single-dish telescopes is $1.8$.
The average value of the ratio is $17$ but is significantly 
skewed by IRAM04191 which has a ratio of $89$ (Andr\'e et al. 1999).  
This result is direct evidence for time-variable accretion.  
As we shall discuss in the next
section, there are several caveats that must be addressed, but
the $L_{acc}/L_{int}$ values plotted are likely lower limits and
should only increase when corrections for optical depth,
excitation, and inclination (Figure 12) are made.

These results may be directly compared to theoretical models of mass accretion.
One such model is an episodic accretion for VeLLOs where accretion bursts 
are short-lived events precipitated by 
instabilities in the protostellar disks.  The recent paper of Dunham \& Vorobyov (2012)
investigate this scenario using hydrodynamic models of disks with gravitational 
instabilities (Vorobyov \& Basu 2010) coupled with radiative transfer models to analyze
the time evolution of the observed protostellar core.  
In reality, the instability does not have to be purely gravitational in origin
but also may be magnetic in origin
(e.g. MRI Armitage et al 2001, Zhu et al. 2010, Simon et al. 2011; Tassis \& Moushovias 2005).
The general nature of the variation
of the mass accretion rate with time is an exponentially decreasing function punctuated 
with short-lived bursts that vary by two orders of magnitude 
(see Figure 6 of Dunham \& Vorobyov 2012).  
The average mass accretion rate in their example model is $3.7 \times 10^{-6}$ \msun /yr
over the first $10^5$ years after the formation of the protostar.  This is a factor of
$4.2$ times higher than the typical accretion rate in their example model at $10^5$ years.  
These numbers are very similar to the observed ratios of the accretion luminosity
derived from the outflow properties to the observed protostellar luminosity in this
paper.  This is only a single example model and models with a range of differing
initial conditions would produce a wider range of average accretion rate ratios.
With the caveats in mind which indicate that the observed $L_{acc}/L_{int}$ and therefore
$<\dot{M}>/\dot{M}_{now}$ is likely underestimated, this result
is consistent with a picture of episodic accretion; but other smoothly varying
accretion models, such as a model with a purely exponential drop in the accretion
rate with time, cannot be ruled out by these observations alone.  Advantages
of the episodic accretion model are that it is physically motivated and that it is also 
successful in explaining the distribution of protostars in a bolometric luminosity
- bolometric temperature diagram (i.e. it physically explains the original
"luminosity problem"; see Figure 8 of Dunham \& Vorobyov 2012).  The results from 
this outflow survey provide additional supporting evidence but not
definitive evidence of the episodic accretion model.

\subsection{Caveats}

Despite our sensitive maps in \co\ and \coo\ J = $2 \rightarrow 1$, it may
still be very difficult to distinguish the terminal bow shock of
the outflow in a complex kinematic environment. Indeed, many of our
protostellar candidates are situated in very confusing regions
where the observable signature of a weak outflow can easily be lost on large
scales.
In this sense, the dynamical time is a lower limit 
to the actual age of the outflows.  
Parker (1991) and Fukui (1993)
have statistically studied outflow ages and compared to the dynamical
time and found an up to order of magnitude discrepancy (see Curtis et al.
2010 for a discussion).  
The impact of this underestimate of dynamical time on the outflow force
is more complicated because more mass would also be swept up by
the larger outflow.  Therefore, the outflow force is unlikely to decrease
in direct proportion to the increase in dynamical time.
The effect of a underestimate of
the outflow terminal bow shock on the estimates of the average accretion luminosity 
is also more complicated.
If the outflow extends beyond the region estimated from our maps,
then the larger outflow age means that more mass has been accreted.
However, that change coupled with an increase in the estimate of the dynamical age 
means that the overall
change in the average accretion luminosity may not change drastically
from what we have estimated in this paper.

This caveat must be weighed against the uncertainty in the
outflow optical depth, excitation temperature,
and inclination which will only increase the derived outflow
quantities.
For instance, if the outflow inclination is increased from $34.3$\degree\ 
to $70$\degree ,
then the outflow force would increase by a factor of 10.
Typical corrections for higher excitation temperatures or
optical depths are additional factors of a few.  Throughout the
calculation of outflow parameters in \S3, the lower limits were consistently
chosen and $L_{acc}$ was calculated for $i = 34.3$\degree\ 
($\frac{sin i}{cos^3 i} \approx 1$) which corresponds
to an $83$\%\ confidence interval that the real outflow inclination is larger.  
Given these factors, it seems very unlikely that the outflow forces plotted for 
VeLLOs in Figure 19 will decrease.

We find at least 
factors of 1.5 and 2 difference between the average accretion rate estimated from outflow 
properties and the observed protostellar luminosity.  
These ratios are significant at the $> 2\sigma$ level for all three sources.
The typical optical depth corrections (factors
of a few) and larger inclinations (also factors of a few) will only tend
to increase this difference.
The accretion luminosity
itself is estimated from very rough values of the outflow jet speed,
entrainment efficiency, mass loss rate in the wind versus
the mass accretion rate, and assumed protostellar radius.  A 
simultaneous change in several of the 
the assumed parameters listed above (e.g. lowering the typical jet speed from
150 km/s to 75 km/s, increasing the ratio of $\mathrm{\frac{\dot{M}_{w}}{\dot{M}_{acc}}=0.2}$,
and increasing the protostellar radius from 3 to 9 $R_{\odot}$) 
would be required to reconcile the two luminosity estimates for typical
optical depth, excitation temperature, and inclination effects.

\section{Conclusions}

We have systematically surveyed 30 VeLLOs and 9 low-luminosity objects
in \co\ and \coo\ J = $2 \rightarrow 1$ with the
HHT from the Dunham et al. (2008) catalog of candidate low-luminosity protostars in 
nearby ($D < 400$ pc) clouds and cores. 
From five-point maps, we identified five sources as new candidates outflows and 
made extended on-the-fly \co\ J=2 $\rightarrow$ 1 maps of these regions.
Only B59 (023) showed evidence of a candiate outflow, despite 
the red wing being obscured by emission from IRAS 17081-2721.
The remaining sources do not show convincing evidence for a large
scale outflow in \co\ maps.  When combined with the grid maps for
all 39 sources, these results indicate that the distinct kinematic signatures of
large-scale outflows are rare ($13$\% single-dish detection rate for VeLLOs) 
toward sources in the Dunham et al. (2008) catalog.
Every candidate protostar from the Dunham catalog with a single-dish detected 
outflow belongs to the highest confidence group classifications (groups 1, 2, and 3).
We also re-mapped two previously detected VeLLO outflows (L673-7 (031) and
L1251-A (045)) to find
the full extent of the outflows and better determine their outflow properties.
Our results indicate that the time averaged accretion luminosity implied
by the outflow properties is at least a factor of two larger than the observed
internal luminosity.  
For larger inclinations of the outflow axis, higher gas excitation temperatures, 
and optical depth corrections this discrepancy will increase. 
This result is consistent with a picture of episodic accretion for
these objects.

In light of the results from our single dish survey, interferometric 
observations are needed to complete the
search for VeLLO outflows. 
Interferometric observations have the advantage of spatially filtering 
out much of the large scale structure that is potentially causing confusion 
in our single-dish survey.  The combined results from an interferometric
survey and this single dish survey will provide a census of outflow properties
and statistics for VeLLOs that will illuminate the history of accretion towards these
sources.

\section*{Acknowledgments}

We thank the staff of the Arizona Radio Observatory including the
operators Bob Moulton and John Downey for their assistance with
the observations.  Partial funding for KS was provided by the
Arizona Space Grant Consortium.   YLS is partially supported by
NSF grant AST-1008577.


\newpage




\begin{figure}
\figurenum{1}
\epsscale{1.0}
\vspace*{16cm}
\includegraphics{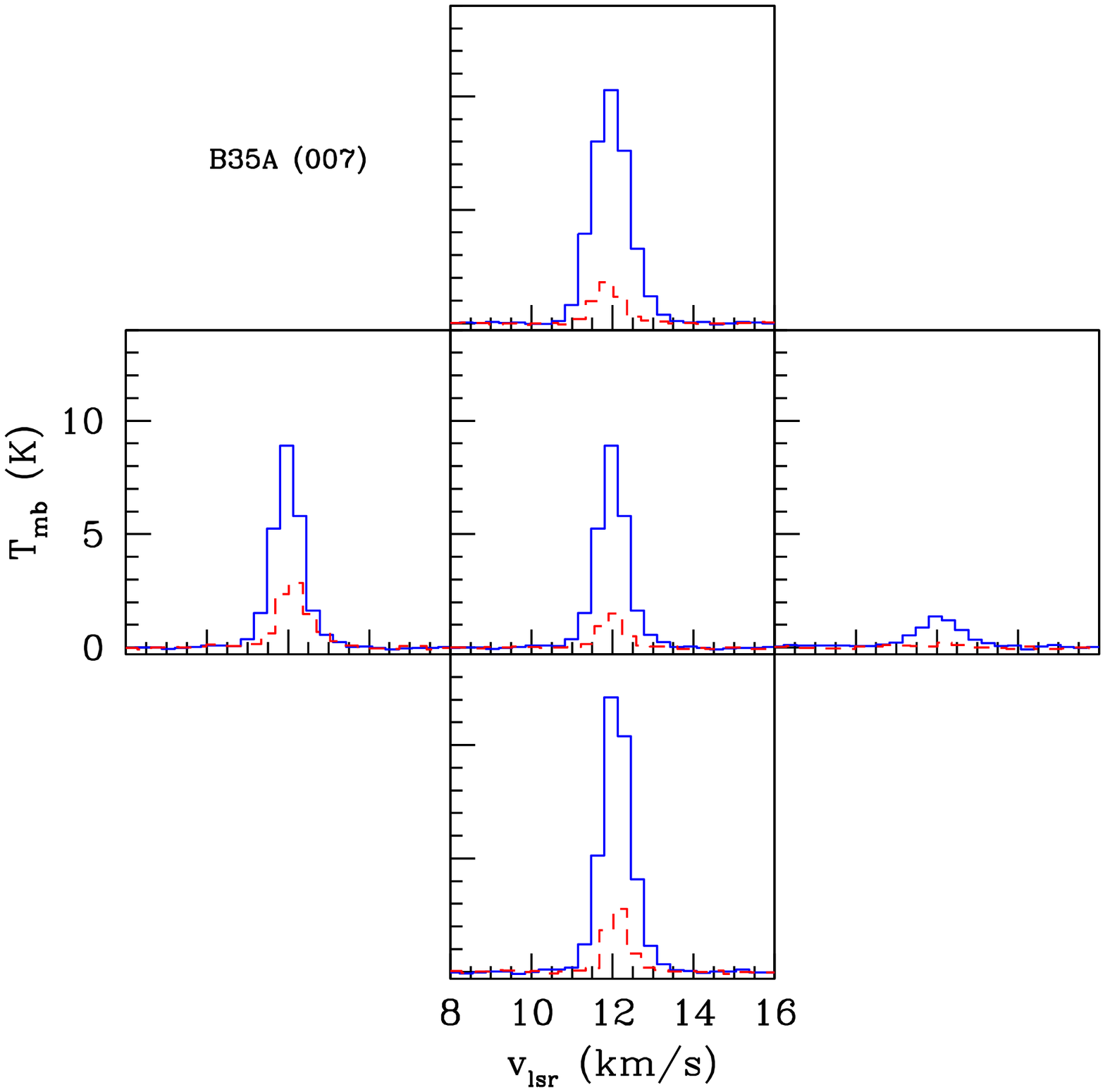}
\includegraphics{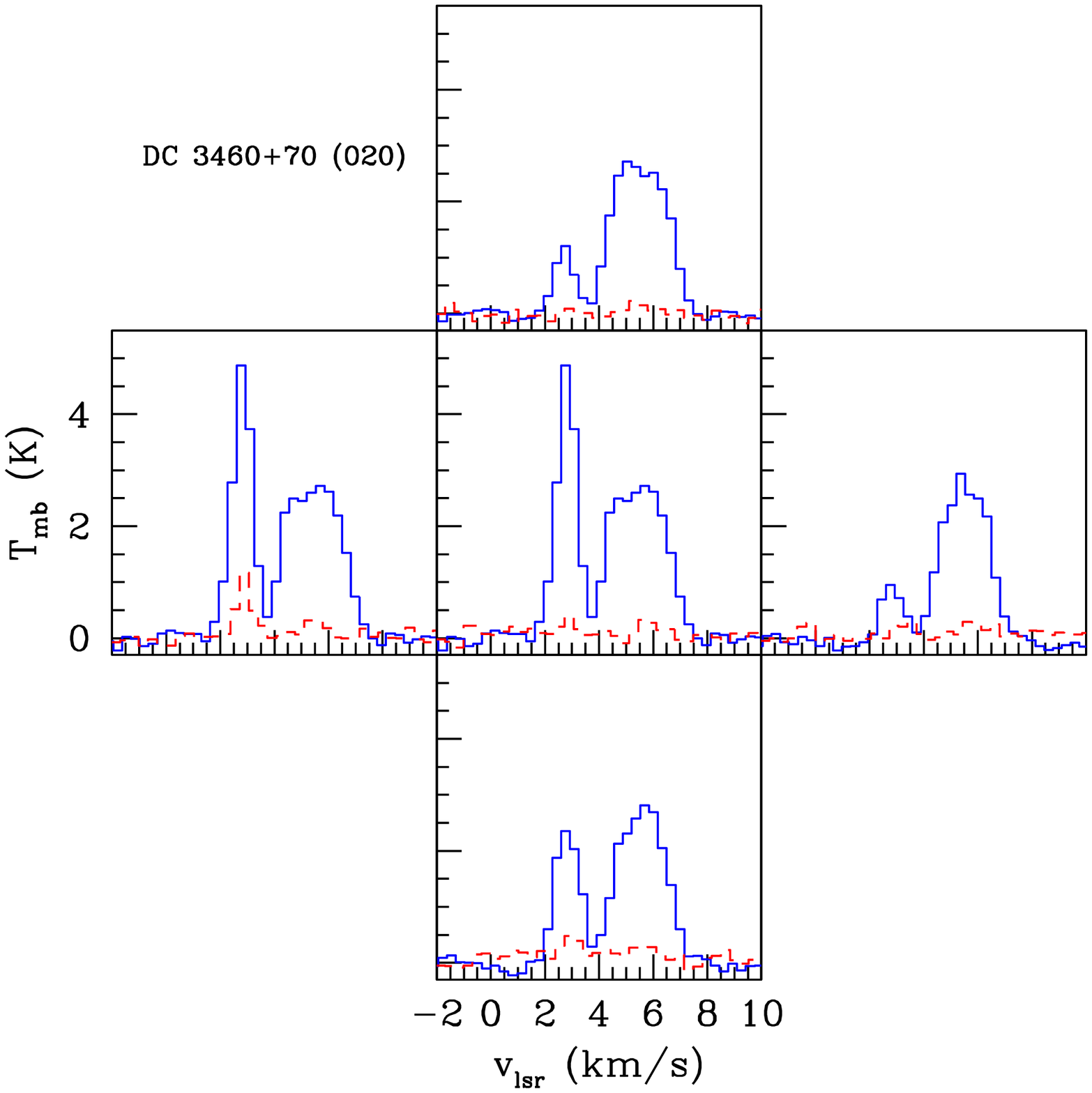}
\includegraphics{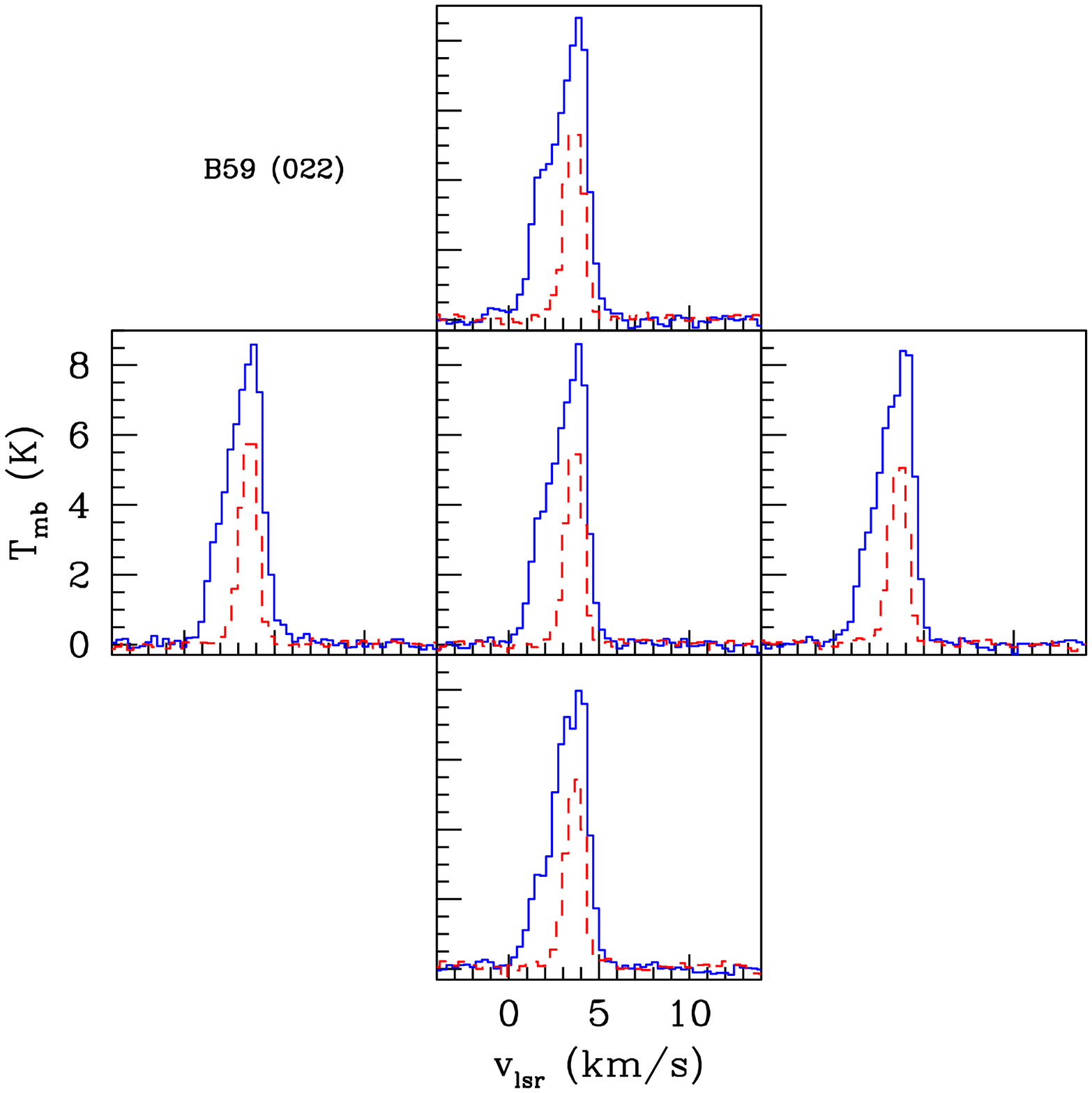}
\includegraphics{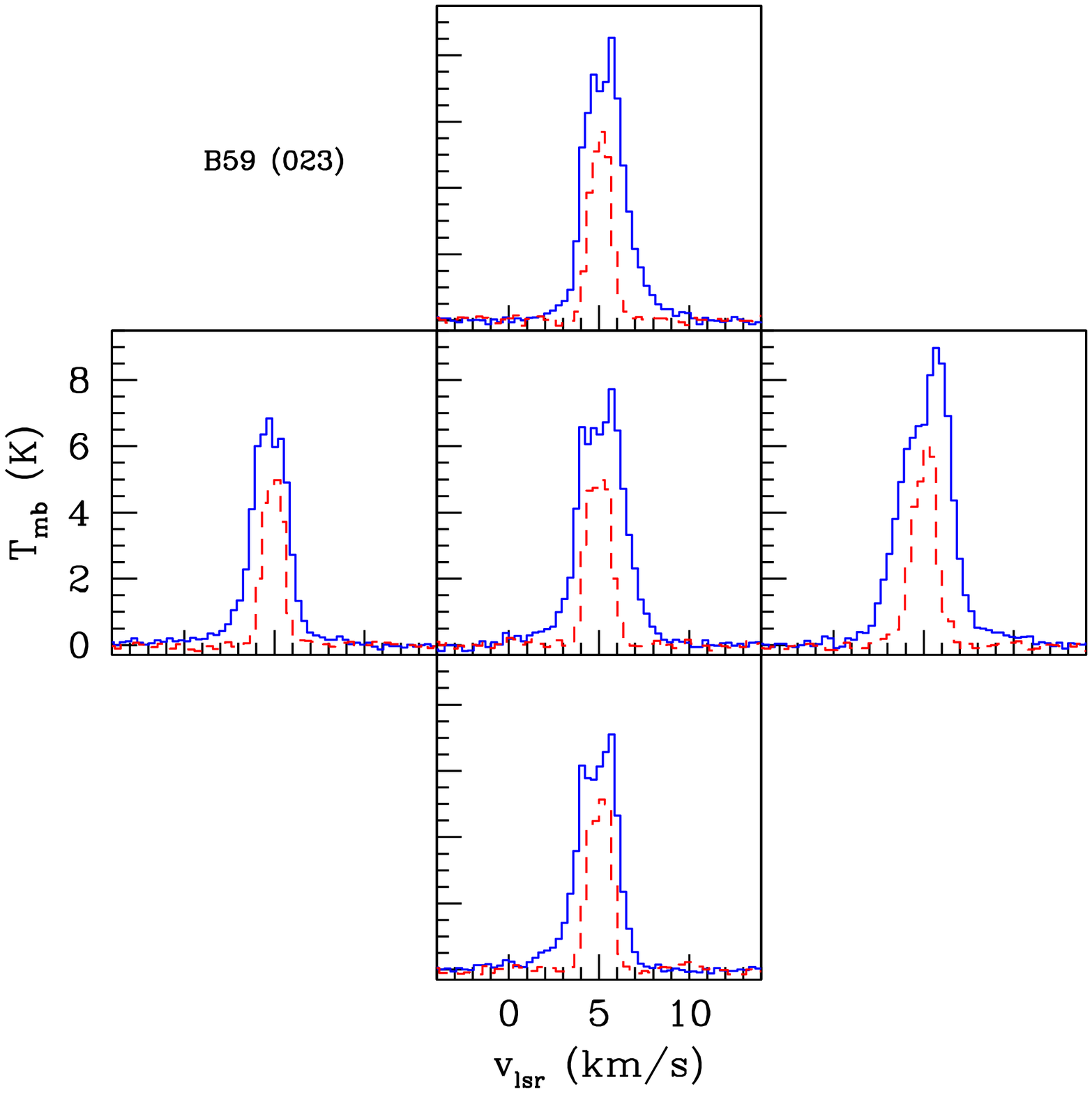}
\figcaption{5-point maps of \co\ (solid line) and \coo\ (dashed line) 
$2-1$ spectra toward the sources.  The offset positions correspond to $30$\as\ shifts
in RA and DEC from the protostar position.  The source name is listed in the upper left of each
5-point map.}
  \end{figure}


\begin{figure}
\figurenum{2}
\epsscale{1.0}
\vspace*{16cm}
\includegraphics{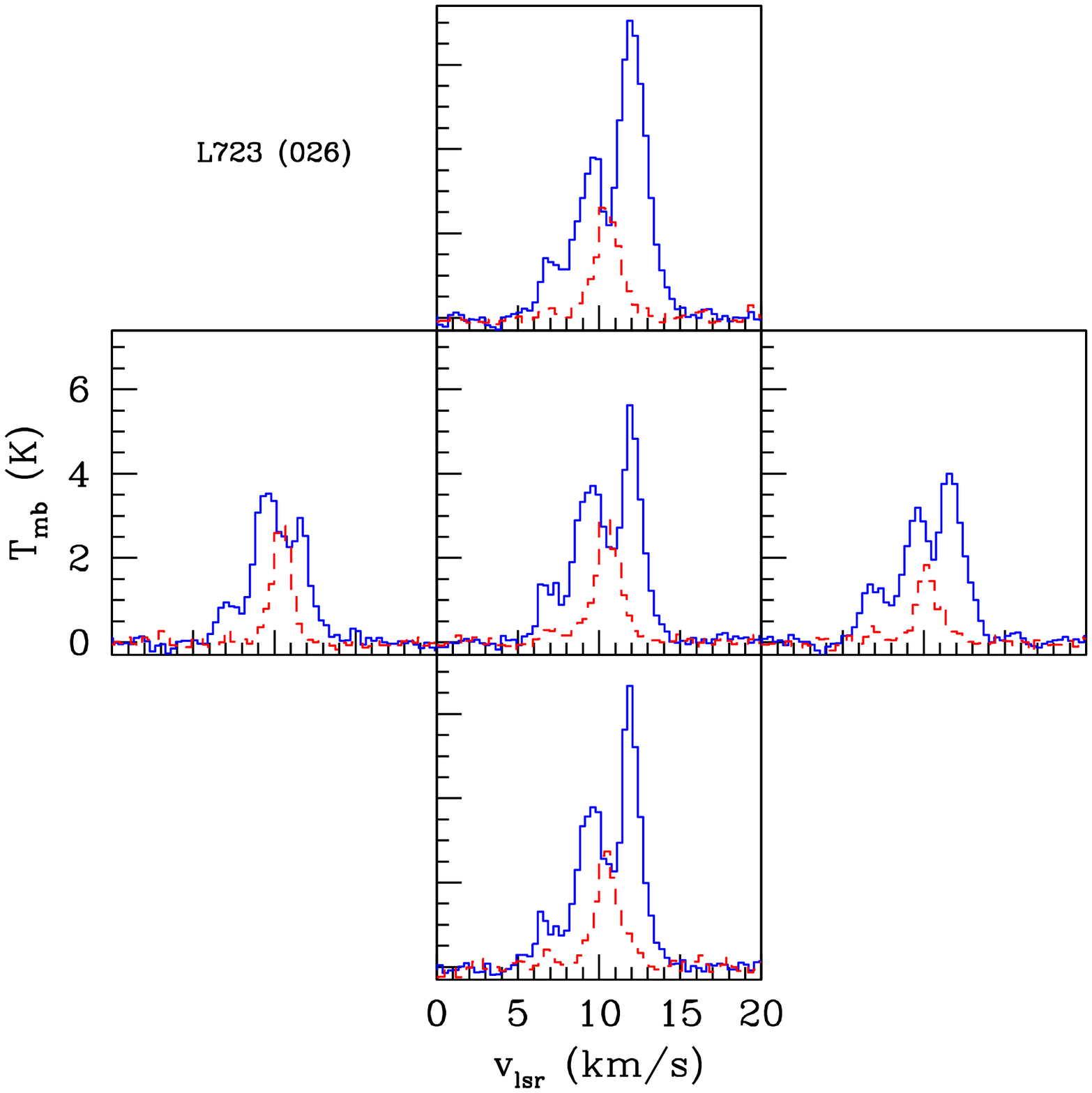}
\includegraphics{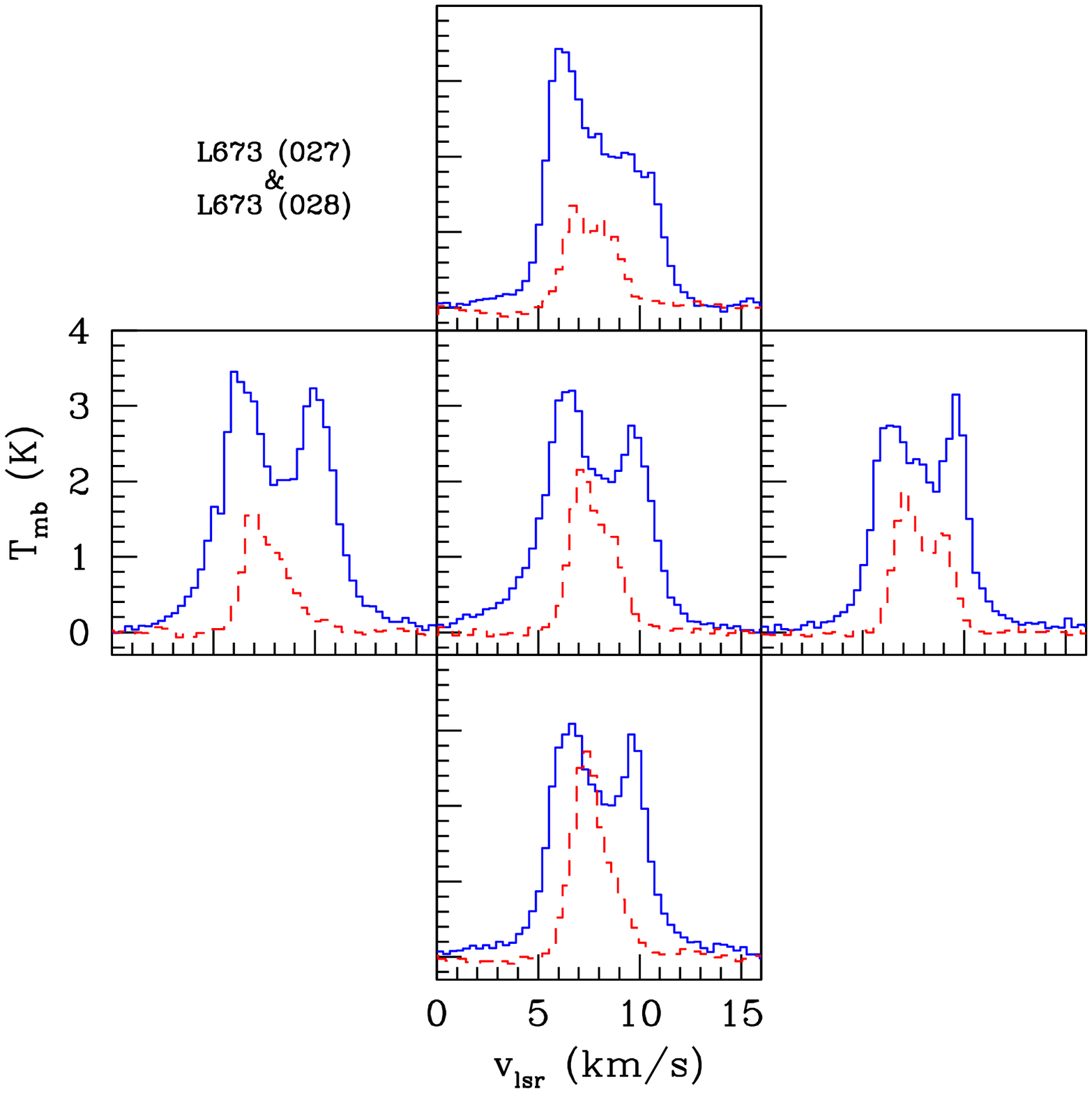}
\includegraphics{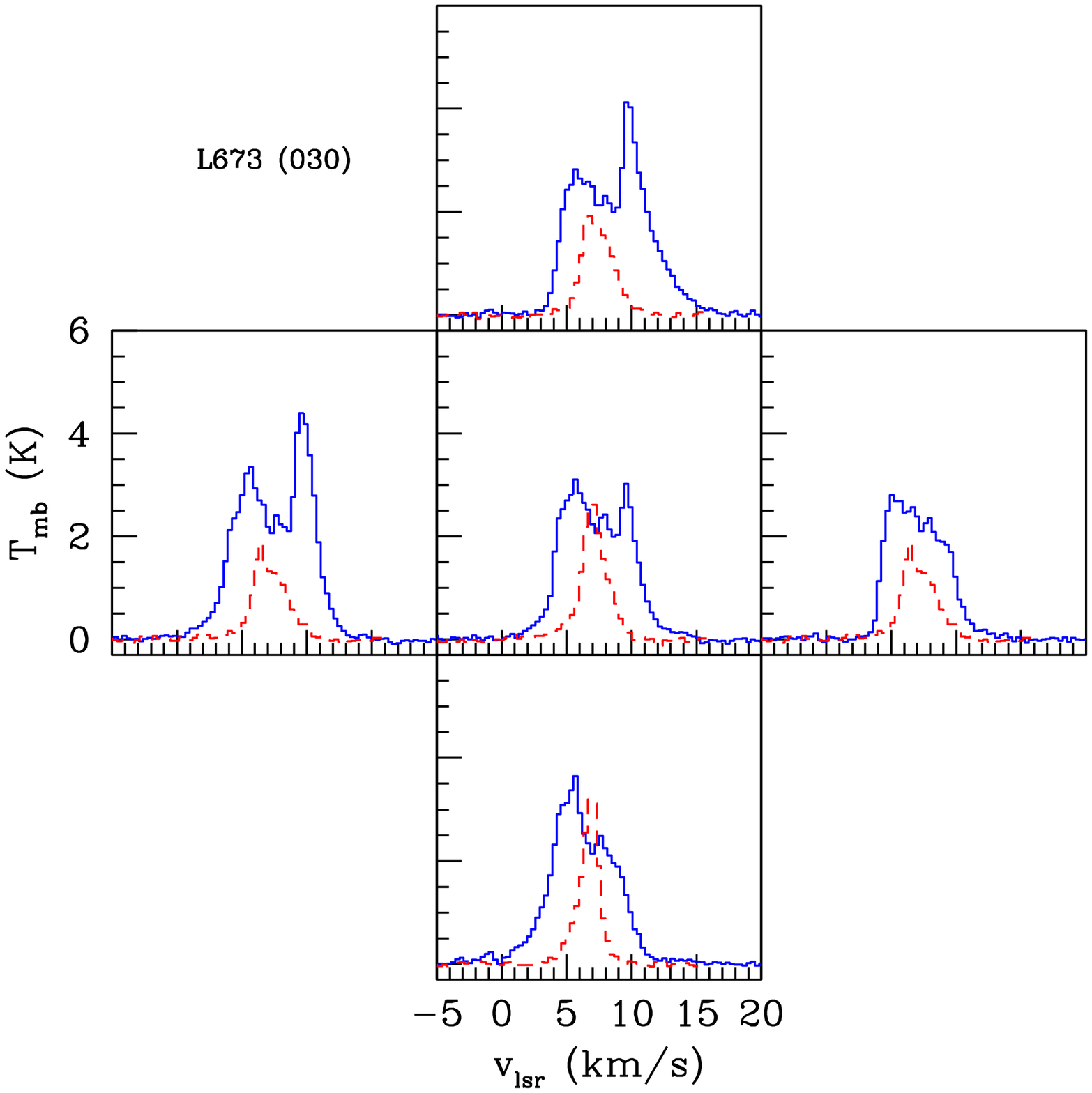}
\includegraphics{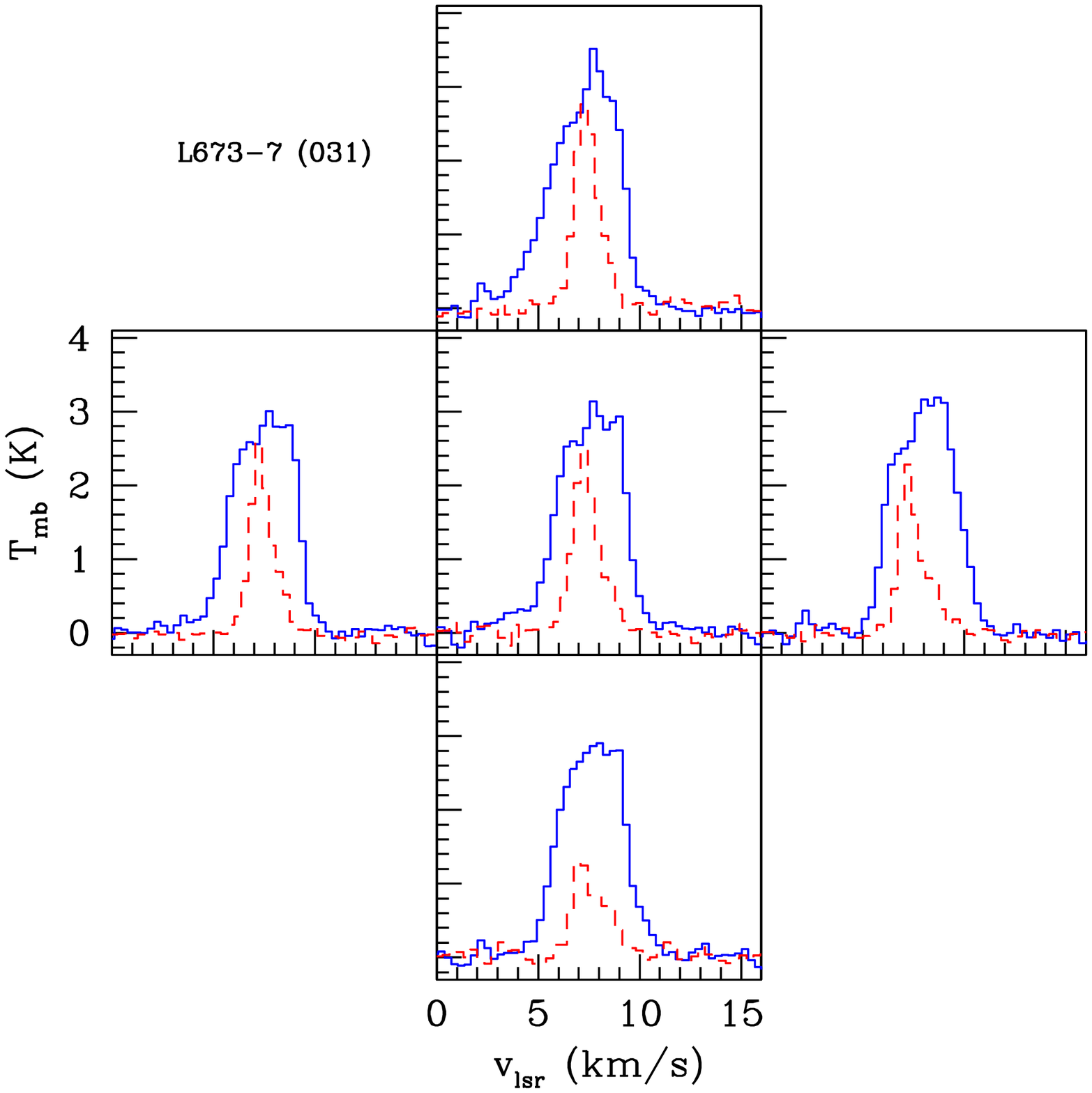}
\figcaption{5-point maps of \co\ (solid line) and \coo\ (dashed line) 
$2-1$ spectra toward the sources.  
The offset positions correspond to $30$\as\ shifts
in RA and DEC from the protostar position.  
The source name is listed in the upper left of each
5-point map.  The sources L673 (027) and L673 (028) are separated by only
9.5\as , therefore only 1 map is shown.}
\end{figure}


\begin{figure}
\figurenum{3}
\epsscale{1.0}
\vspace*{16cm}
\includegraphics{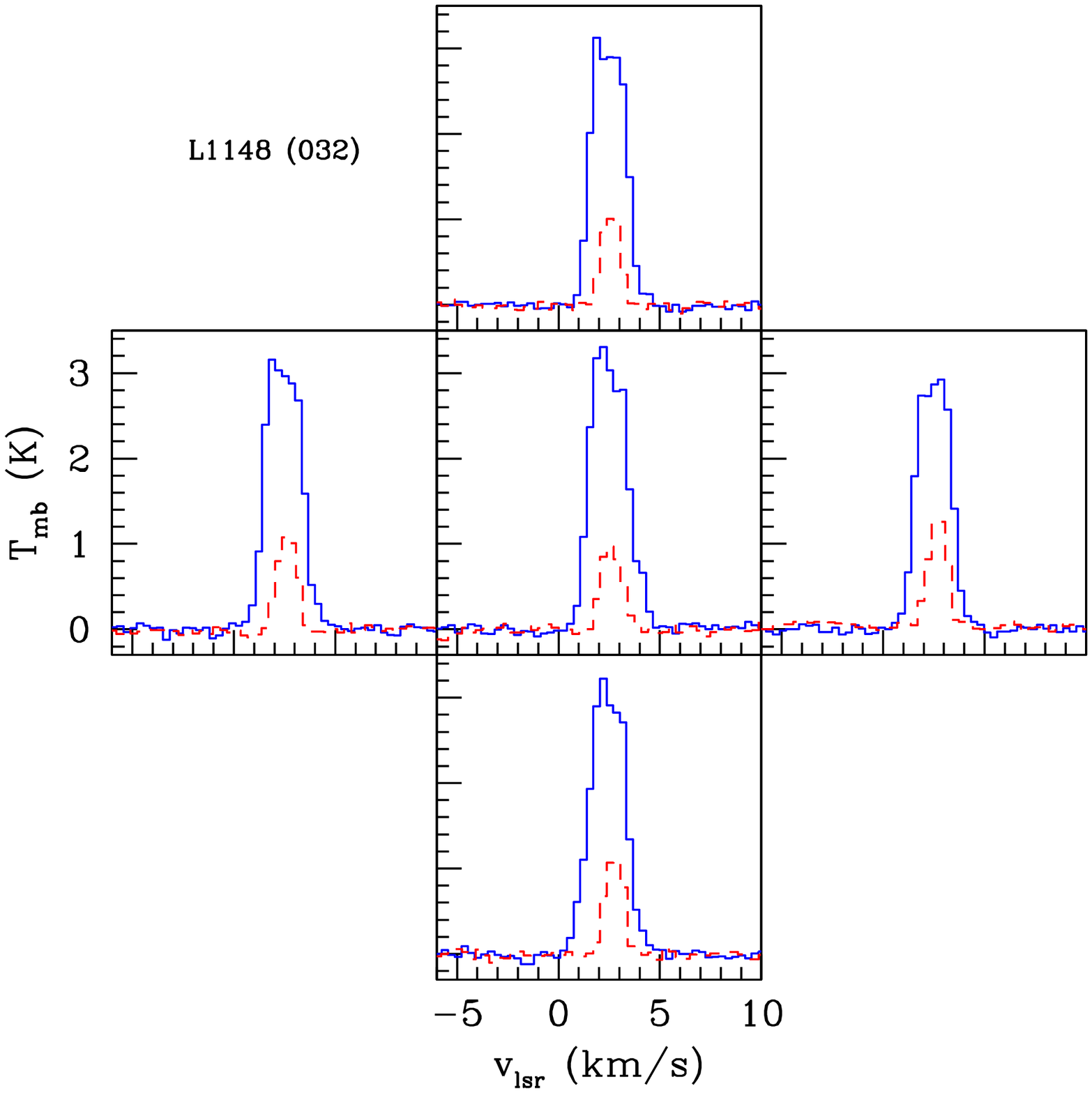}
\includegraphics{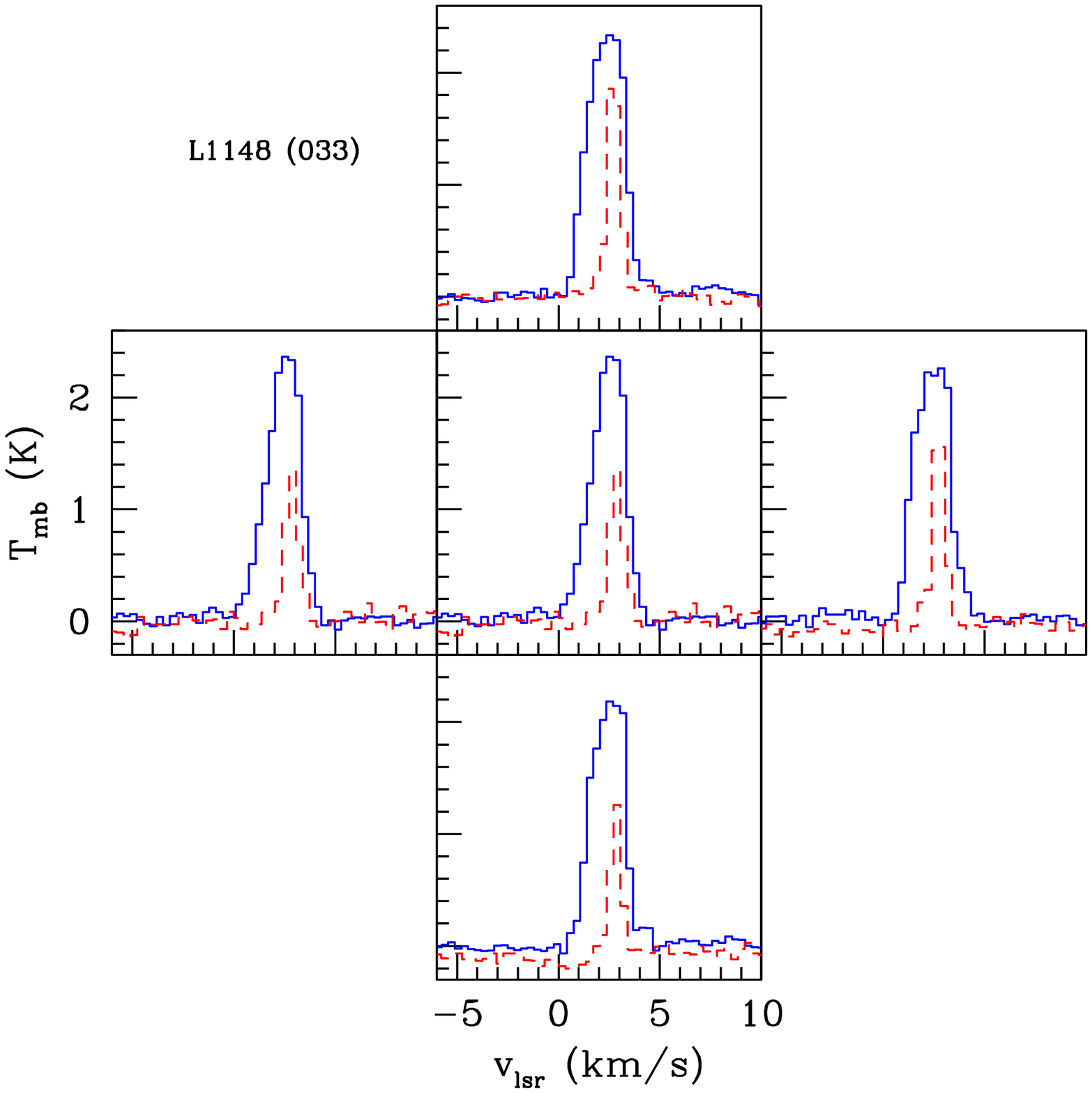}
\includegraphics{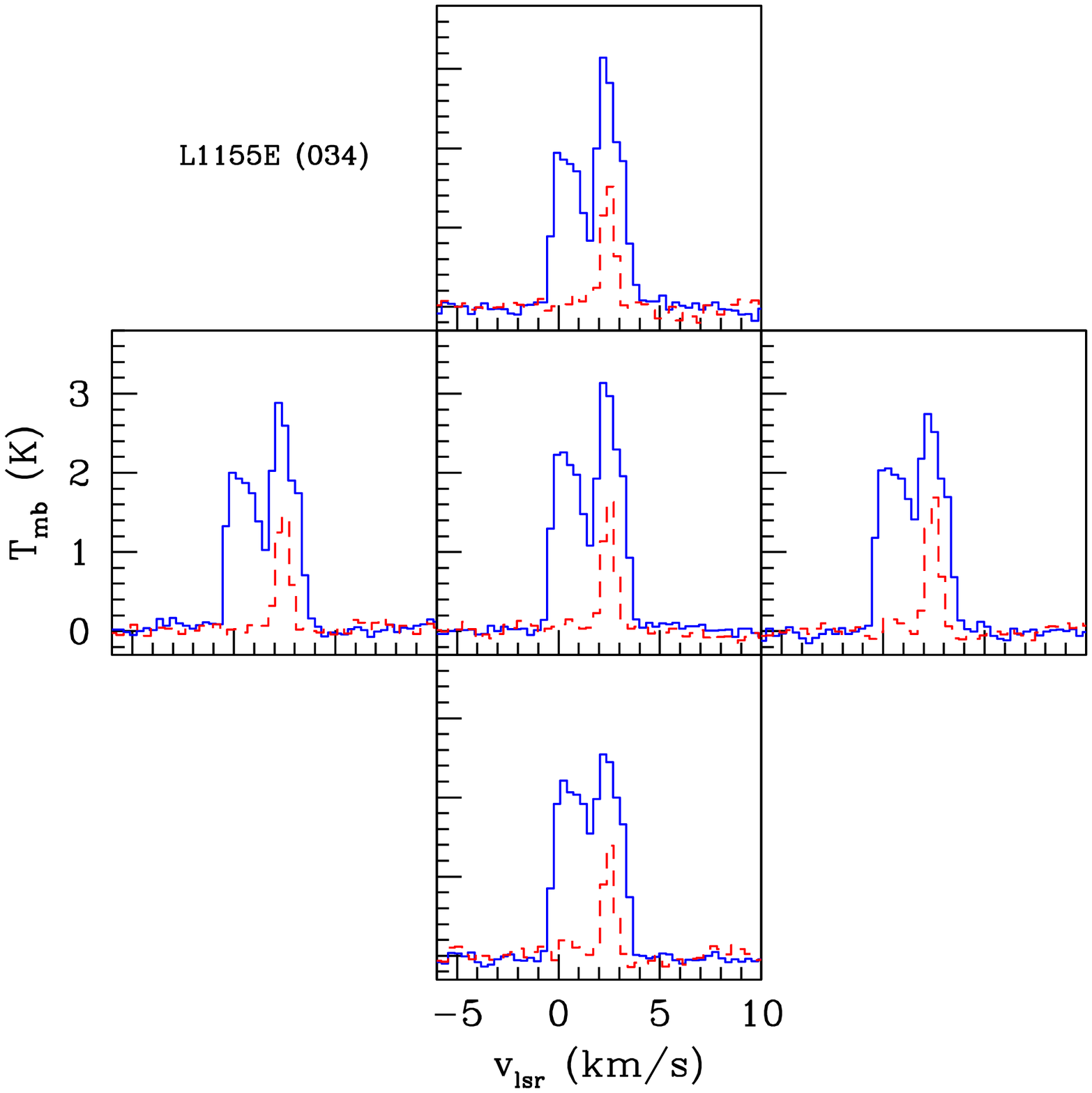}
\includegraphics{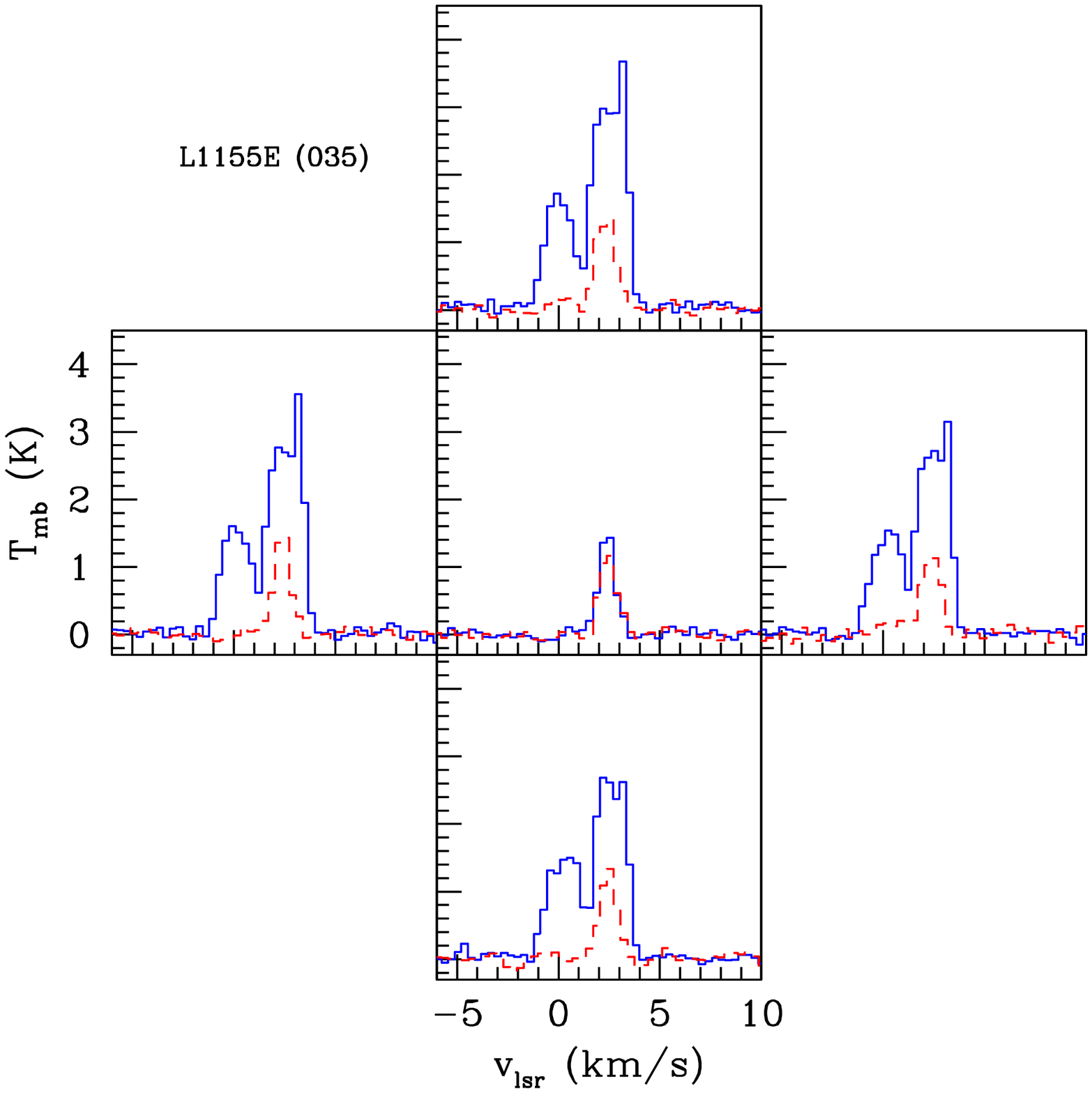}
\figcaption{5-point maps of \co\ (solid line) and \coo\ (dashed line) 
$2-1$ spectra toward the sources.  The offset positions correspond to $30$\as\ shifts
in RA and DEC from the protostar position.  The source name is listed in the upper left of each
5-point map.}
\end{figure}


\begin{figure}
\figurenum{4}
\epsscale{1.0}
\vspace*{16cm}
\includegraphics{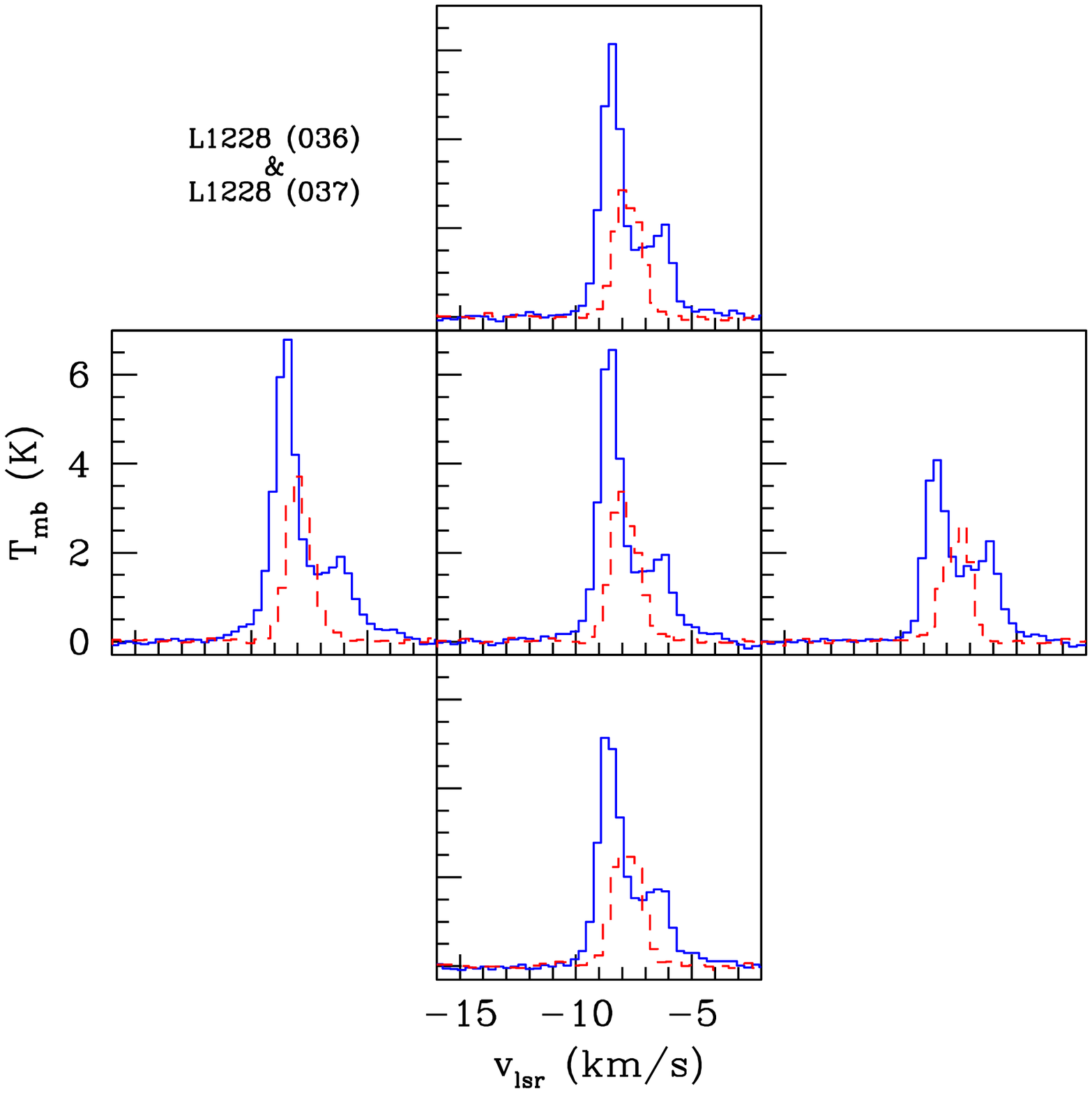}
\includegraphics{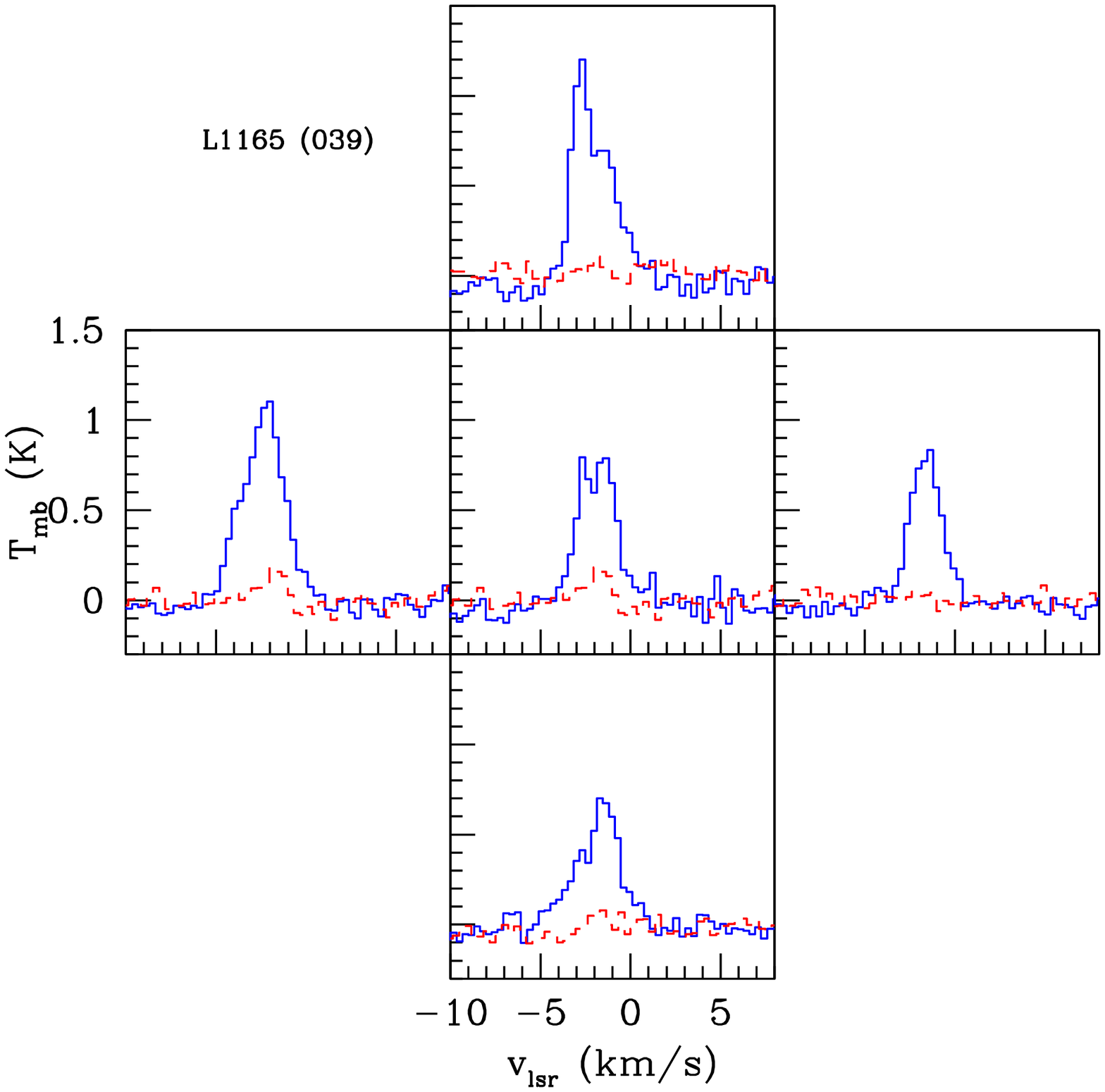}
\includegraphics{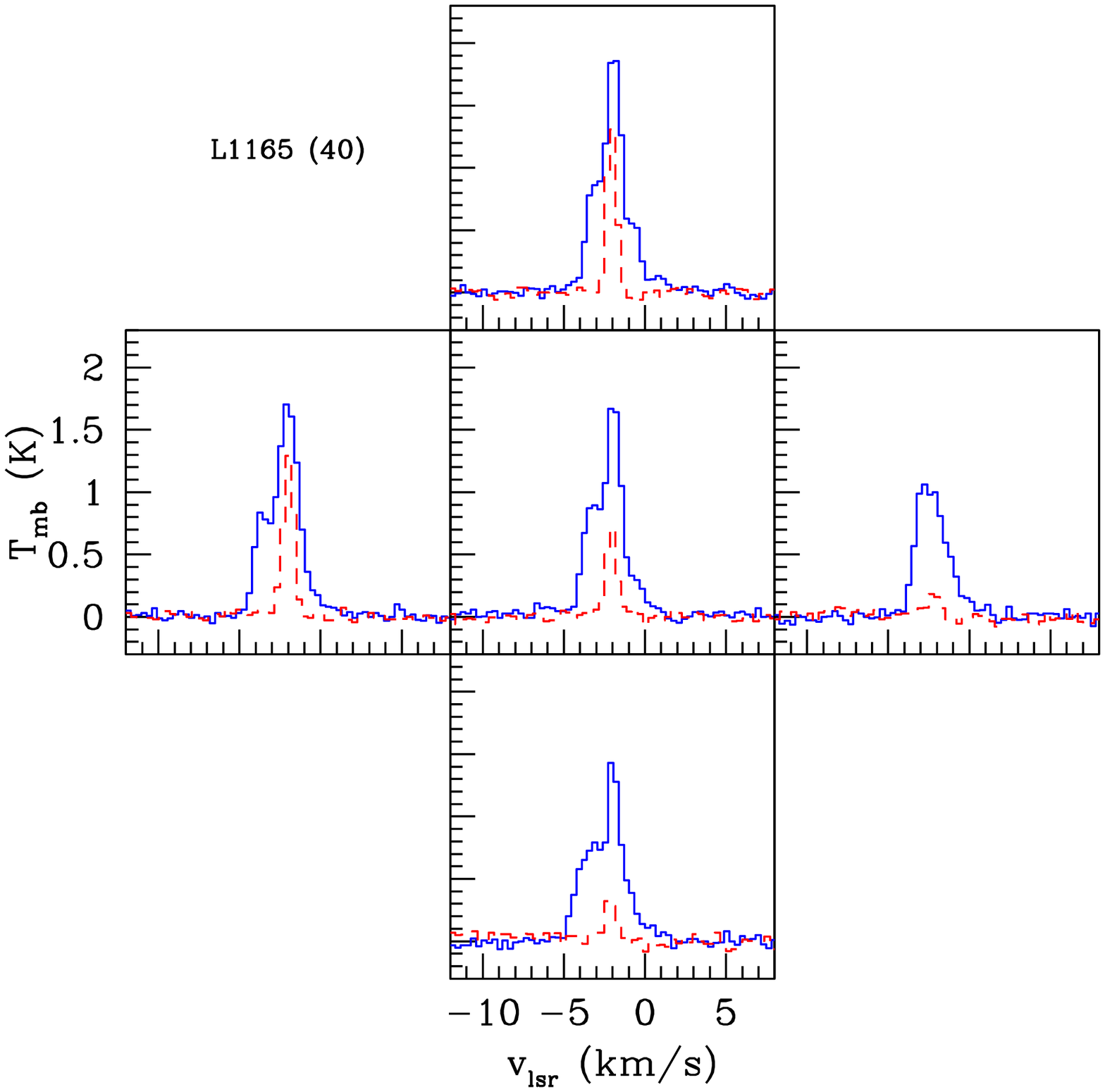}
\includegraphics{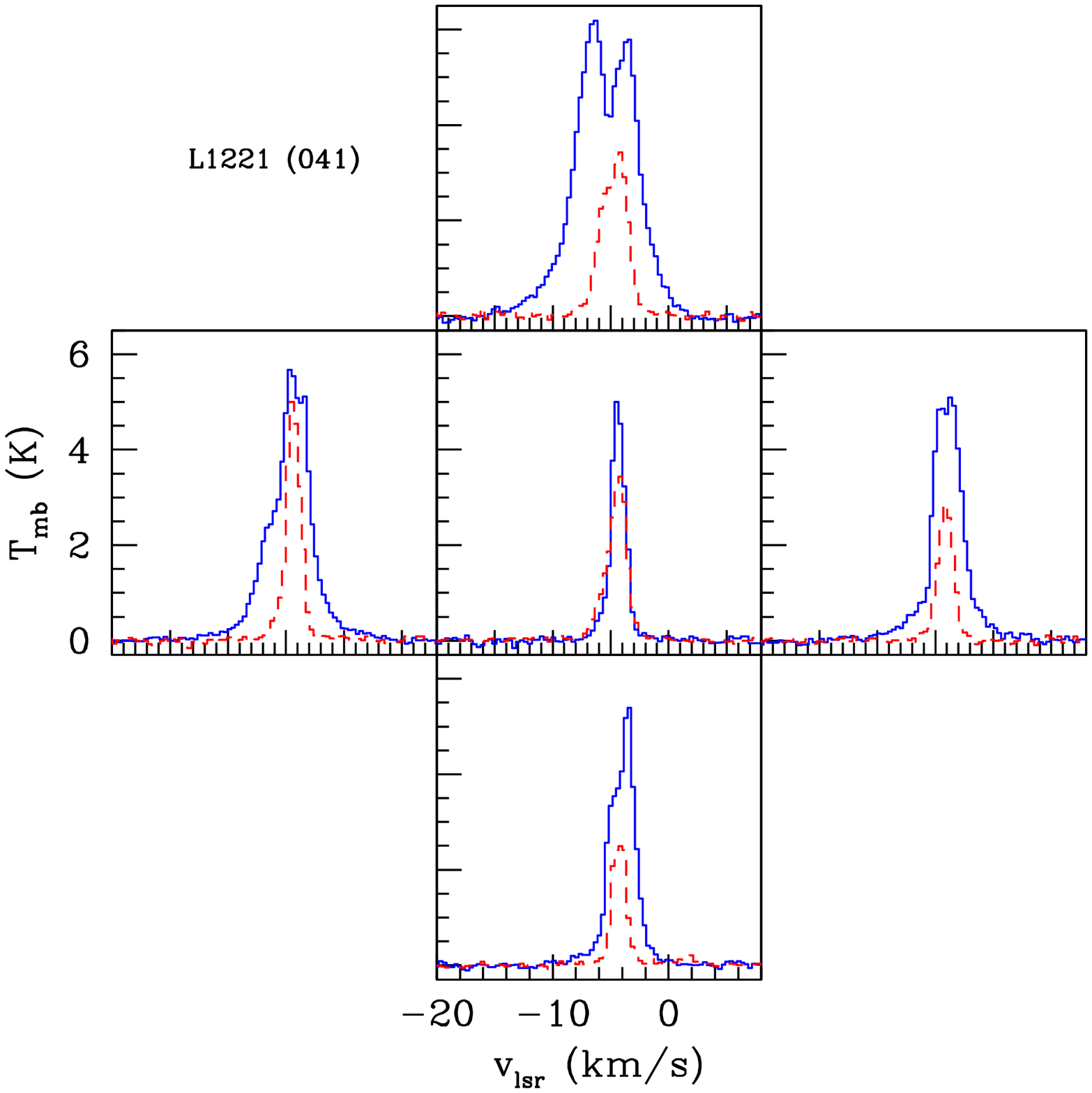}
\figcaption{5-point maps of \co\ (solid line) and \coo\ (dashed line) 
$2-1$ spectra toward the sources.  The offset positions correspond to $30$\as\ shifts
in RA and DEC from the protostar position.  The source name is listed in the upper left of each
5-point map. The sources L673 (027) and L673 (028) are separated by only
3.8\as , therefore only 1 map is shown.}
\end{figure}


\begin{figure}
\figurenum{5}
\epsscale{1.0}
\vspace*{16cm}
\includegraphics{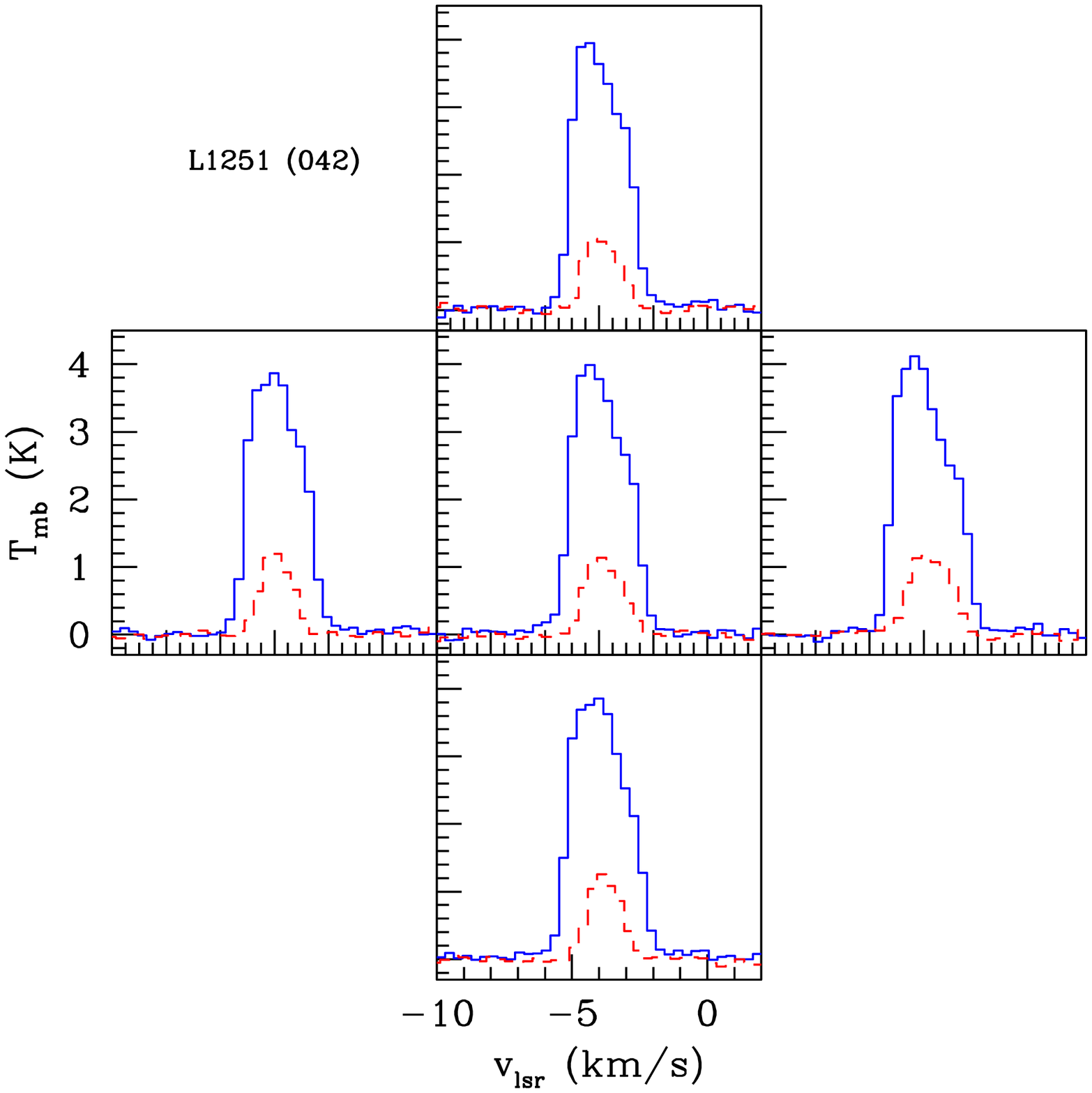}
\includegraphics{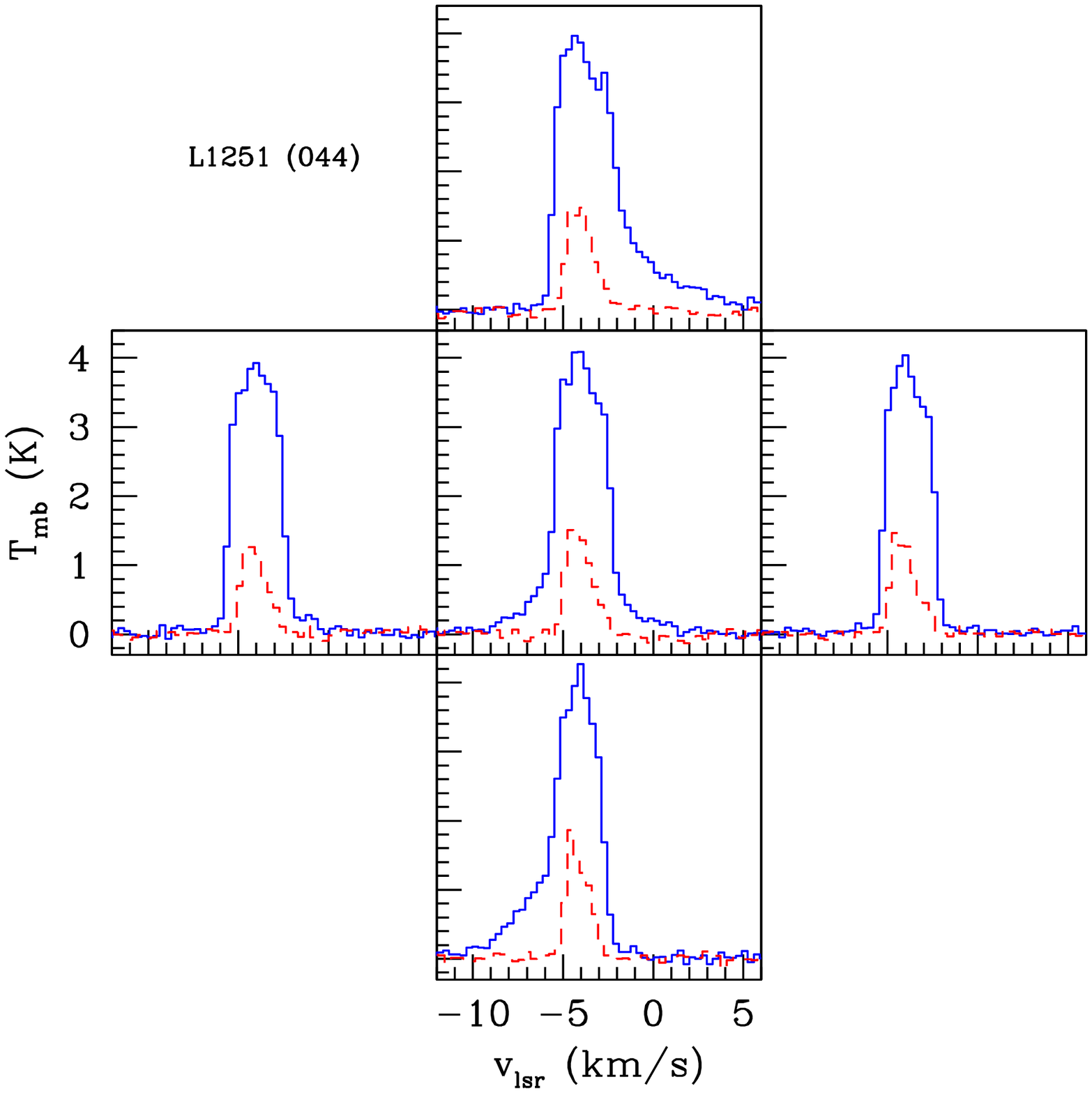}
\includegraphics{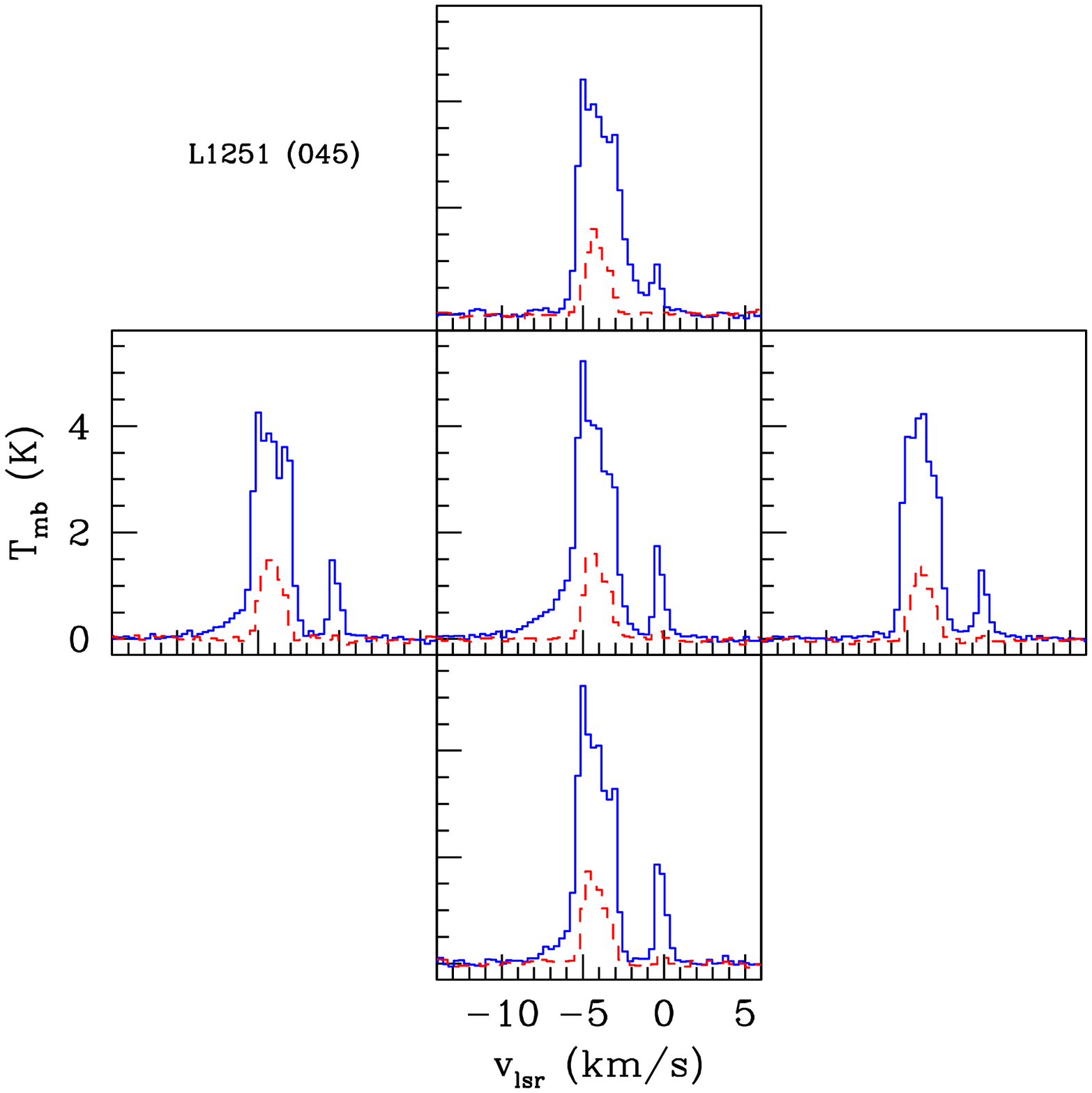}
\includegraphics{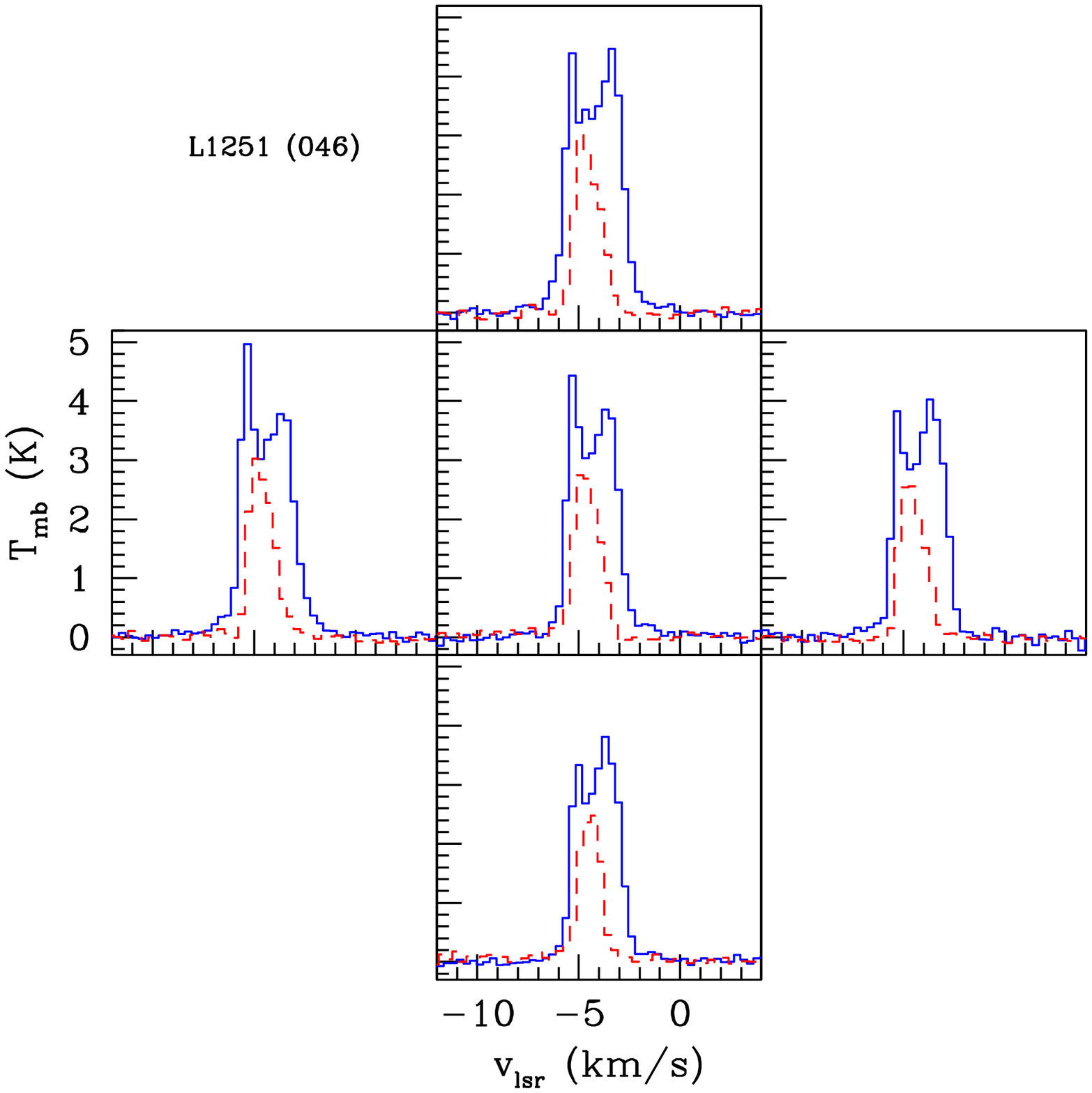}
\figcaption{5-point maps of \co\ (solid line) and \coo\ (dashed line) 
$2-1$ spectra toward the sources.  The offset positions correspond to $30$\as\ shifts
in RA and DEC from the protostar position.  The source name is listed in the upper left of each
5-point map. L1251 (044) and L1251 (045) are previously detected outflows.}
\end{figure}


\begin{figure}
\figurenum{6}
\epsscale{1.0}
\vspace*{16cm}
\includegraphics{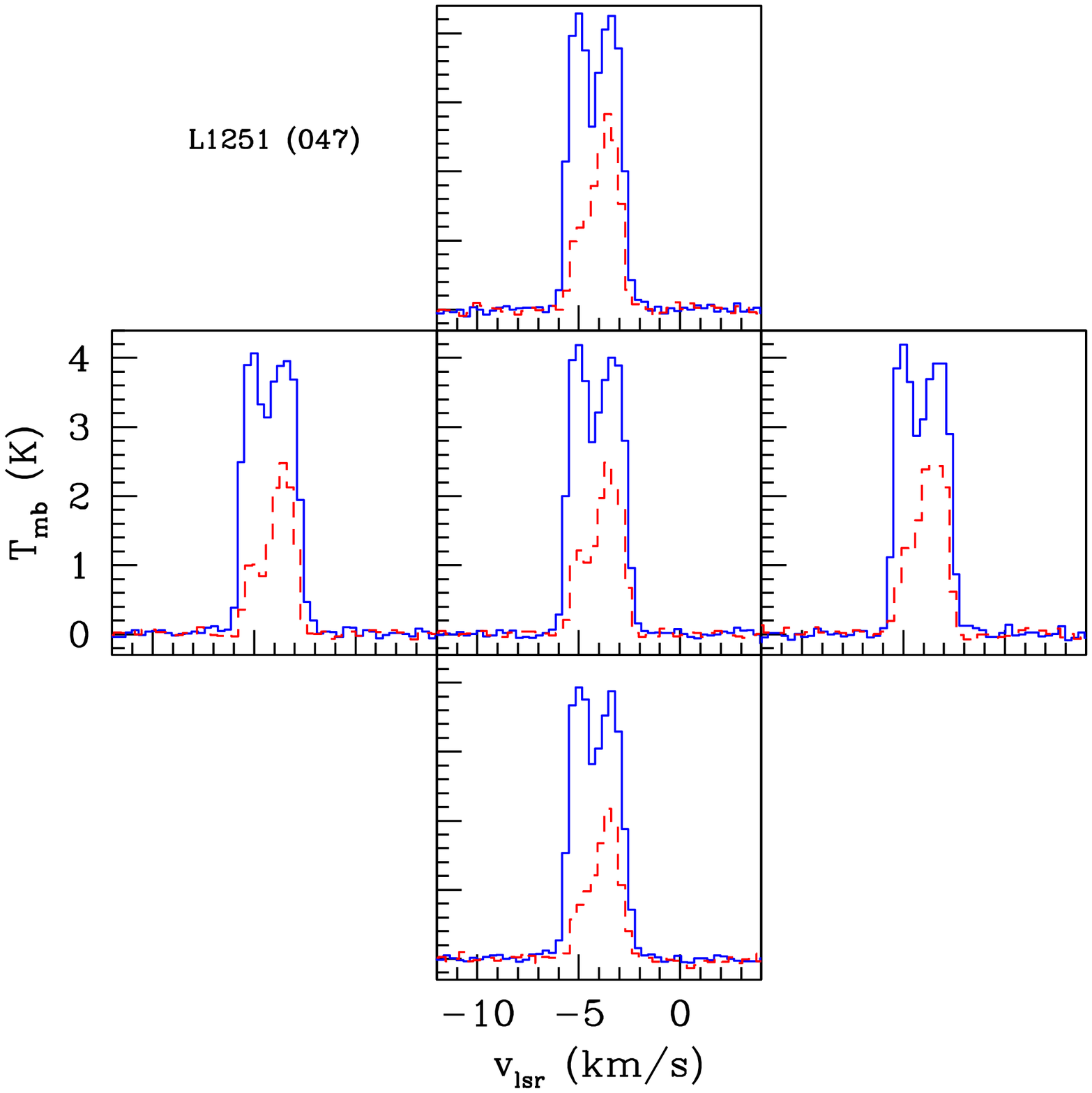}
\includegraphics{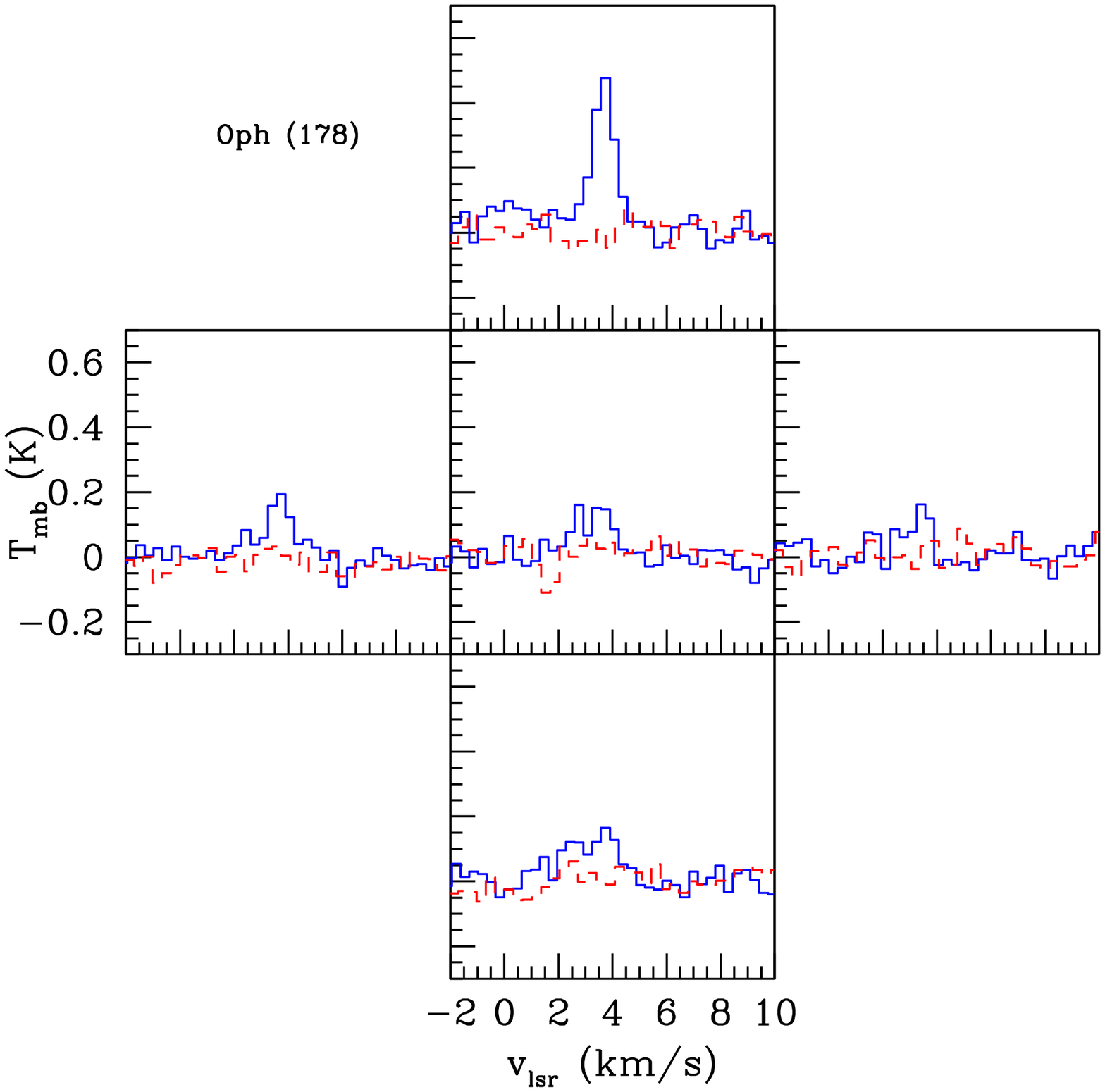}
\includegraphics{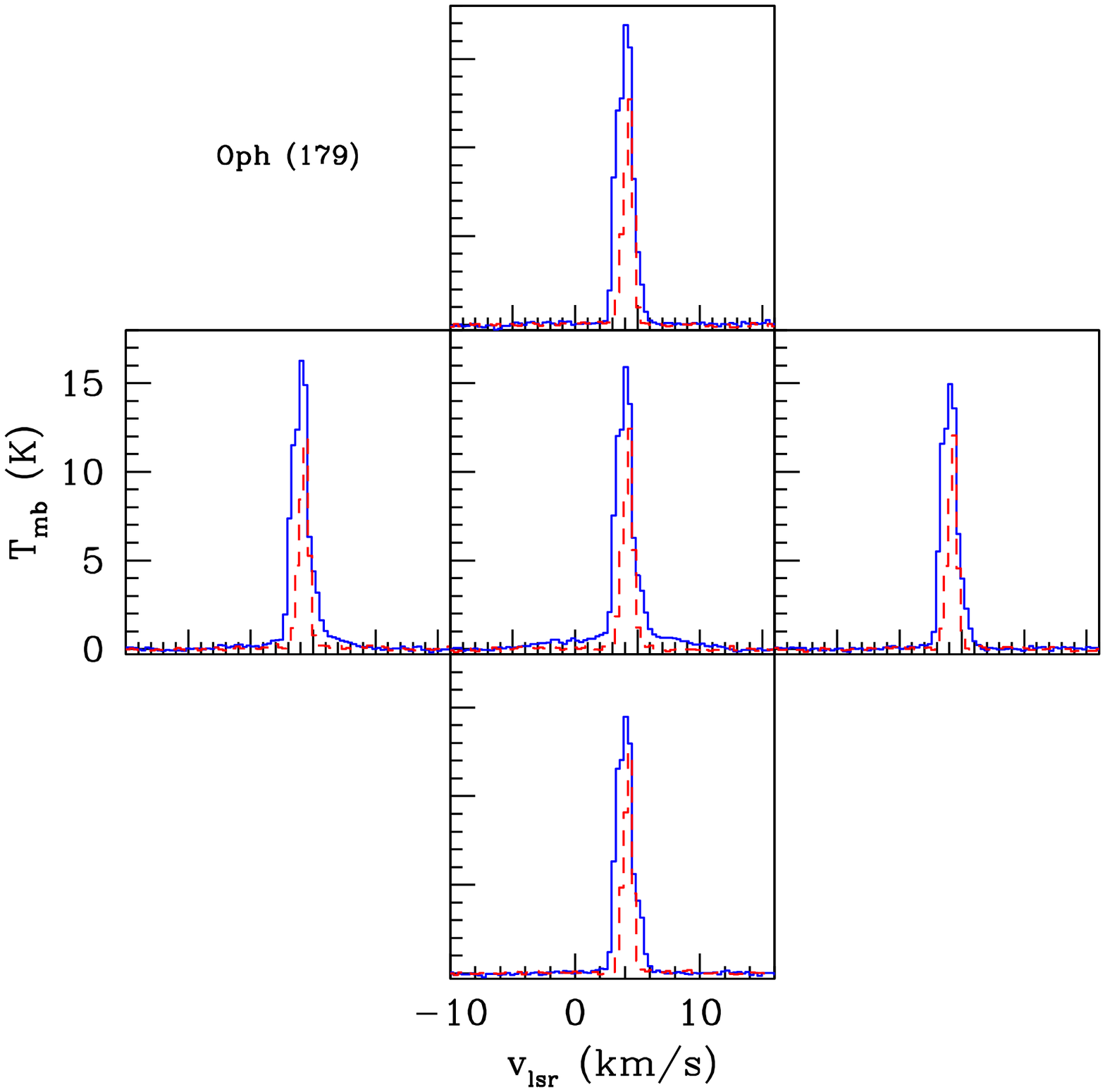}
\includegraphics{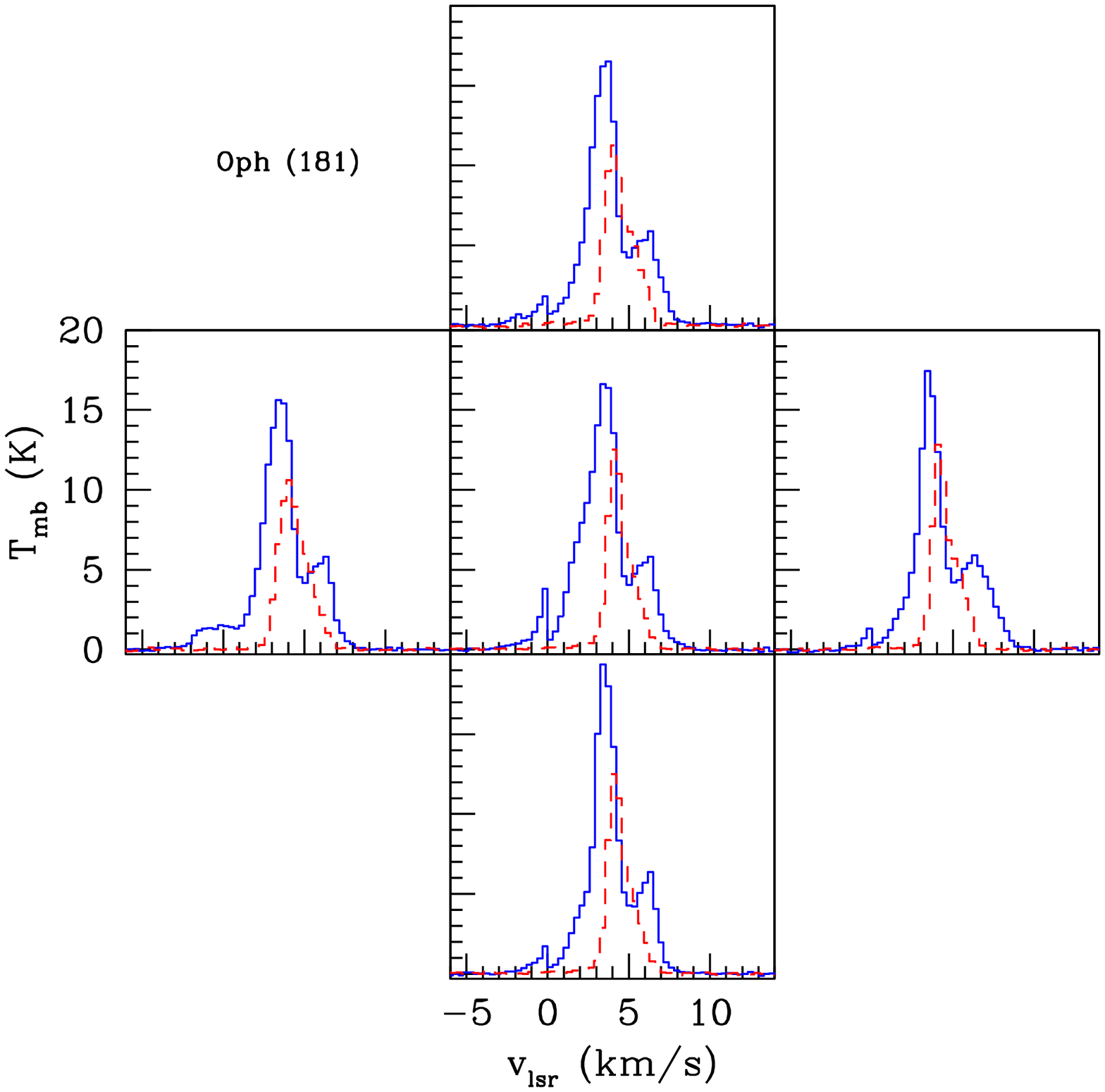}
\figcaption{5-point maps of \co\ (solid line) and \coo\ (dashed line) 
$2-1$ spectra toward the sources.  The offset positions correspond to $30$\as\ shifts
in RA and DEC from the protostar position.  The source name is listed in the upper left of each
5-point map.}
\end{figure}


\begin{figure}
\figurenum{7}
\epsscale{1.0}
\vspace*{16cm}
\includegraphics{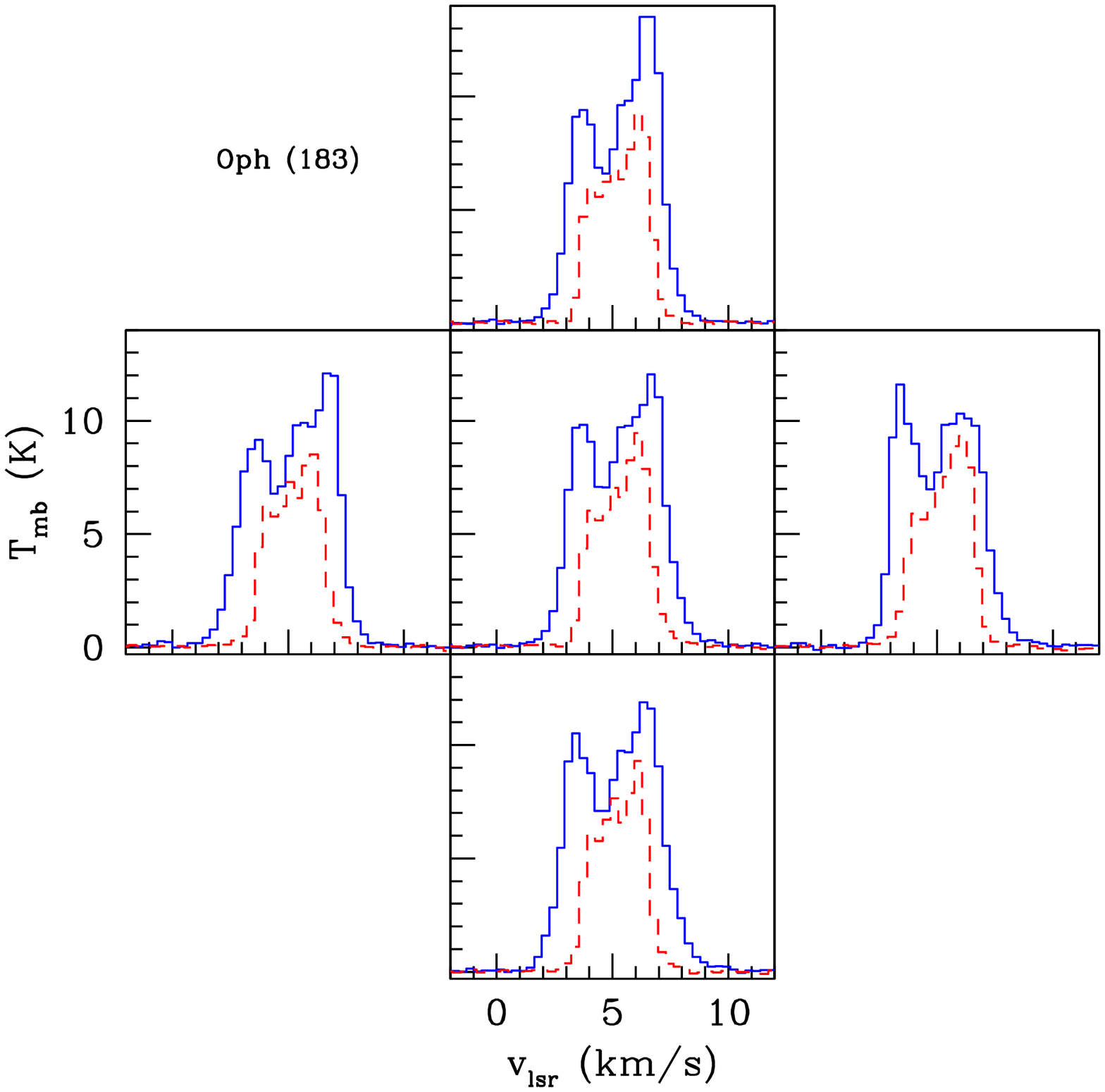}
\includegraphics{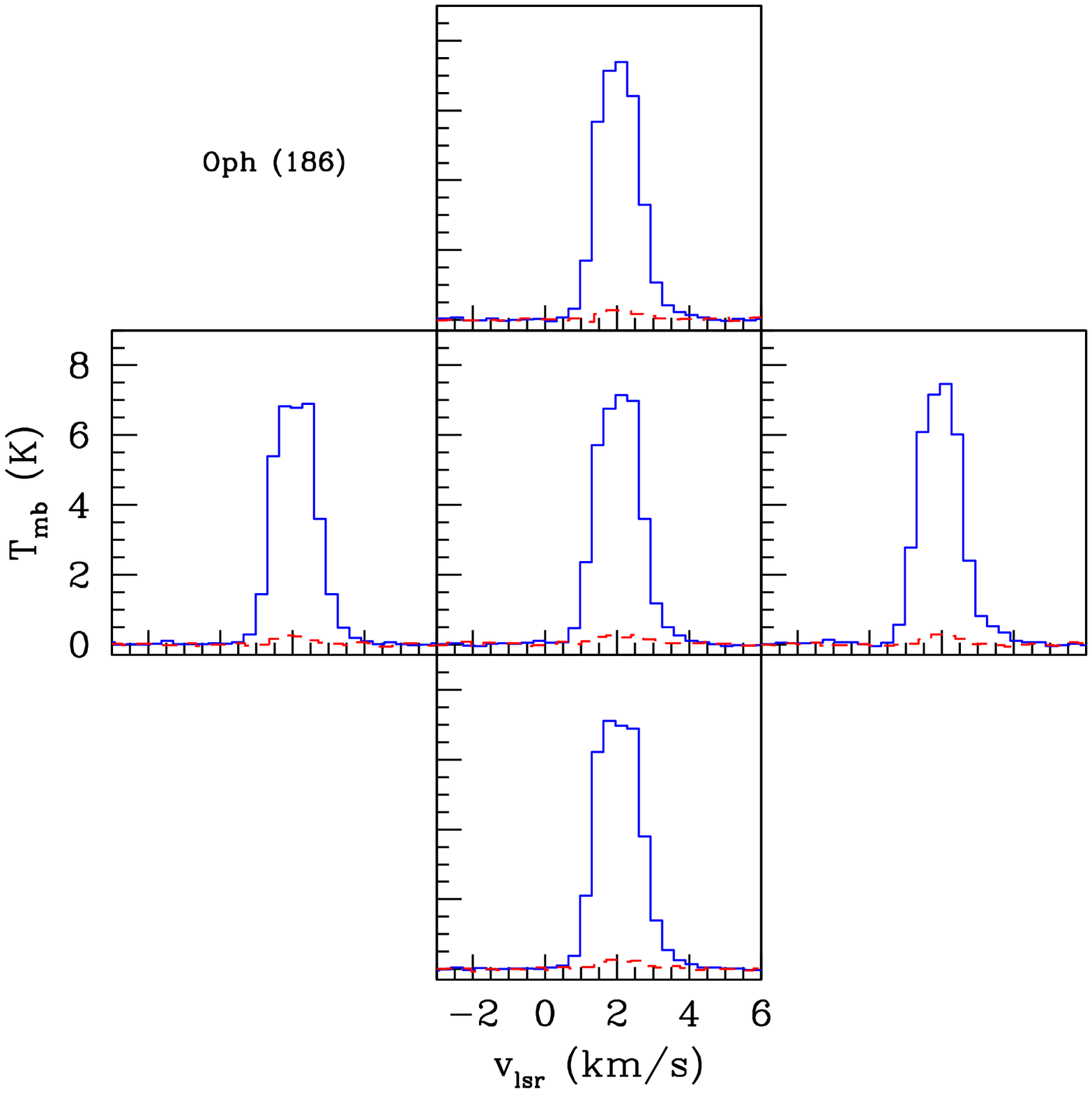}
\includegraphics{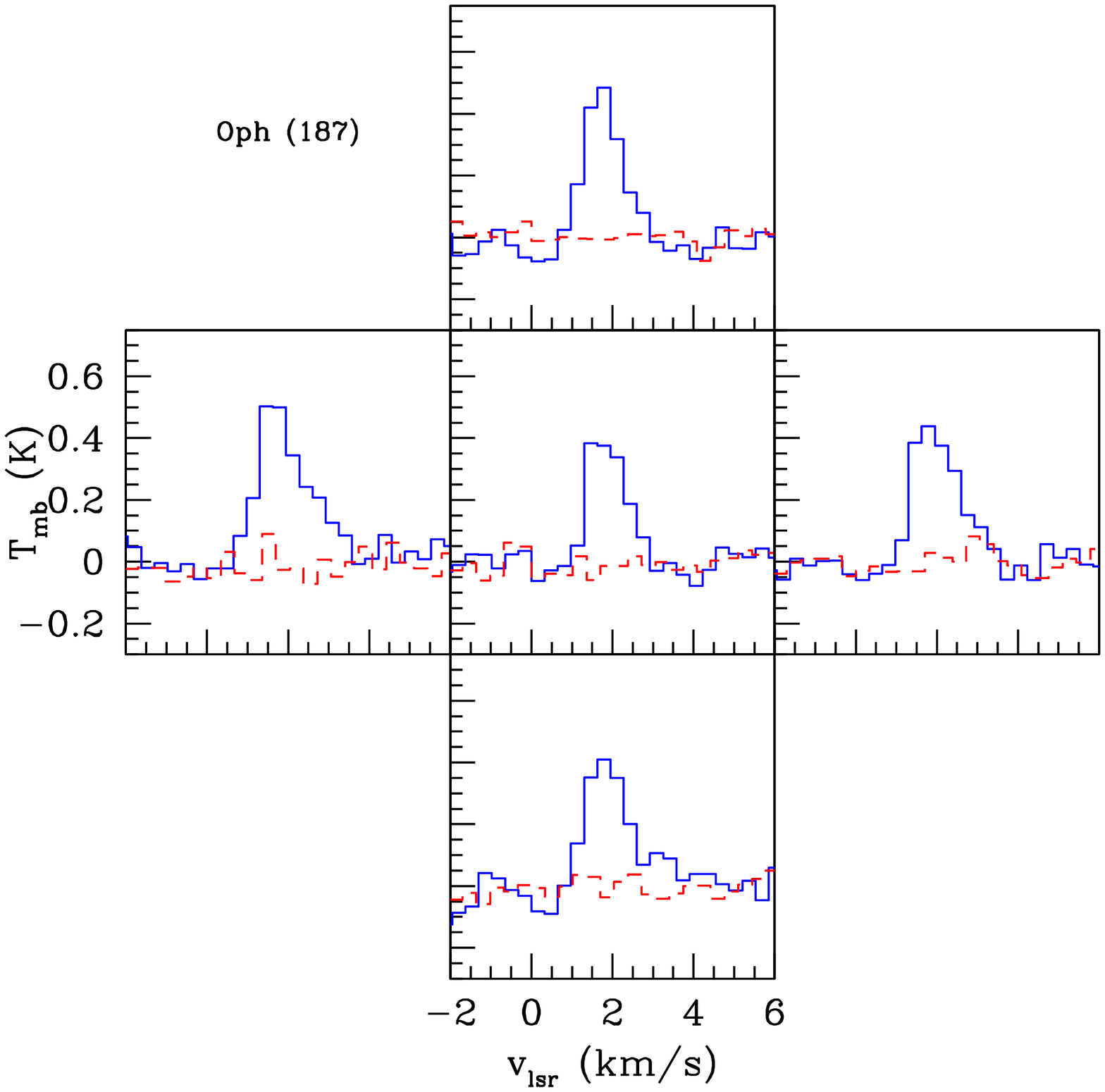}
\includegraphics{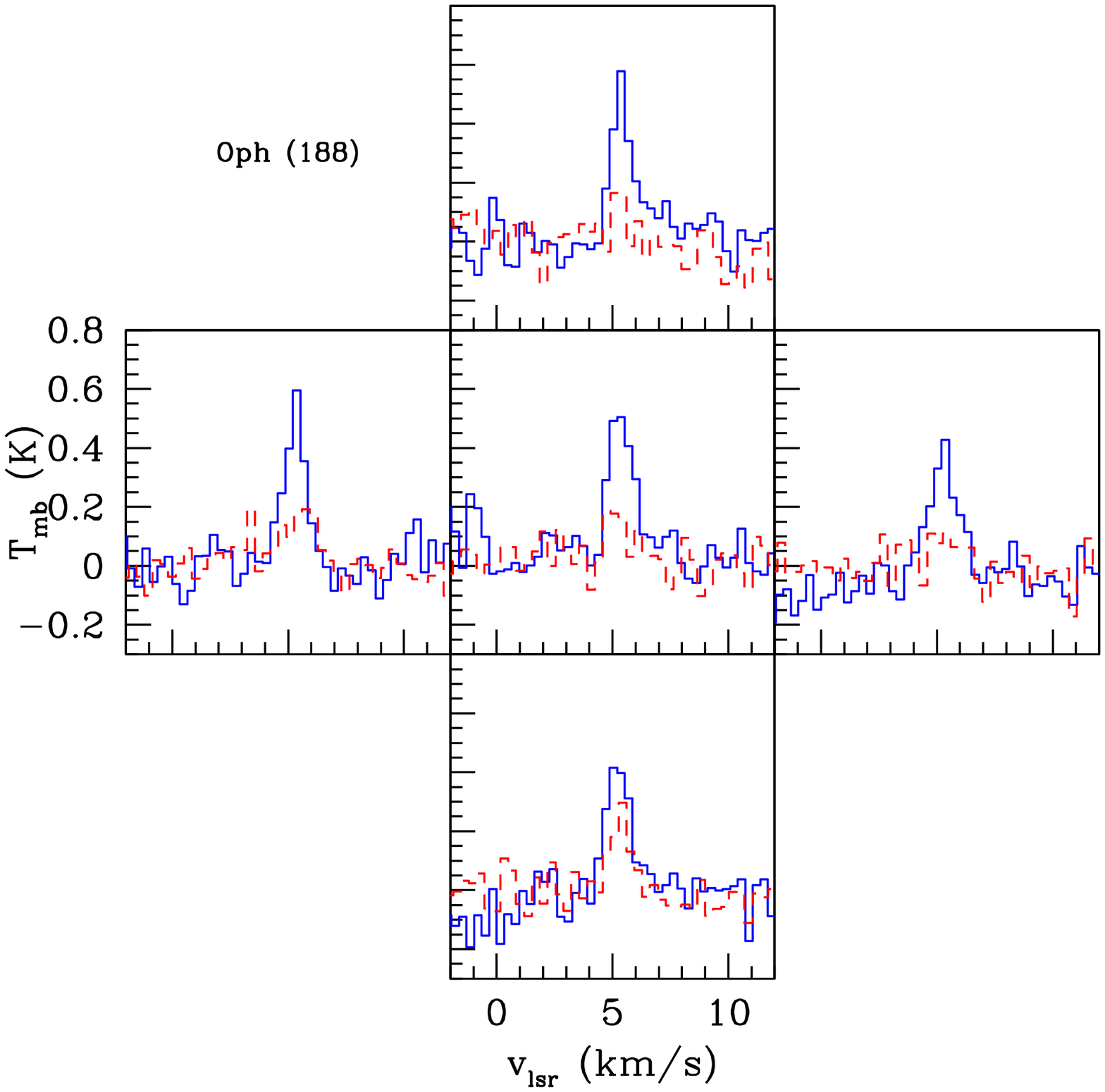}
\figcaption{5-point maps of \co\ (solid line) and \coo\ (dashed line) 
$2-1$ spectra toward the sources.  The offset positions correspond to $30$\as\ shifts
in RA and DEC from the protostar position.  The source name is listed in the upper left of each
5-point map.}
\end{figure}


\begin{figure}
\figurenum{8}
\epsscale{1.0}
\vspace*{16cm}
\includegraphics{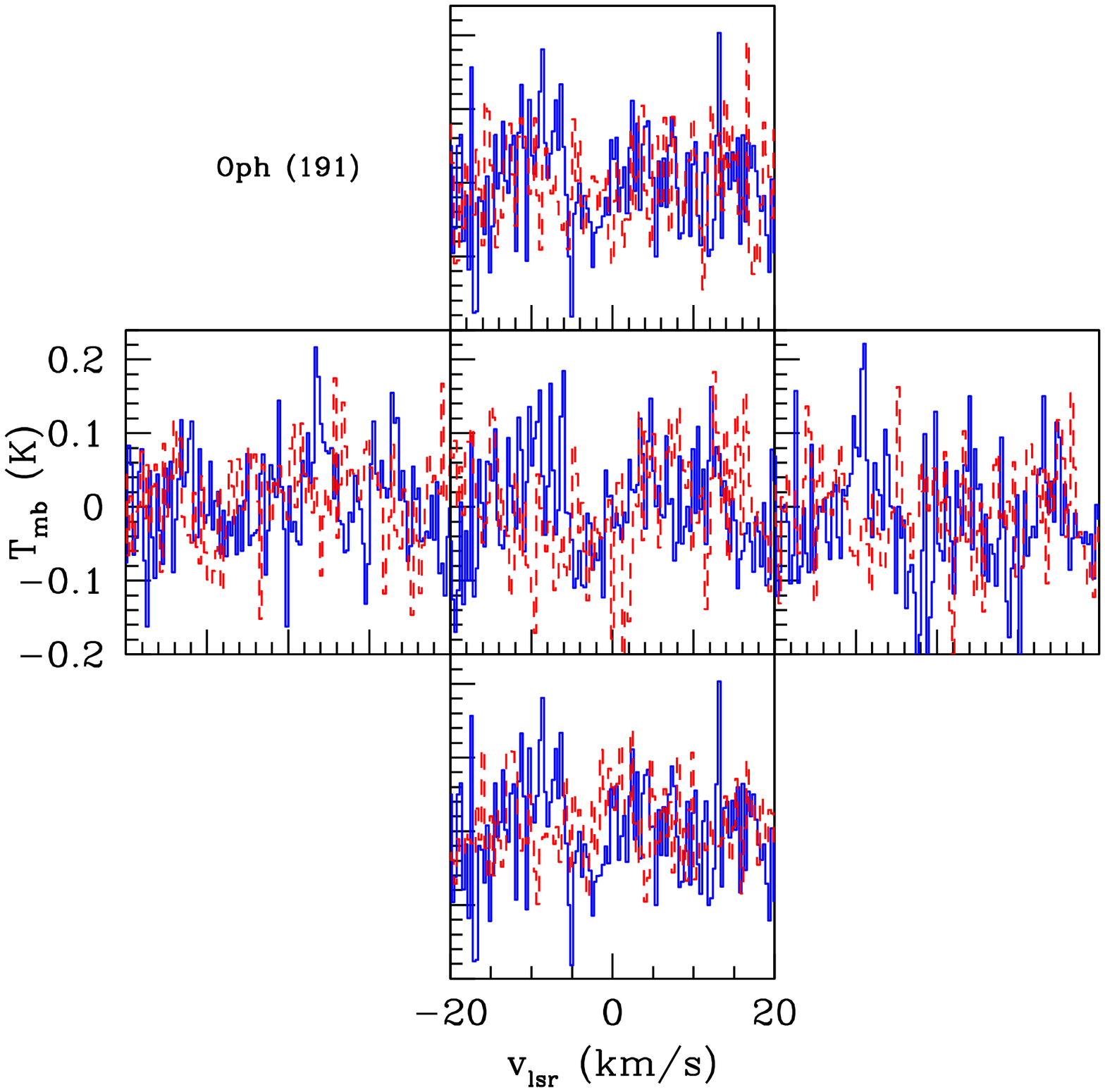}
\includegraphics{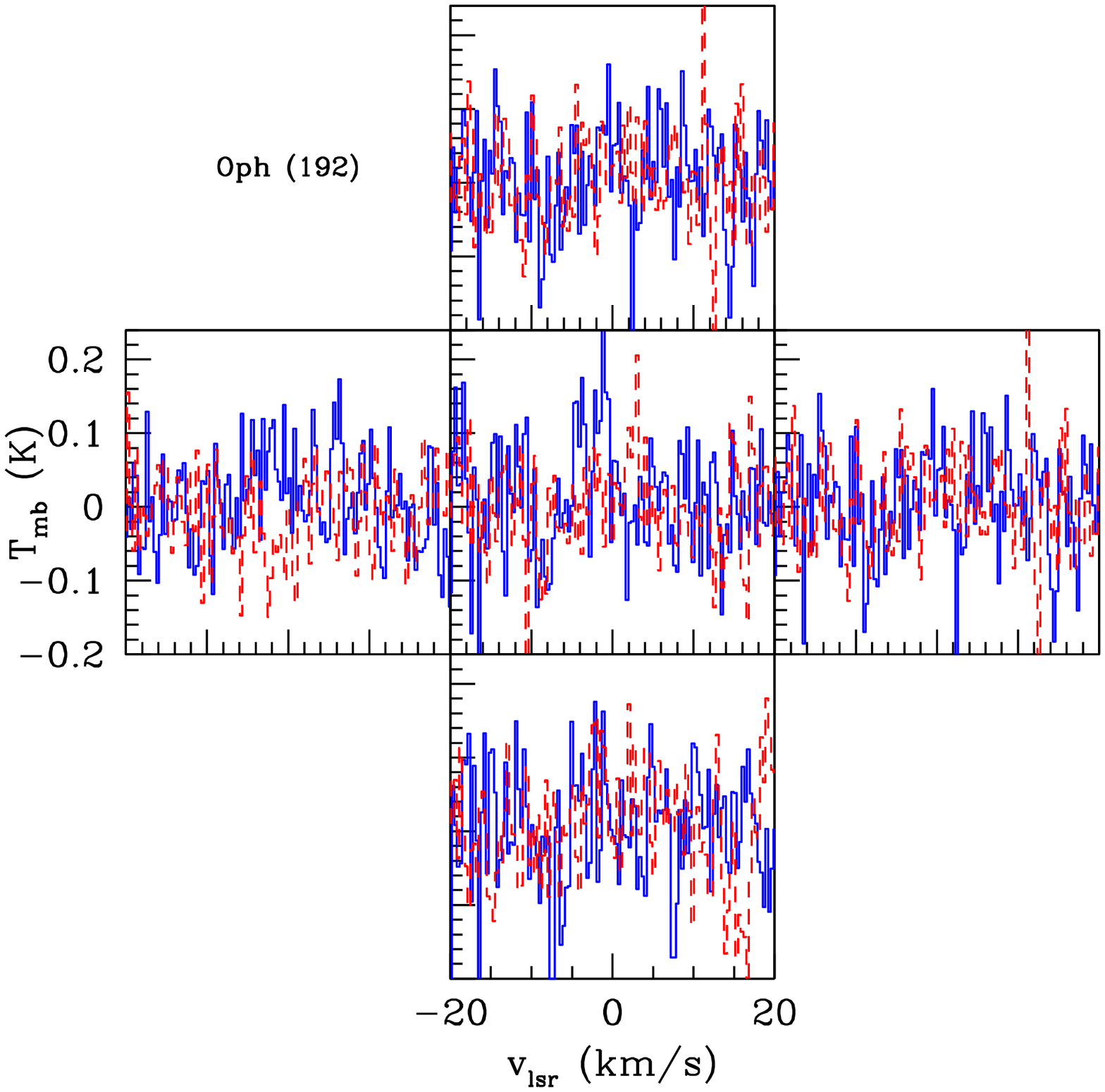}
\includegraphics{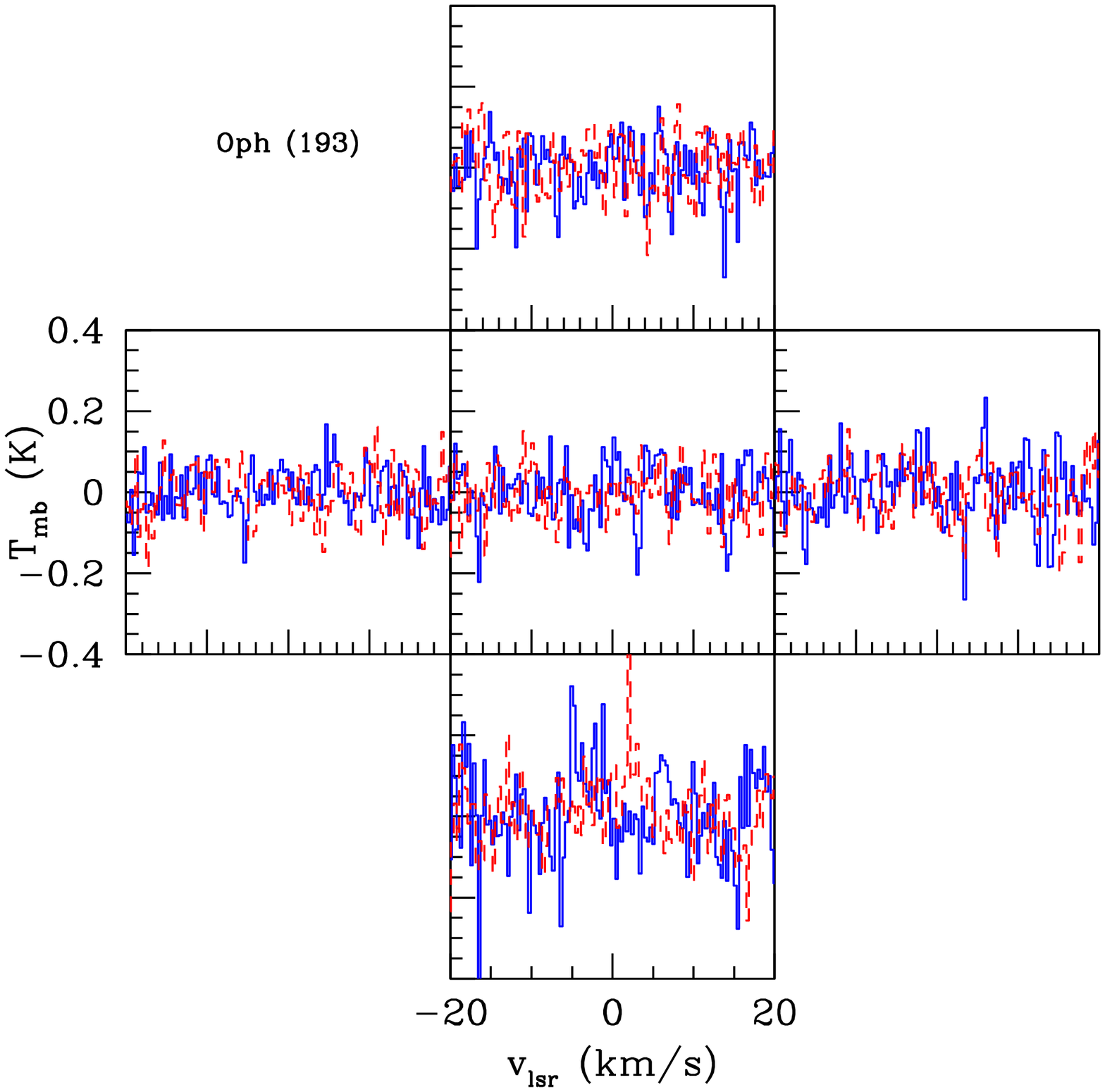}
\includegraphics{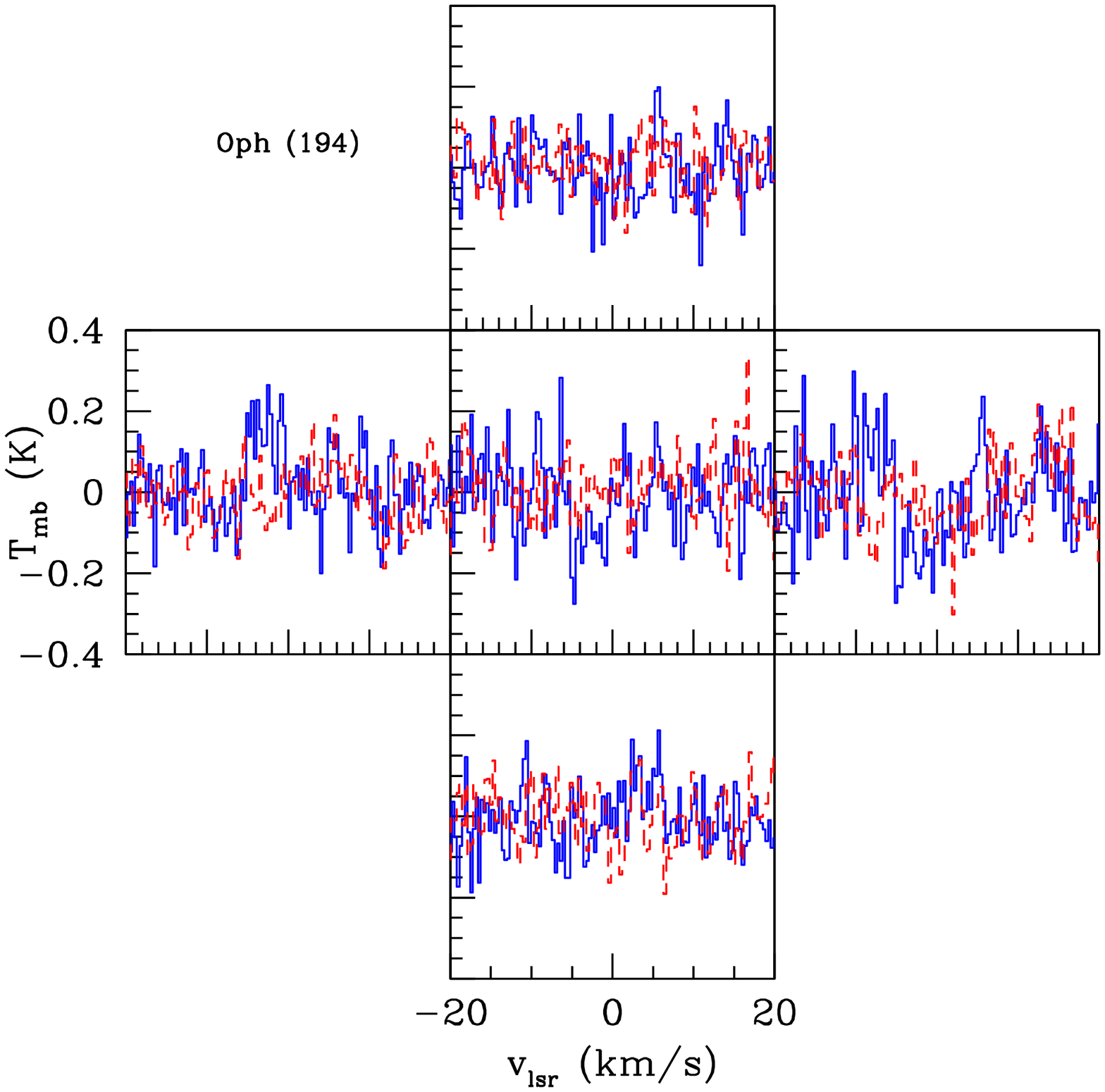}
\figcaption{5-point maps of \co\ (solid line) and \coo\ (dashed line) 
$2-1$ spectra toward the sources.  The offset positions correspond to $30$\as\ shifts
in RA and DEC from the protostar position.  The source name is listed in the upper left of each
5-point map.}
\end{figure}


\begin{figure}
\figurenum{9}
\epsscale{1.0}
\vspace*{16cm}
\includegraphics{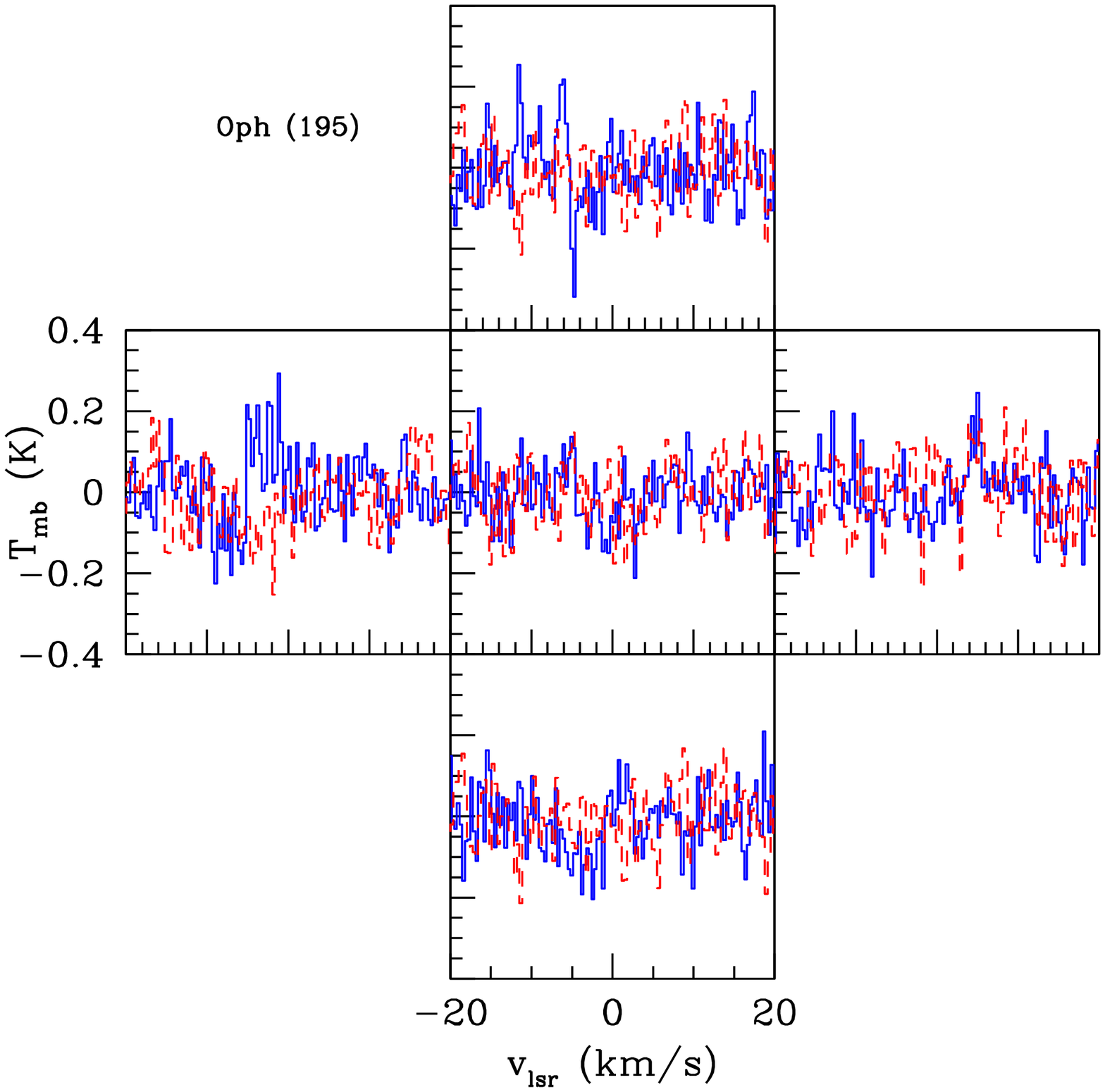}
\includegraphics{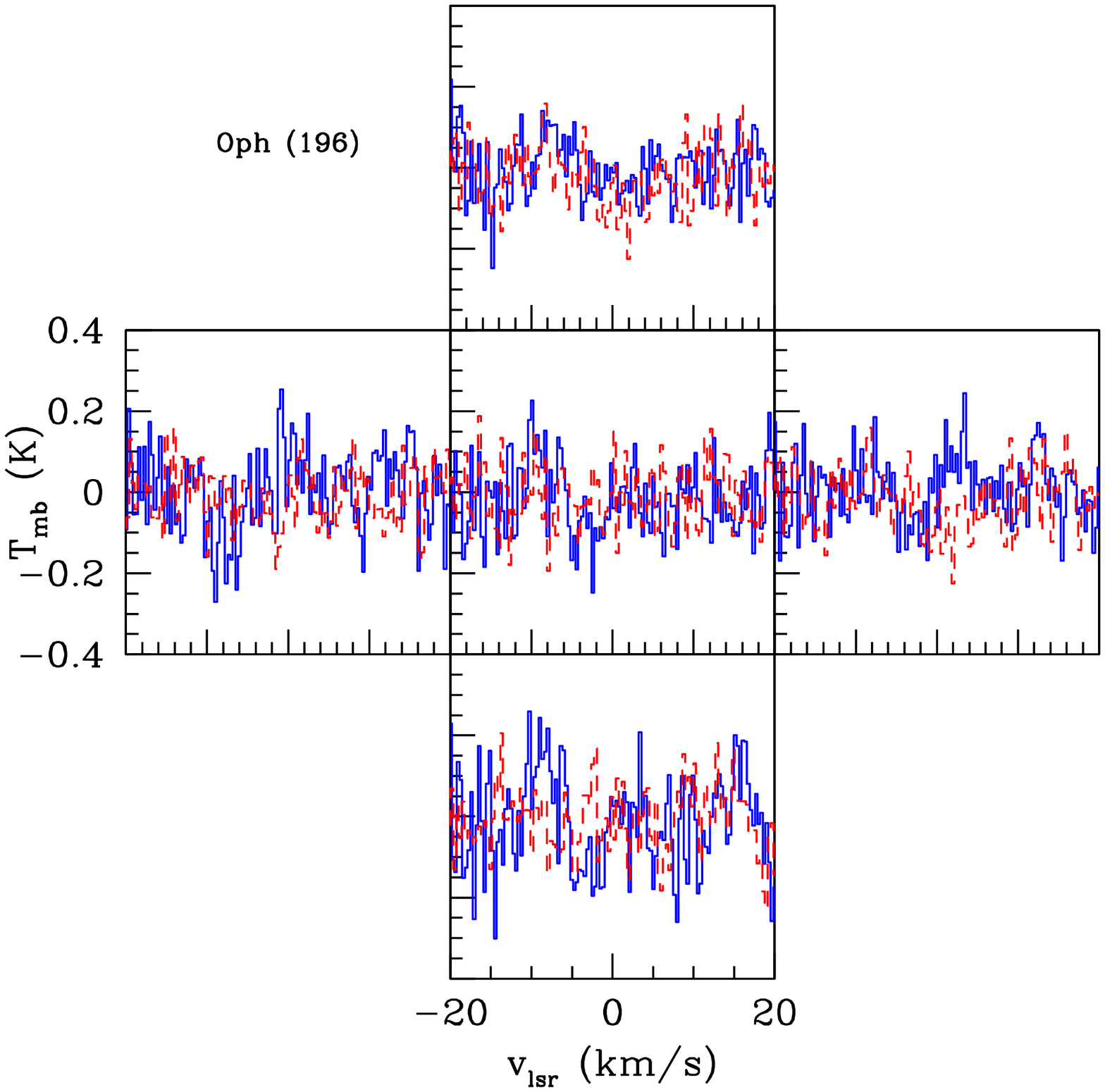}
\includegraphics{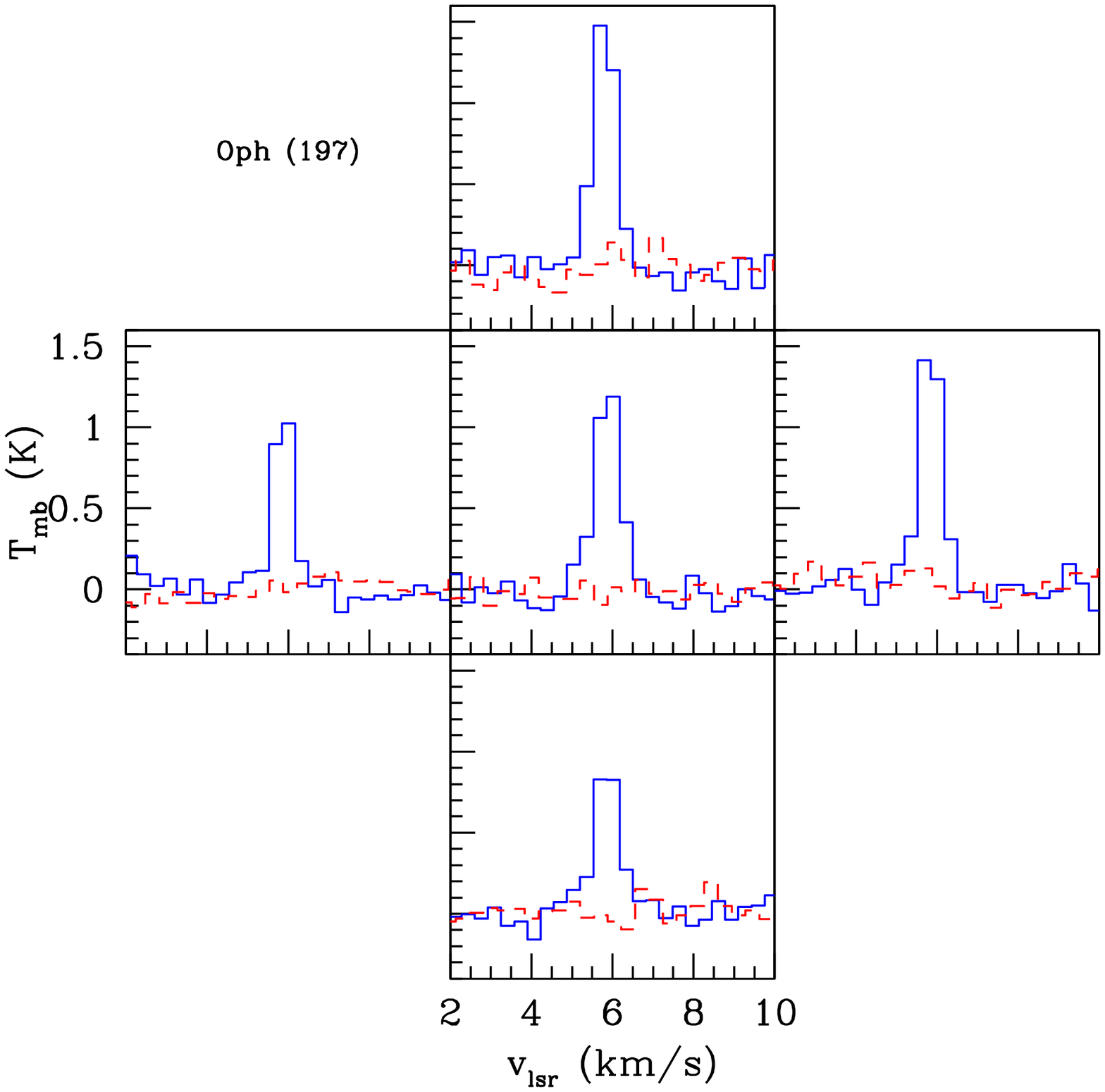}
\includegraphics{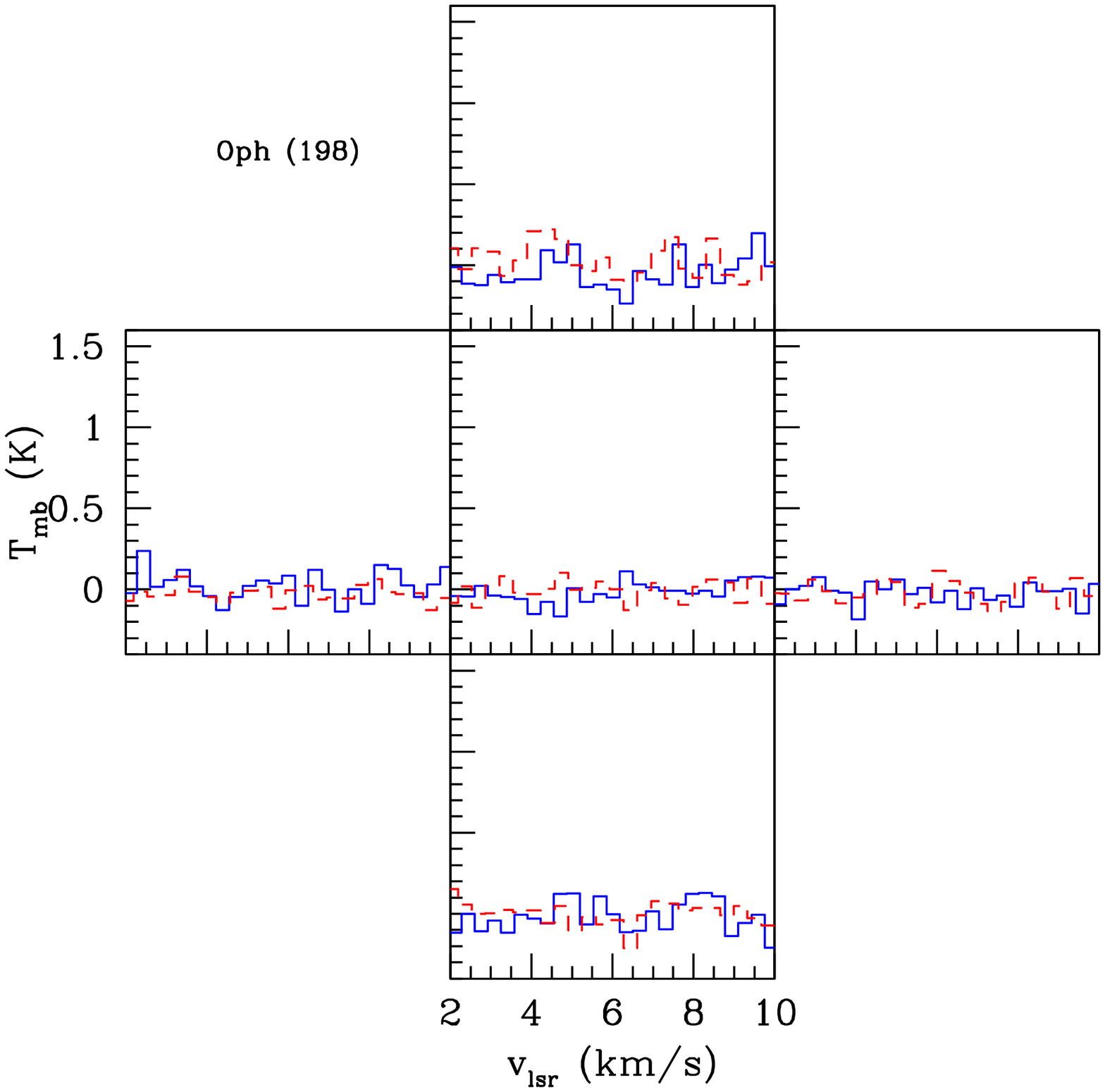}
\figcaption{5-point maps of \co\ (solid line) and \coo\ (dashed line) 
$2-1$ spectra toward the sources.  The offset positions correspond to $30$\as\ shifts
in RA and DEC from the protostar position.  The source name is listed in the upper left of each
5-point map.}
\end{figure}


\begin{figure}
\figurenum{10}
\epsscale{1.0}
\vspace*{16cm}
\includegraphics{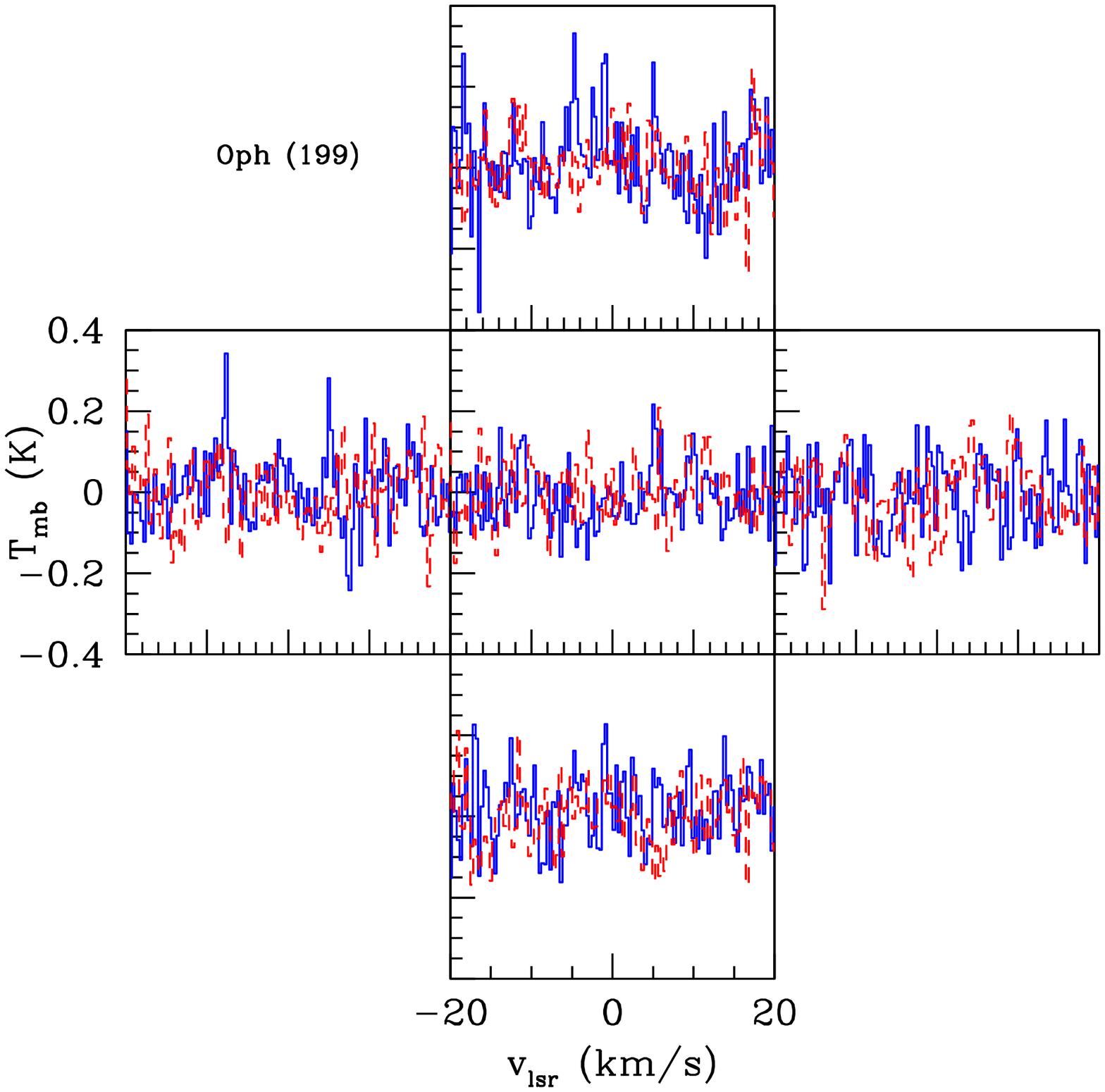}
\includegraphics{./f10b.eps}
\figcaption{5-point maps of \co\ (solid line) and \coo\ (dashed line) 
$2-1$ spectra toward the sources.  The offset positions correspond to $30$\as\ shifts
in RA and DEC from the protostar position.  The source name is listed in the upper left of each
5-point map.}
\end{figure}


\begin{figure}
\figurenum{11}
\epsscale{1.0}
\vspace*{16cm}
\includegraphics{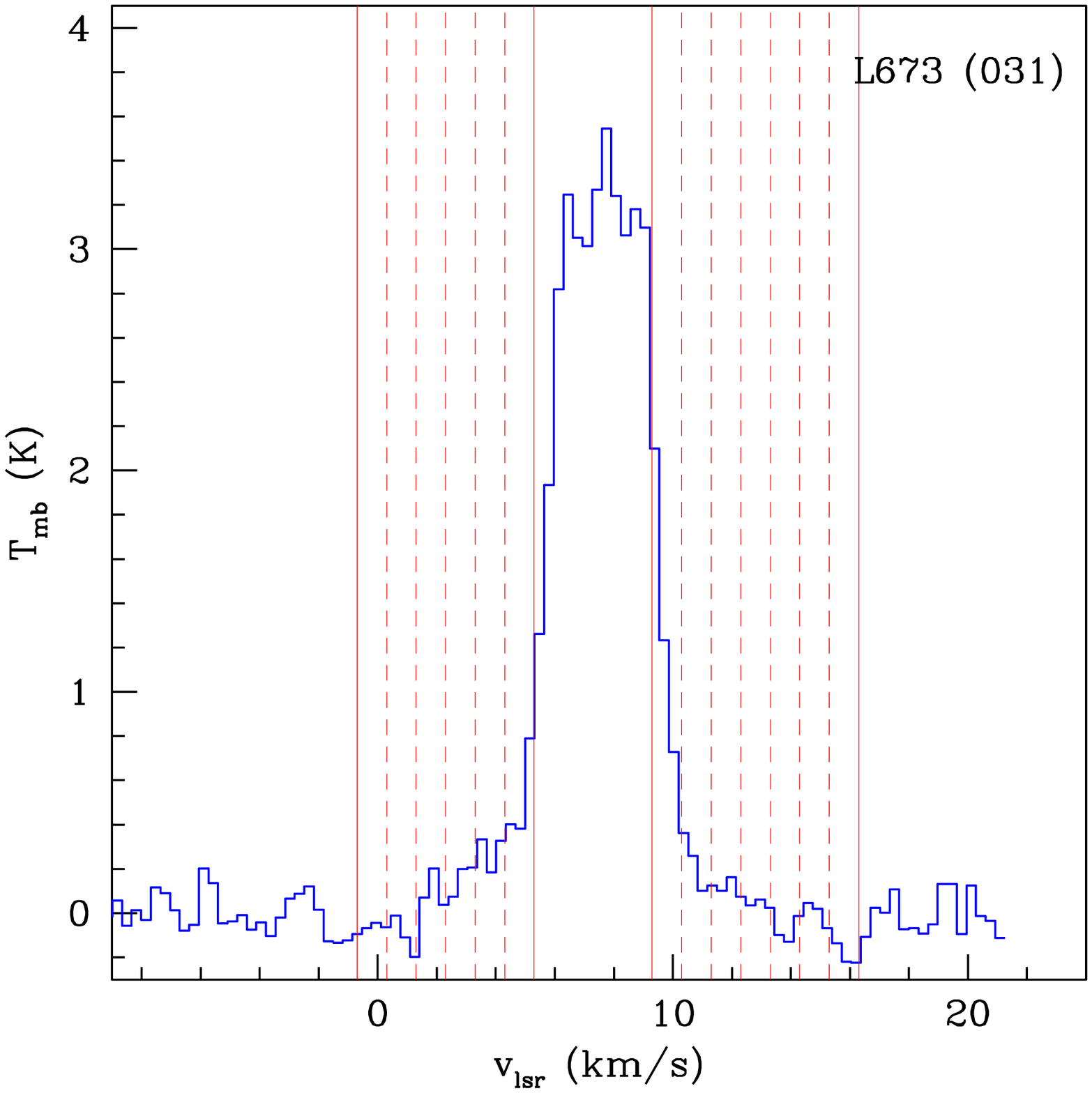}
\includegraphics{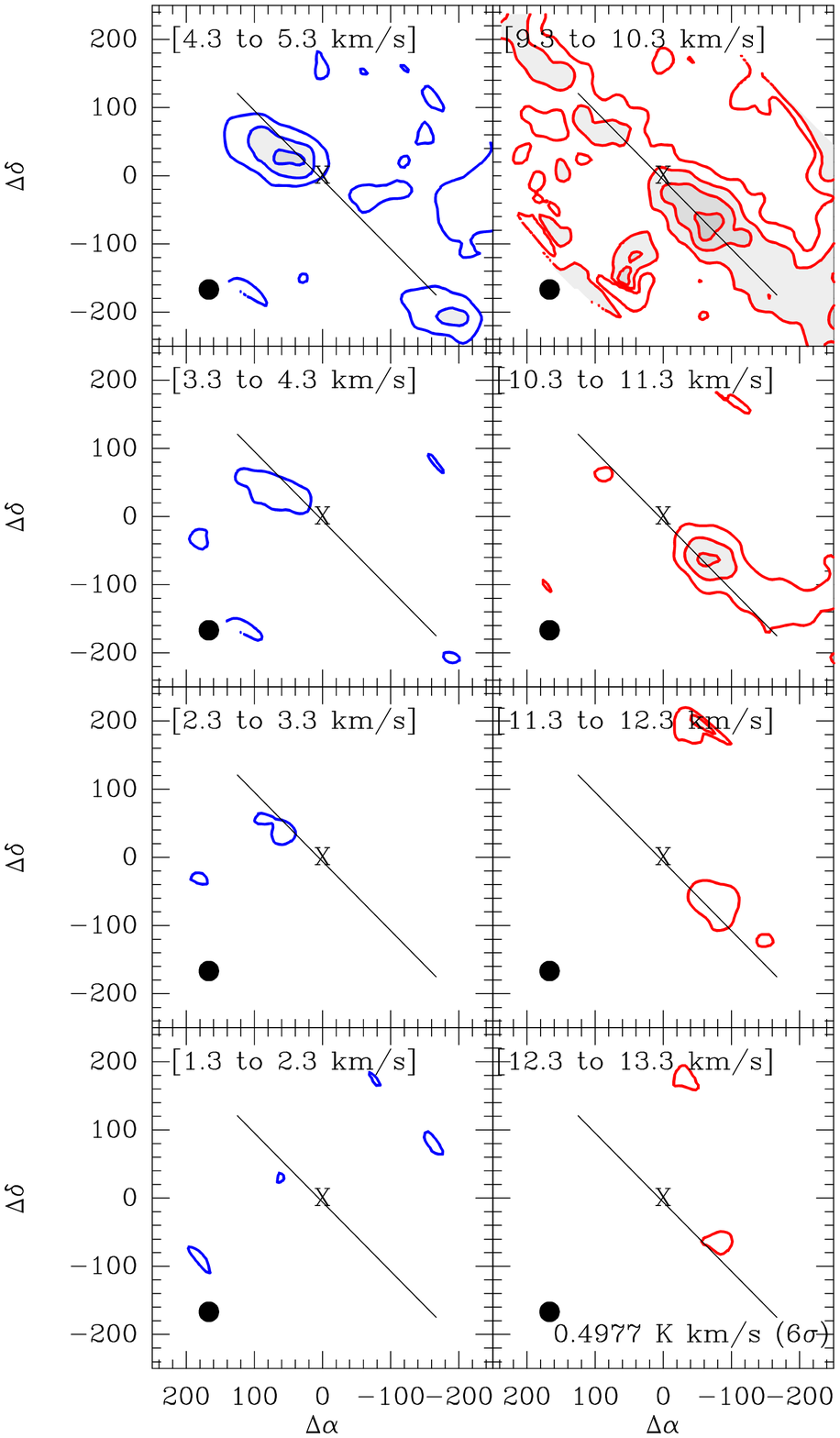}
\includegraphics{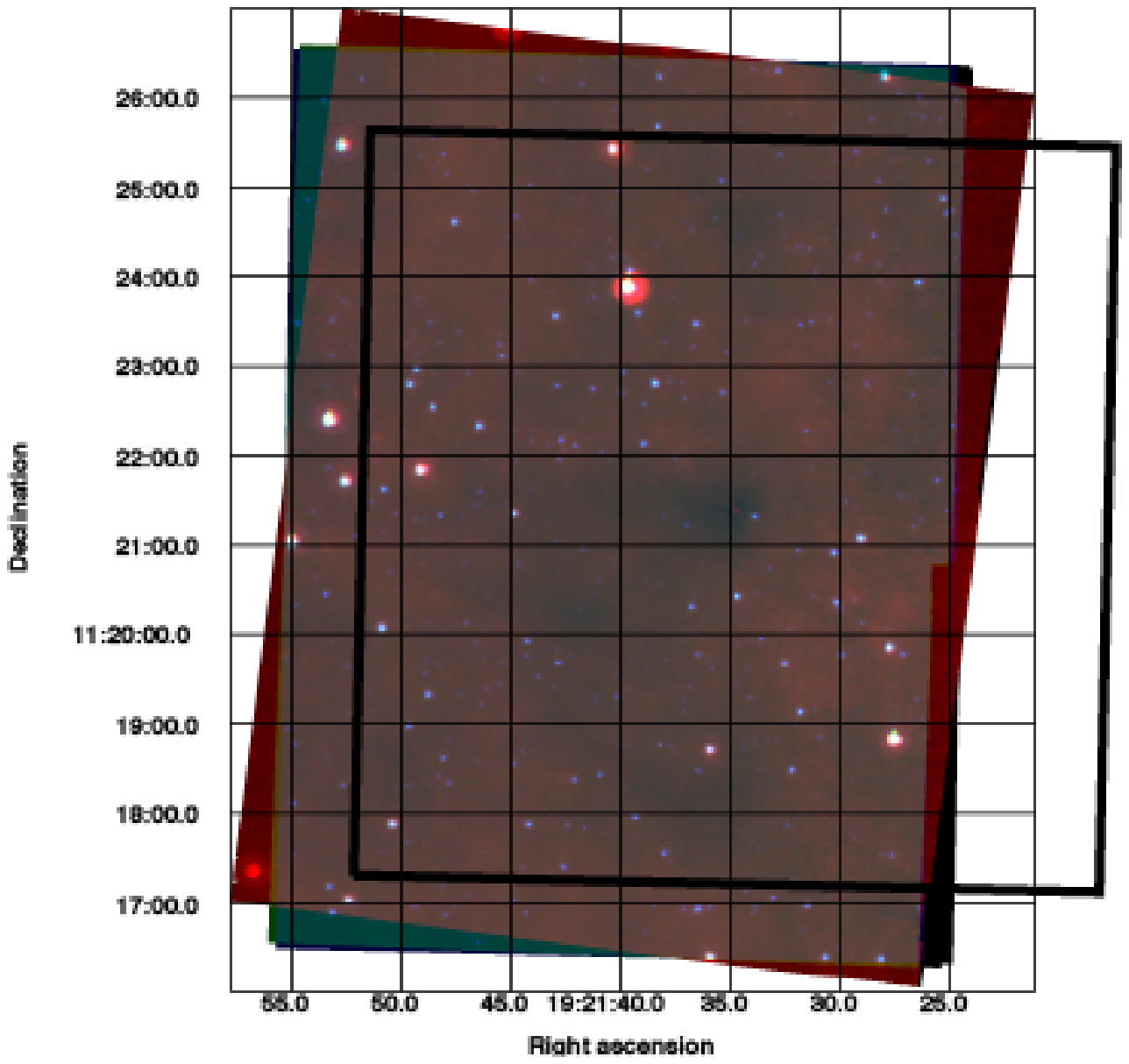}
\figcaption{LEFT: Channel map of outflow emission from L673-7 (031). The left column
shows emission from the blue wing while the right column shows emission form 
the red wing with velocity intervals listed at the top
of each panel. Contours start at 3$\sigma_I$. The direction of the outflow
and position of the source are indicated. UPPER RIGHT: 
Velocity channels used to calculate outflow properties shown for the central 
spectrum.  LOWER RIGHT:  \textit{Spitzer} 3 color image (4.5, 8.0, 24.0 $\mu$m)
centered on L673-7.  The solid box delineates the mapped region.}
\label{L673maps}
\end{figure}
\clearpage


\begin{figure}
\figurenum{12}
\epsscale{.80}
\plotone{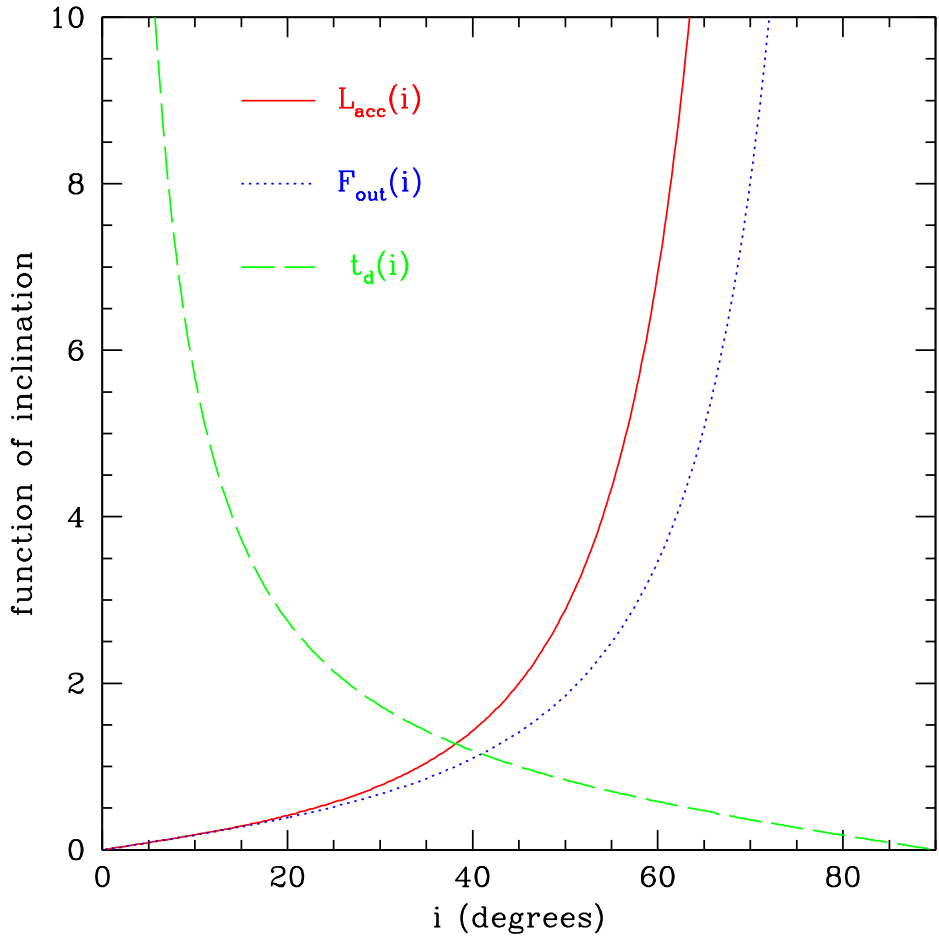}
\caption{The dependence of inferred protostellar accretion luminosity $L_{acc}$, 
outflow force $F_{out}$, and the dynamical time $t_d$ of the outflow
with inclination of the outflow.  Inclination is defined as the angle from
the line-of-sight to the observer ($i = 0$\degree\ is an outflow aligned with 
the line-of-sight while $i = 90$\degree\ is perpendicular to the line-of-sight).
The dynamical time of the outflow may change by a factor of 7.6 over a
reasonable inclination range ($20 - 70$\degree ) while the outflow force 
and inferred protostellar accretion luminosity may change by factors 20.6 
and 57.3 respectively over the same range.
}
\label{i}
\end{figure}
\clearpage


\begin{figure}
\figurenum{13}
\epsscale{1.0}
\vspace*{16cm}
\includegraphics{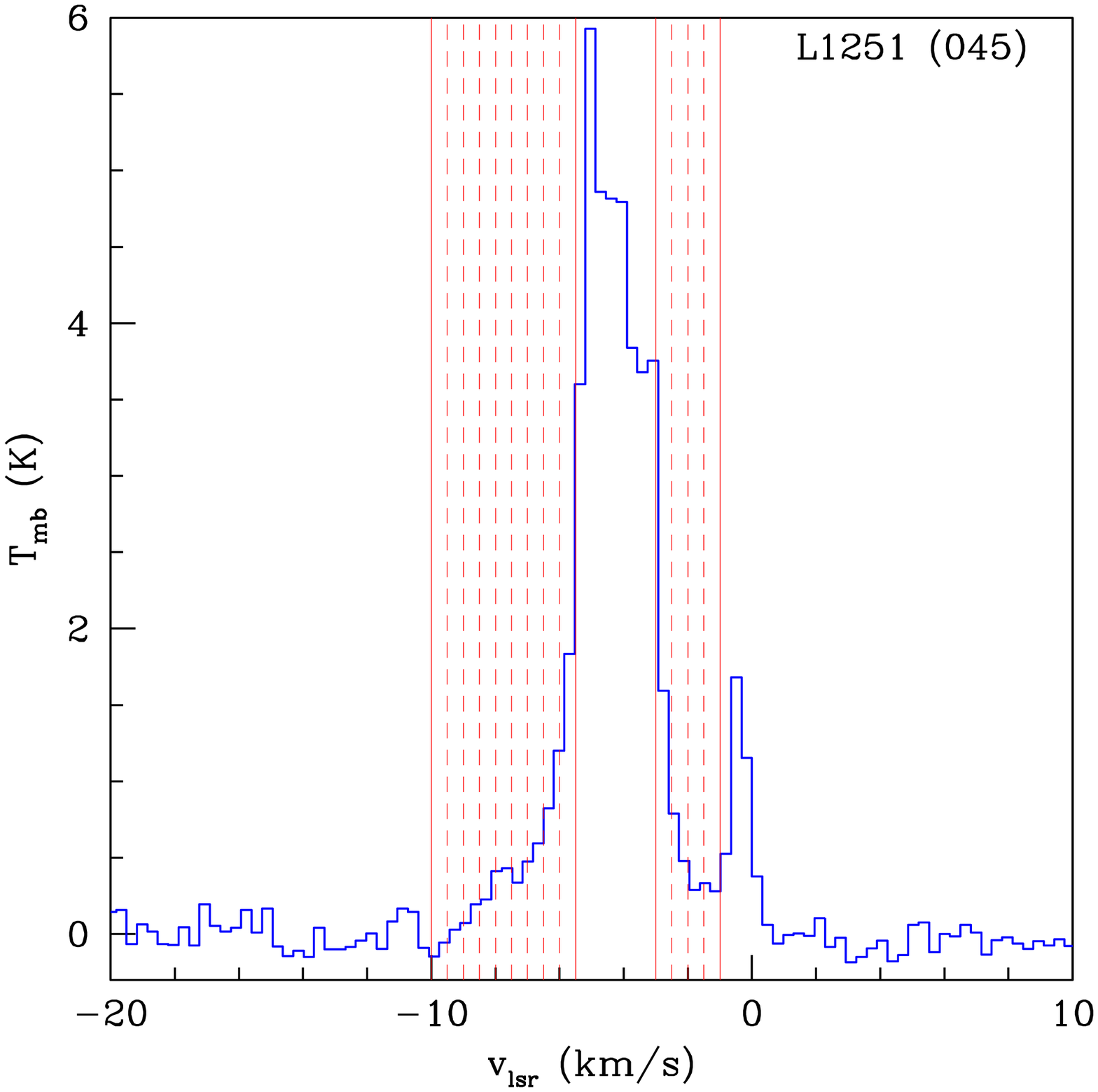}
\includegraphics{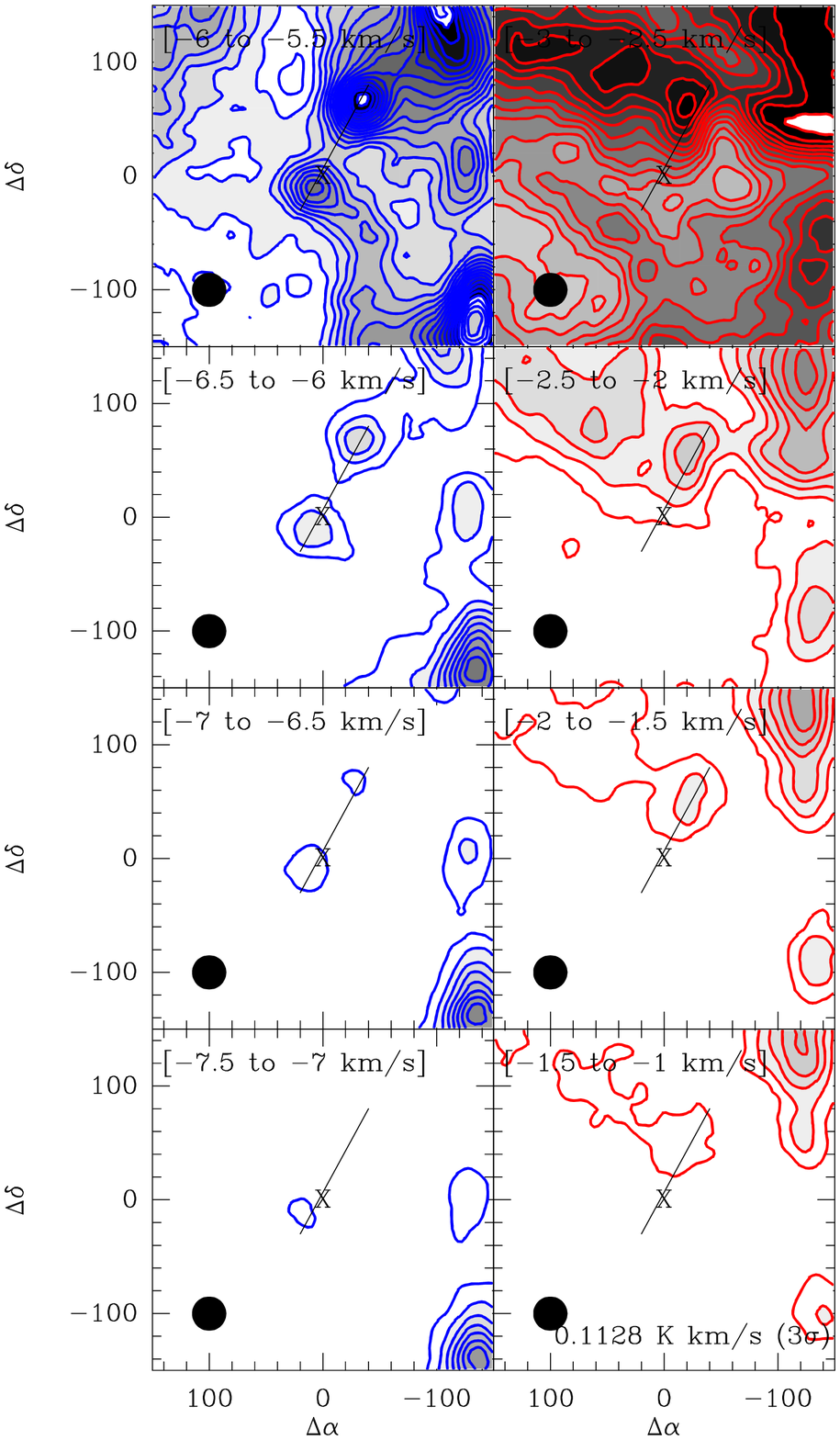}
\includegraphics{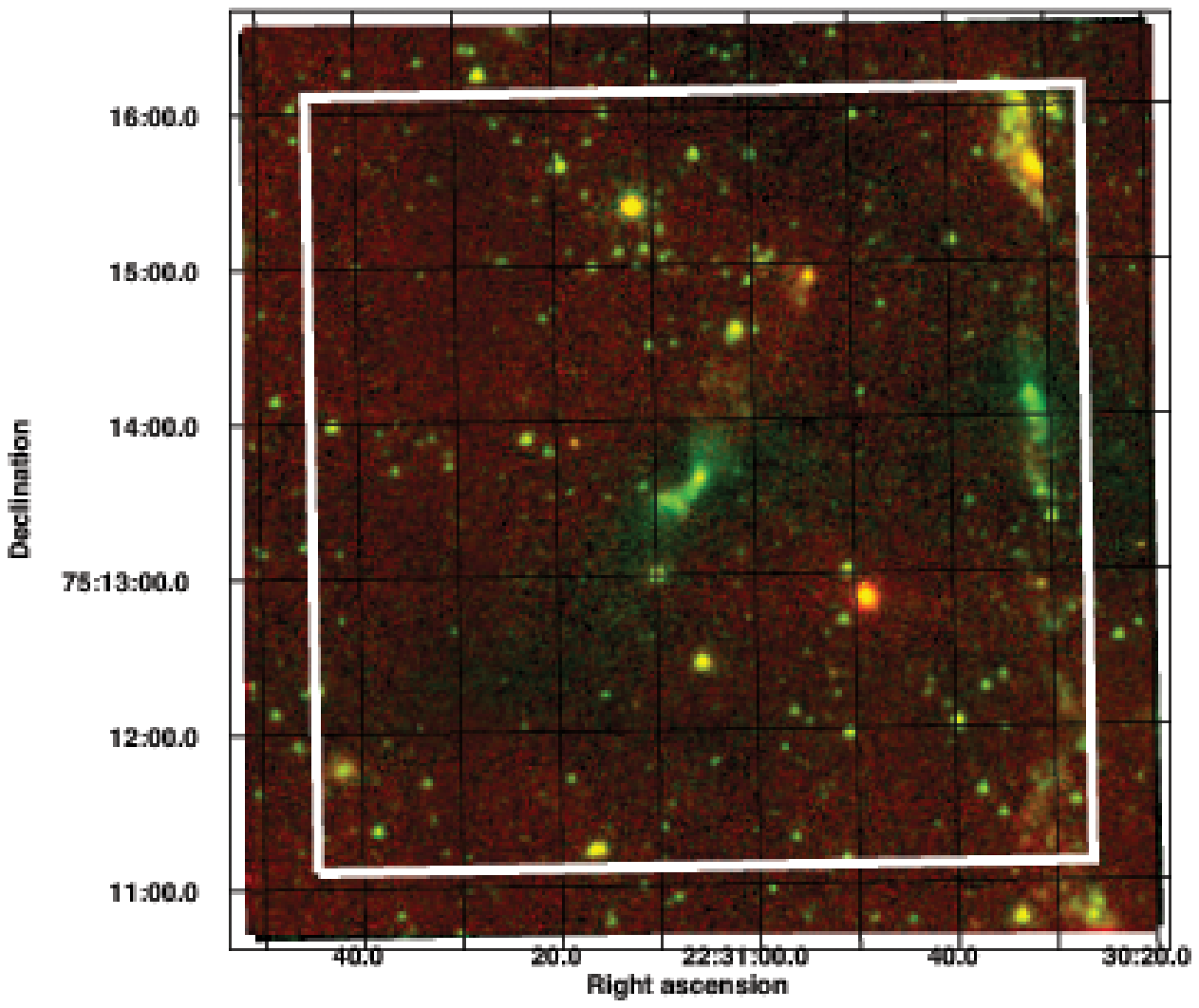}
\figcaption{LEFT: Channel map of outflow emission from L1251-A. The left column
shows emission from the blue wing while the right column shows emission form 
the red wing with velocity intervals listed at the top
of each panel. Contours start at 3$\sigma_I$. The direction of the outflow
and position of the source are indicated.  The strong outflow source
L1251-A (044) is also seen along the eastern side of each panel. UPPER RIGHT: 
Velocity channels used to calculate outflow properties shown for the central 
spectrum.  LOWER RIGHT:  \textit{Spitzer} 2 color image (4.6, 8.0 $\mu$m)
centered on L1251A (045).  The solid box delineates the mapped region.  Shocked
$4.6$ $\mu$m emission (green) traces the directions of the outflow.}
\label{L1251maps}
\end{figure}
\clearpage


\begin{figure}
\figurenum{14}
\epsscale{1.0}
\vspace*{16cm}
\includegraphics{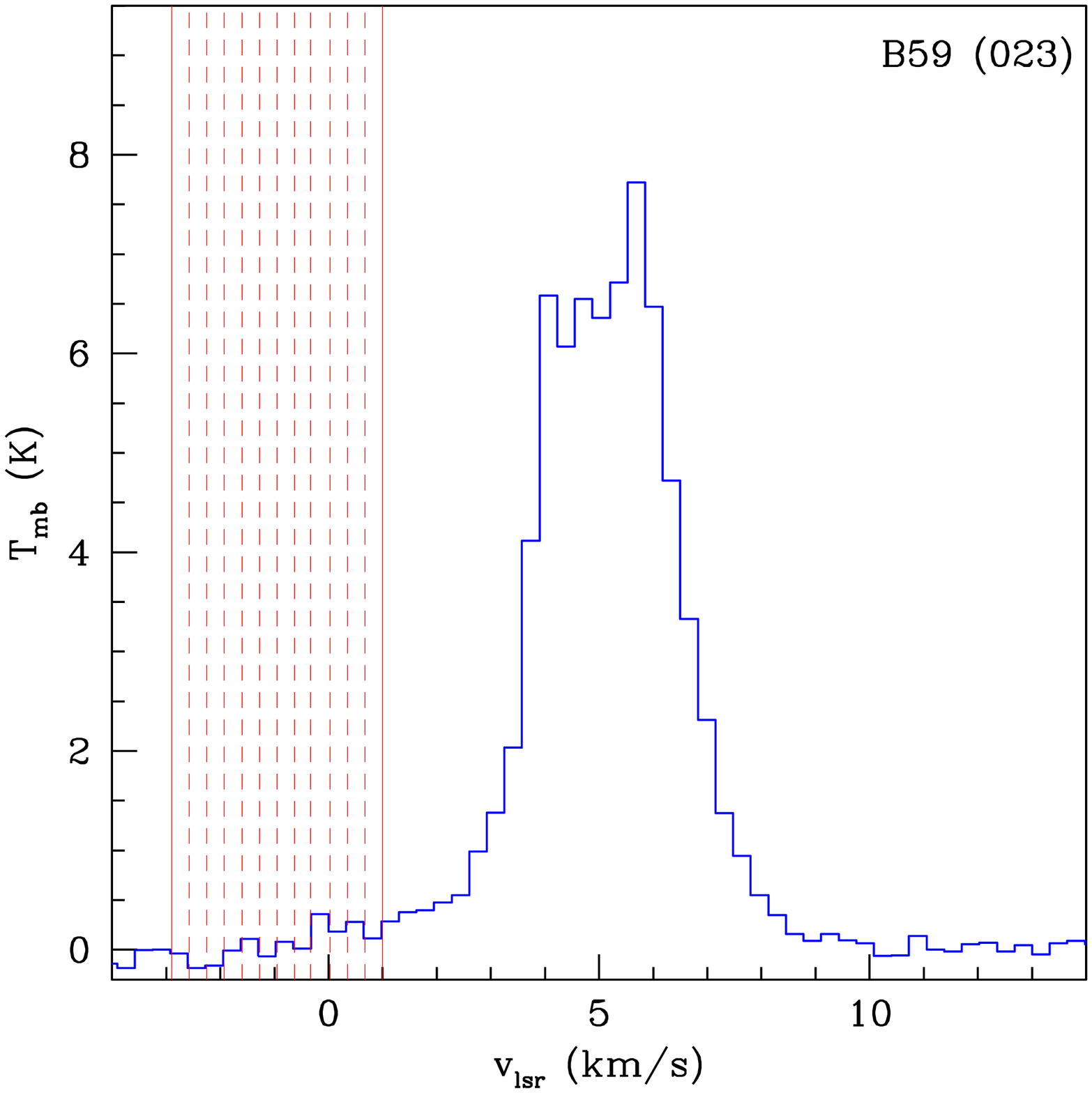}
\includegraphics{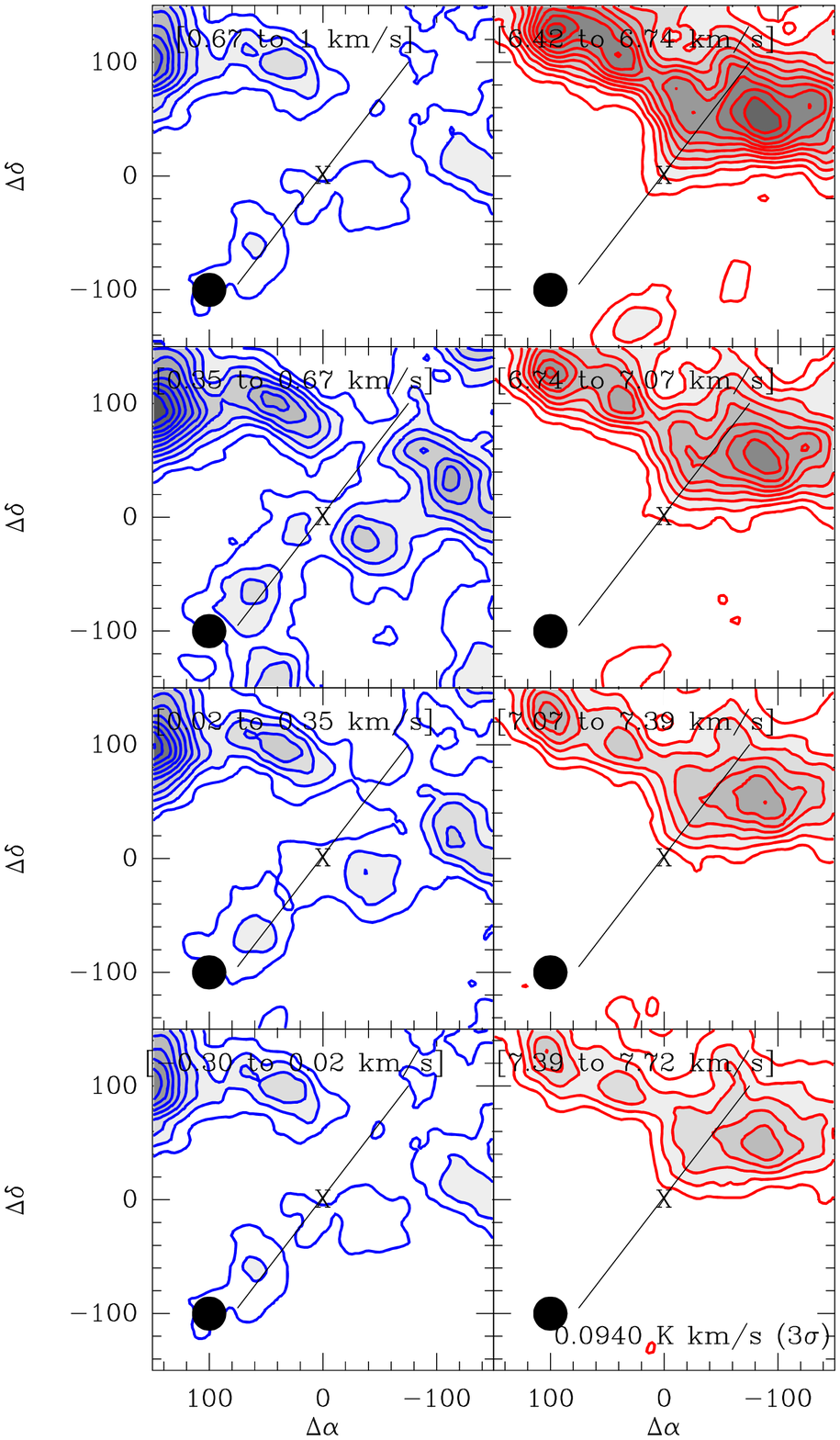}
\includegraphics{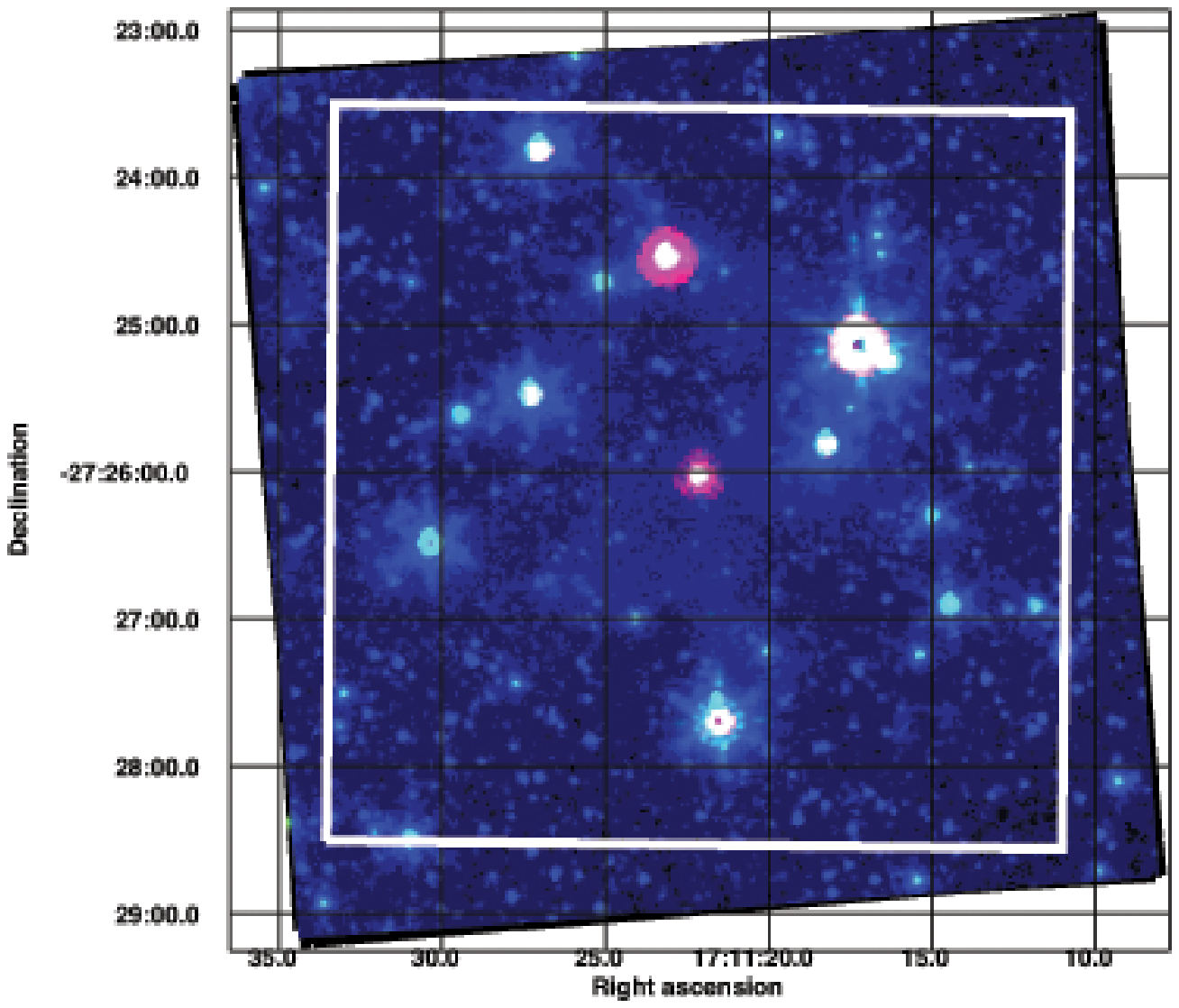}
\figcaption{LEFT:Channel map of outflow emission from B59 (023). 
The left column
shows emission from the blue wing with velocity intervals listed 
at the top
of each panel.  The right column shows emission from the red wing.  
Contours
start at 3$\sigma_I$. The strong outflow source
IRAS17081-2721 is also seen along the northern side of each panel and may 
obscure the red lobe of the outflow. UPPER RIGHT: 
Velocity channels used to calculate outflow properties shown for the 
central 
spectrum.  LOWER RIGHT:  \textit{Spitzer} 2 color image (4.6, 8.0 $\mu$m)
centered on B59 (023).  
The solid box delineates the mapped region.  }
\label{B59maps}
\end{figure}
\clearpage


\begin{figure}
\figurenum{15}
\epsscale{1.0}
\vspace*{16cm}
\includegraphics{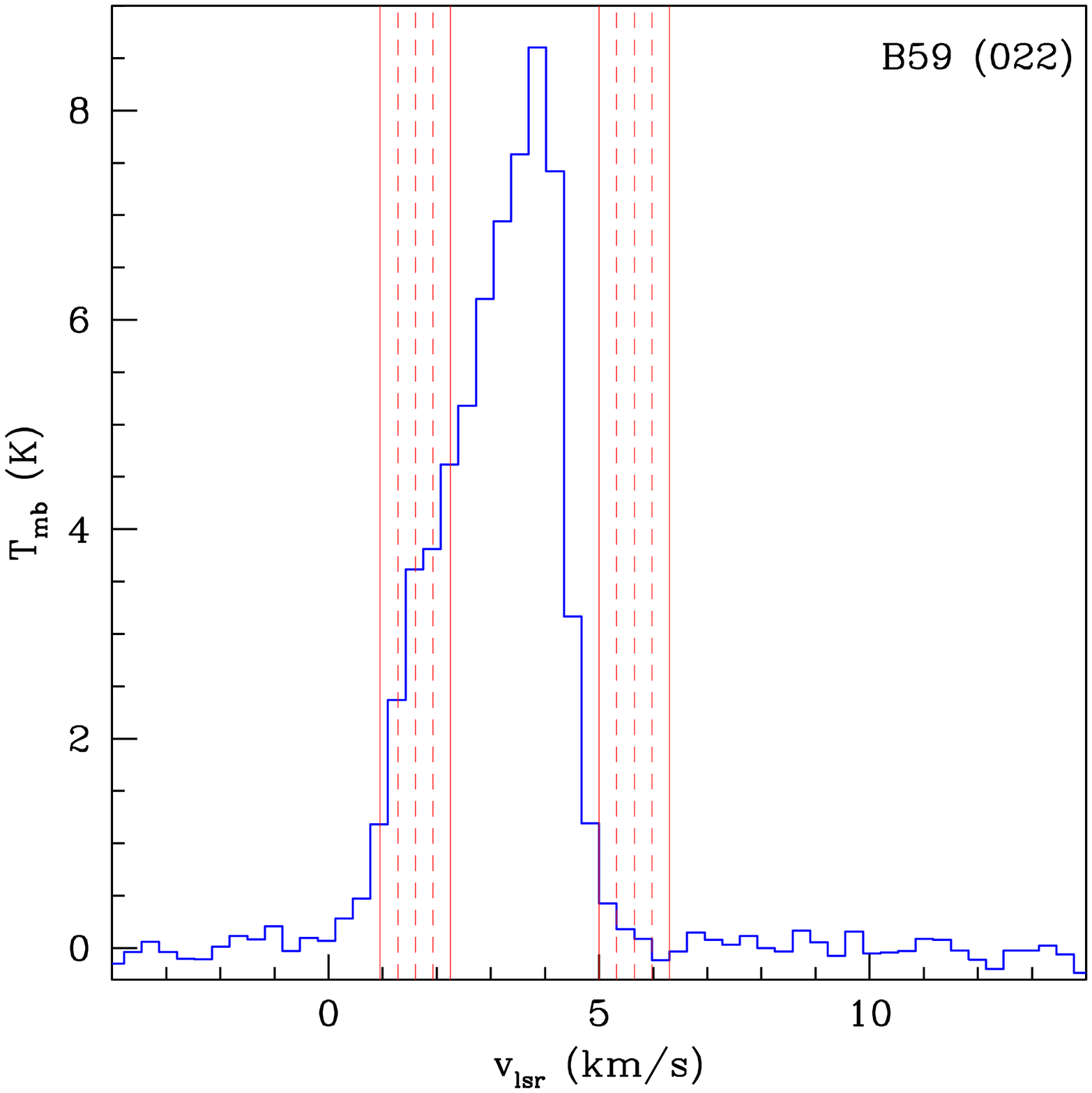}
\includegraphics{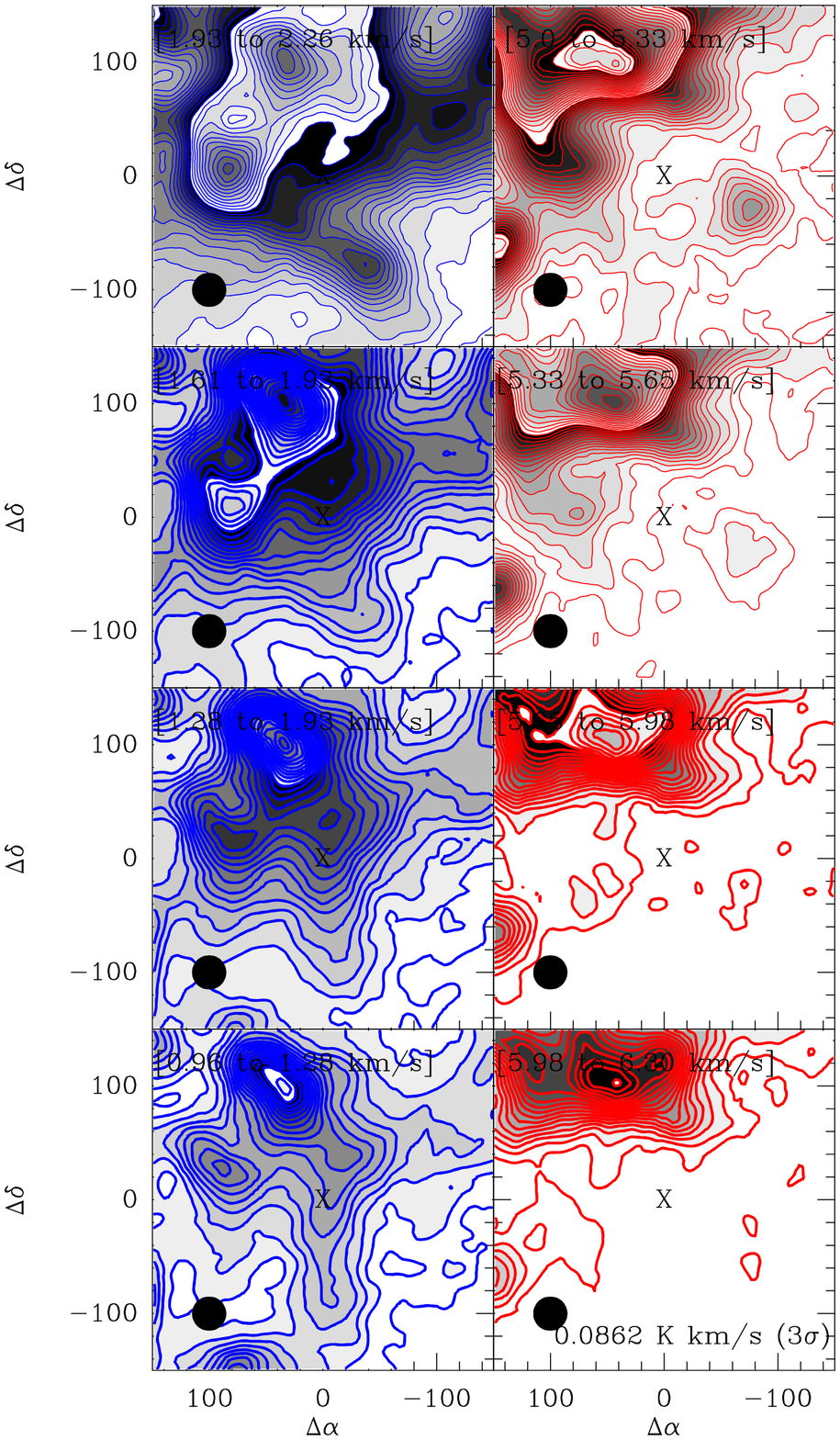}
\includegraphics{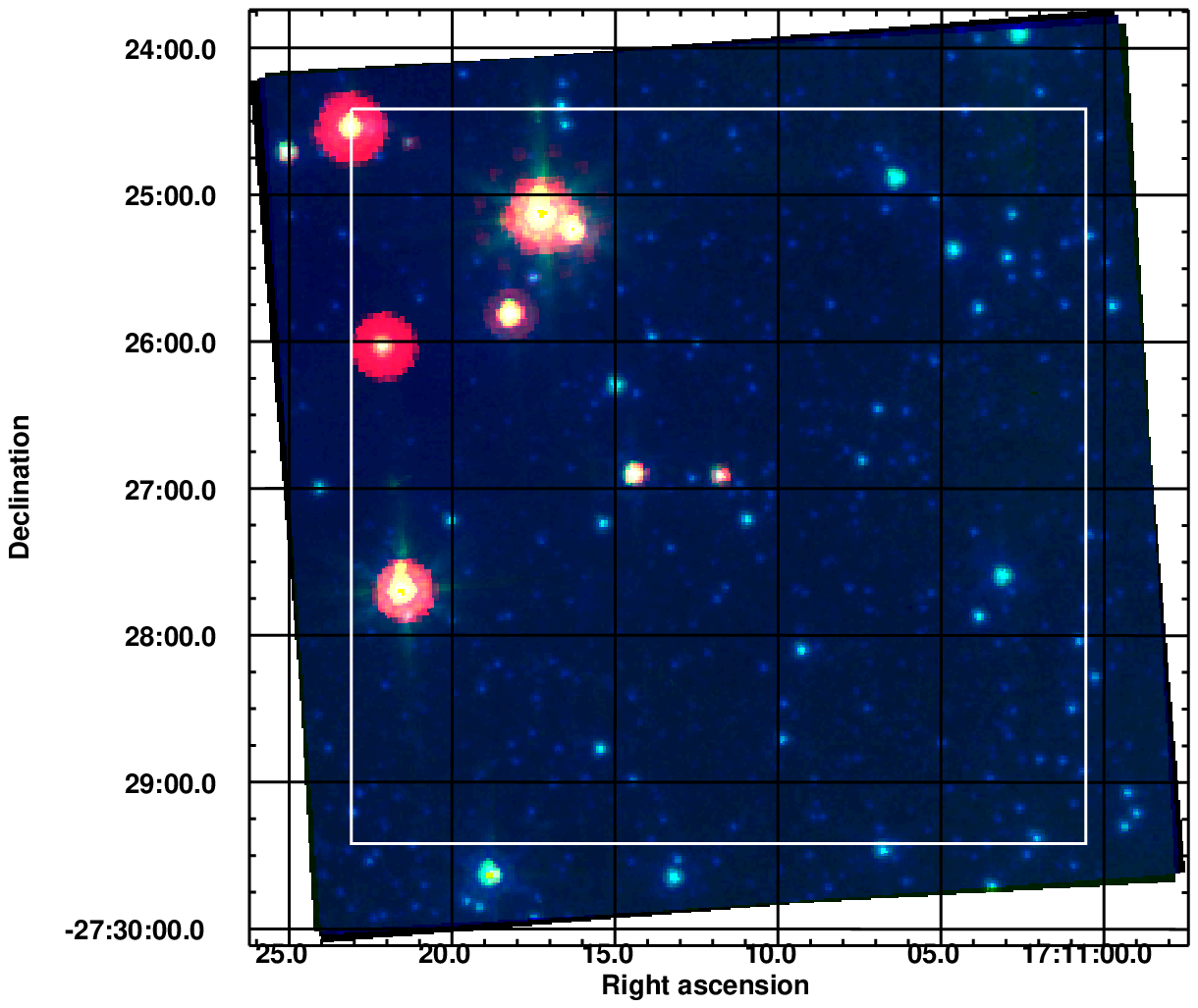}
\figcaption{LEFT:  Channel map of outflow emission from B59 (022). No outflow was detected around this source.
The left column
shows emission from the blue wing with velocity intervals listed 
at the top
of each panel.  The right column shows emission from the red wing.  
Contours
start at 3$\sigma_I$. The strong outflow source
IRAS17081-2721 is also seen along the northern side of each panel. UPPER RIGHT: 
Velocity channels used to calculate outflow properties shown for the 
central 
spectrum.  LOWER RIGHT:  Three color (4.6, 8.0, 24.0 $\mu$m) \textit{Spitzer} image centered on
B59 (022). 
The solid box delineates the mapped region. }
\label{B5922maps}
\end{figure}
\clearpage


\begin{figure}
\figurenum{16}
\epsscale{1.0}
\vspace*{16cm}
\includegraphics{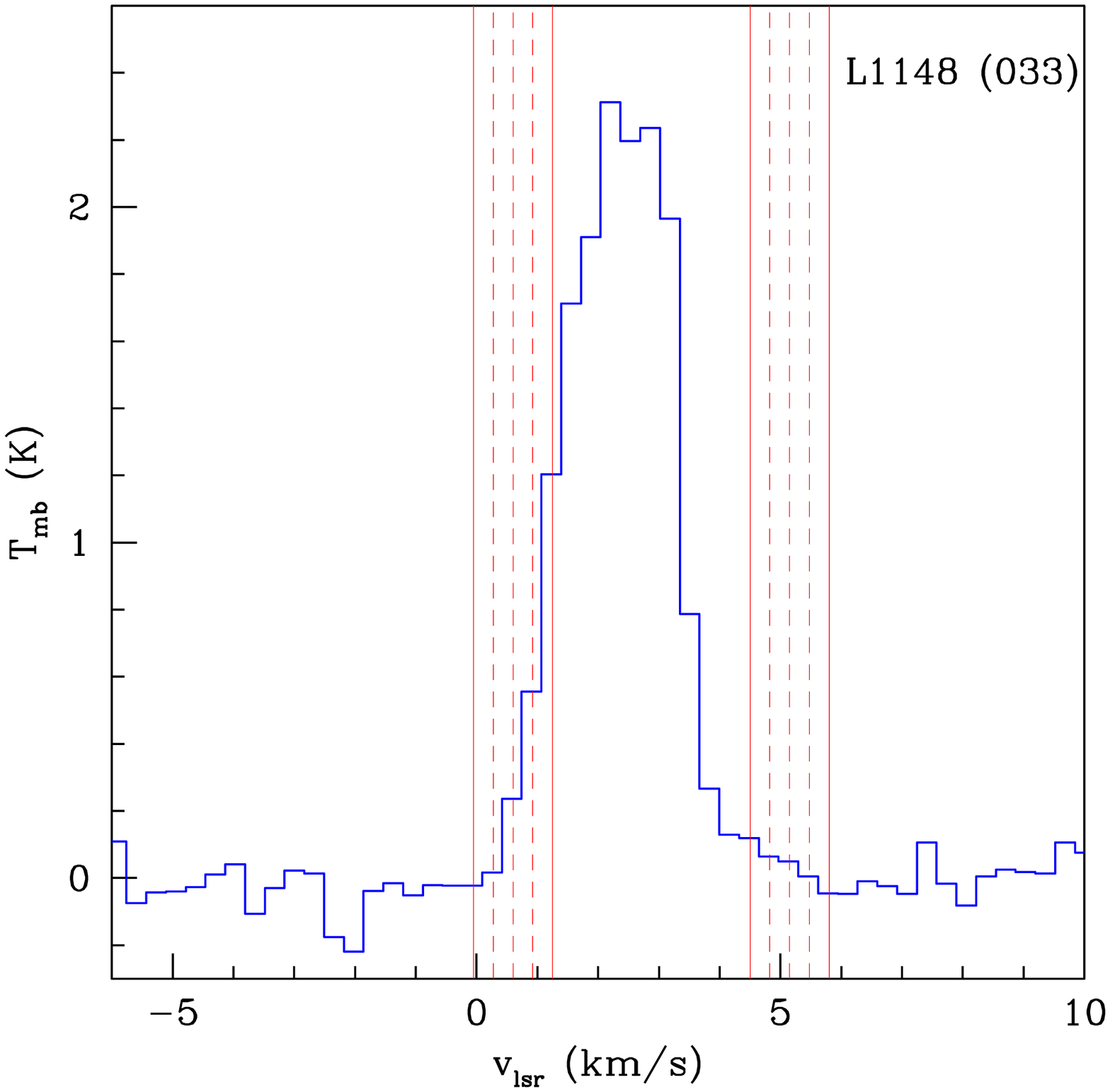}
\includegraphics{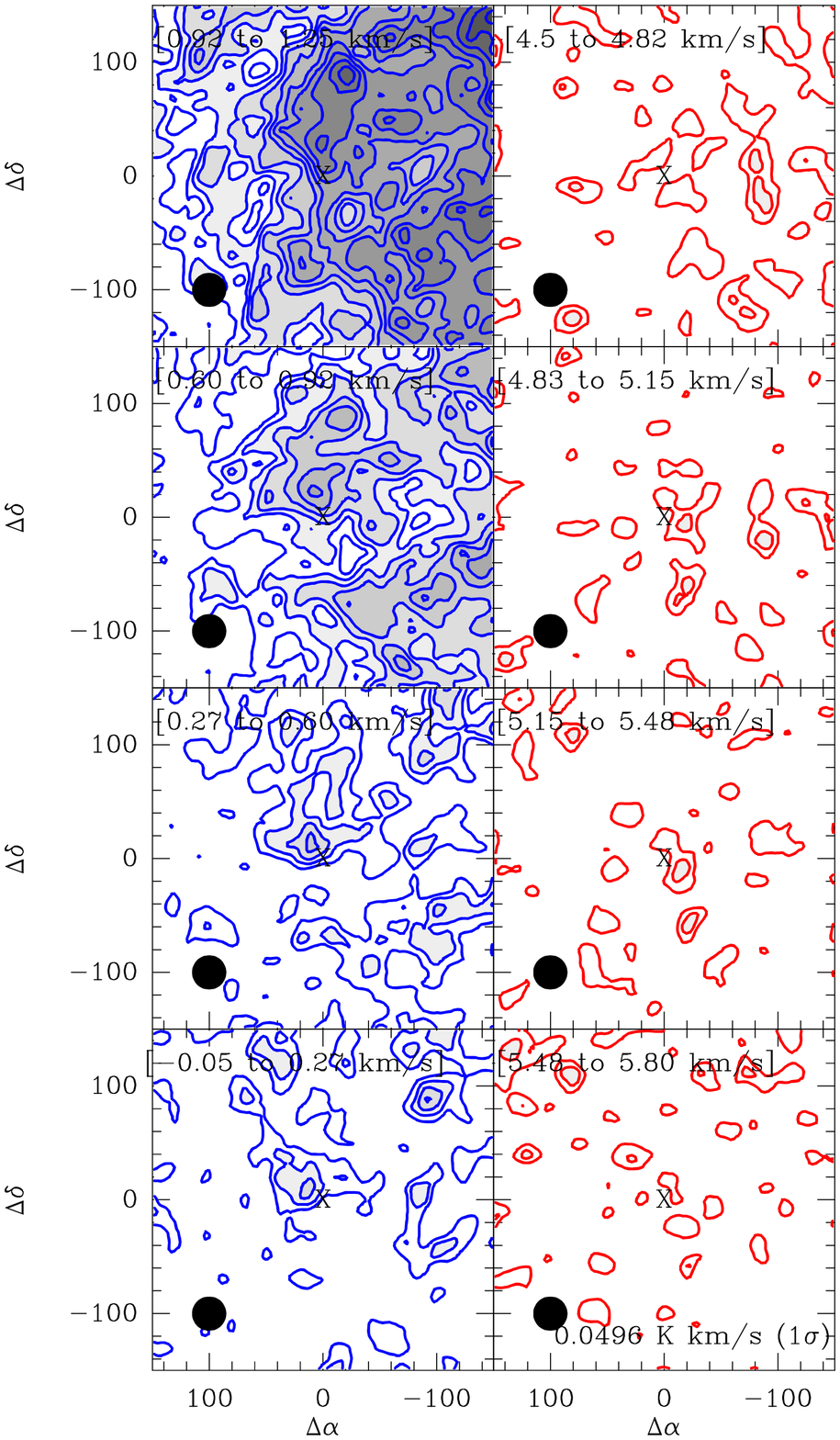}
\includegraphics{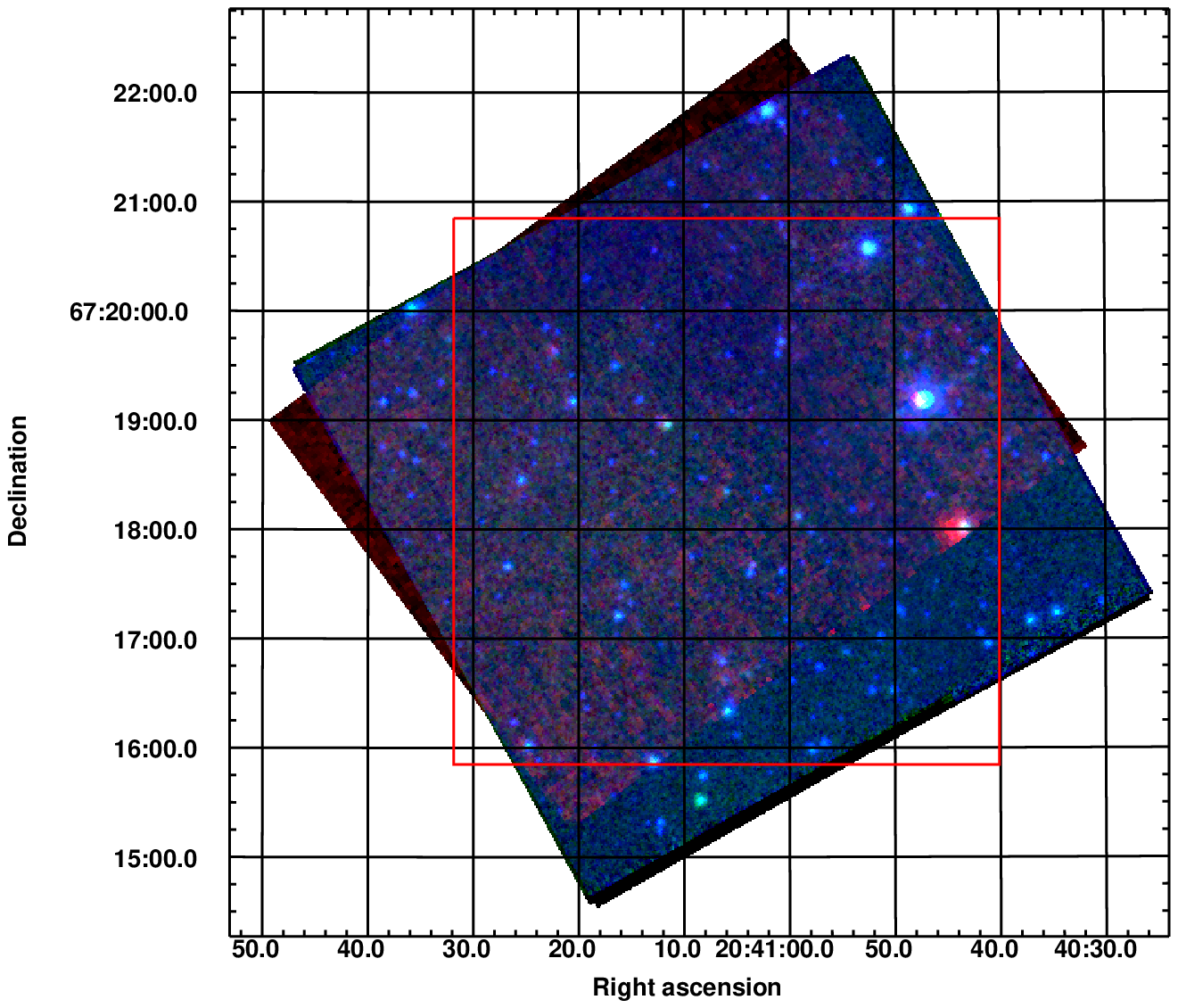}
\figcaption{LEFT: Channel map of outflow emission from L1148 (033). No outflow was detected around this source.
The left column
shows emission from the blue wing with velocity intervals listed 
at the top
of each panel.  The right column shows emission from the red wing.  
Contours
start at 3$\sigma_I$. There is no distinctive outflow signature
from the candidate protostar is seen. UPPER RIGHT: 
Velocity channels used to plot channel maps shown for the 
central 
spectrum.  LOWER RIGHT: Three color (4.6, 8.0, 24.0 $\mu$m) \textit{Spitzer} image centered on
L1148 (033). The solid box delineates the mapped region. }
\label{L1148maps}
\end{figure}
\clearpage


\begin{figure}
\figurenum{17}
\epsscale{1.0}
\vspace*{16cm}
\includegraphics{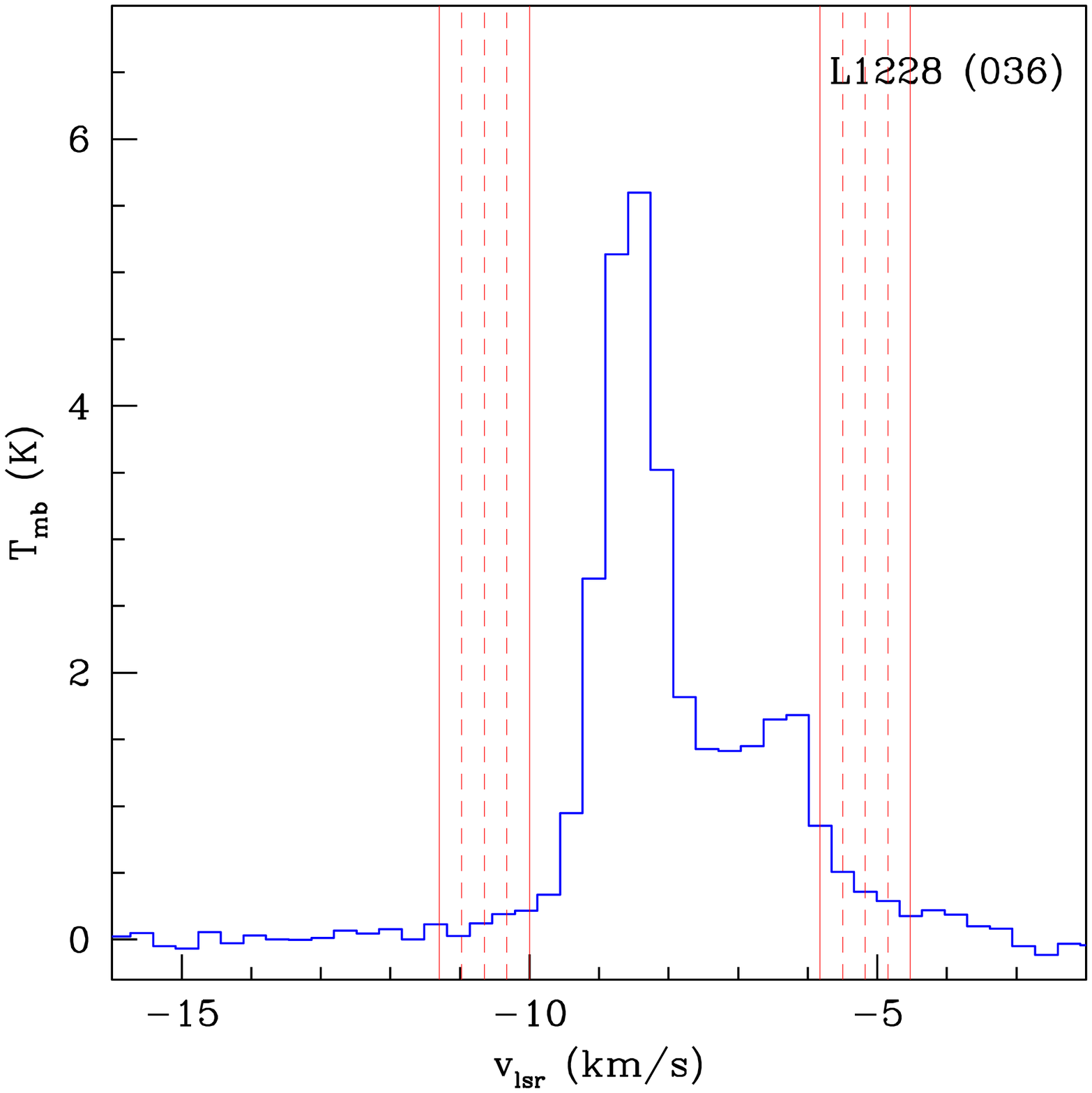}
\includegraphics{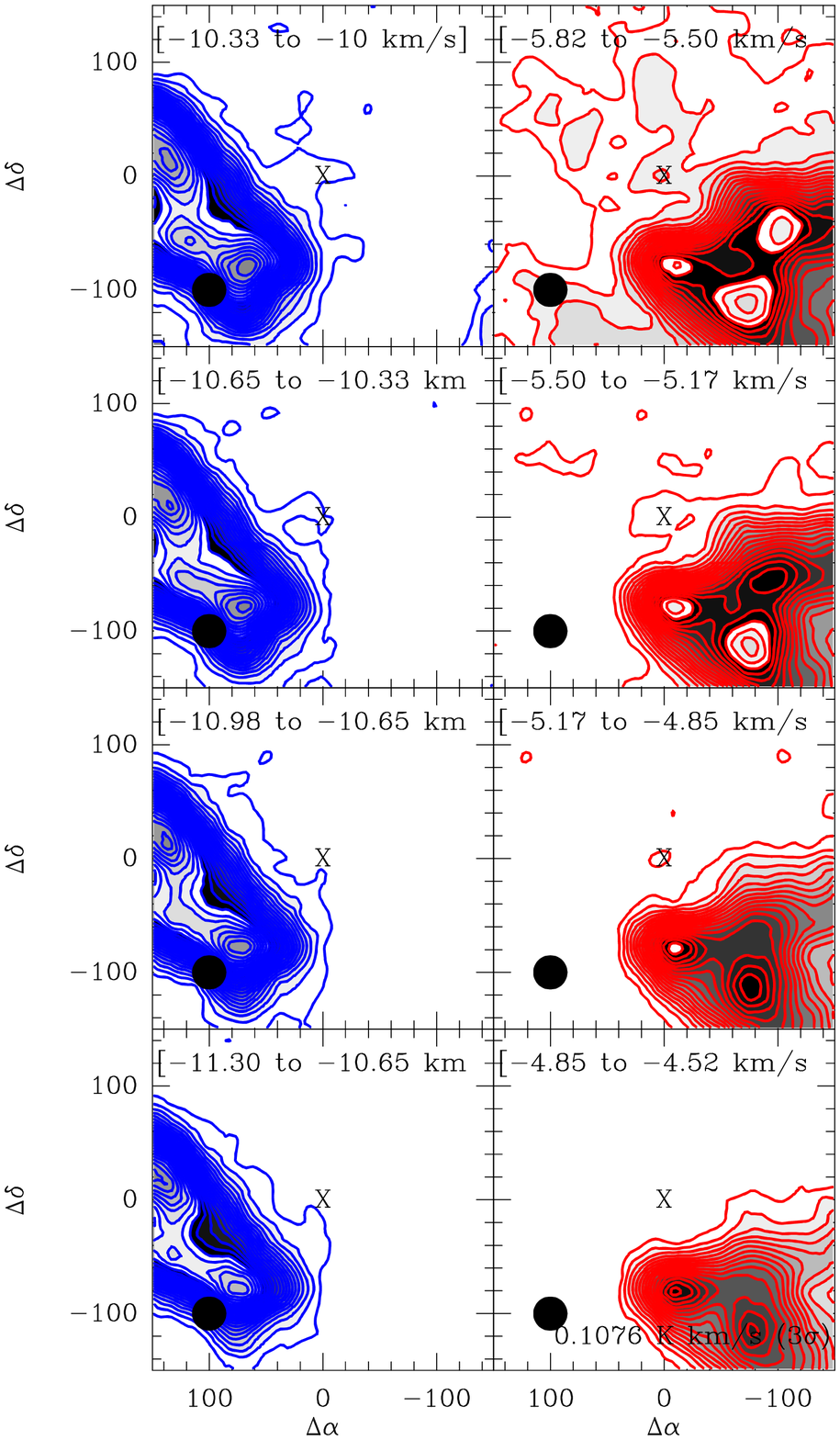}
\includegraphics{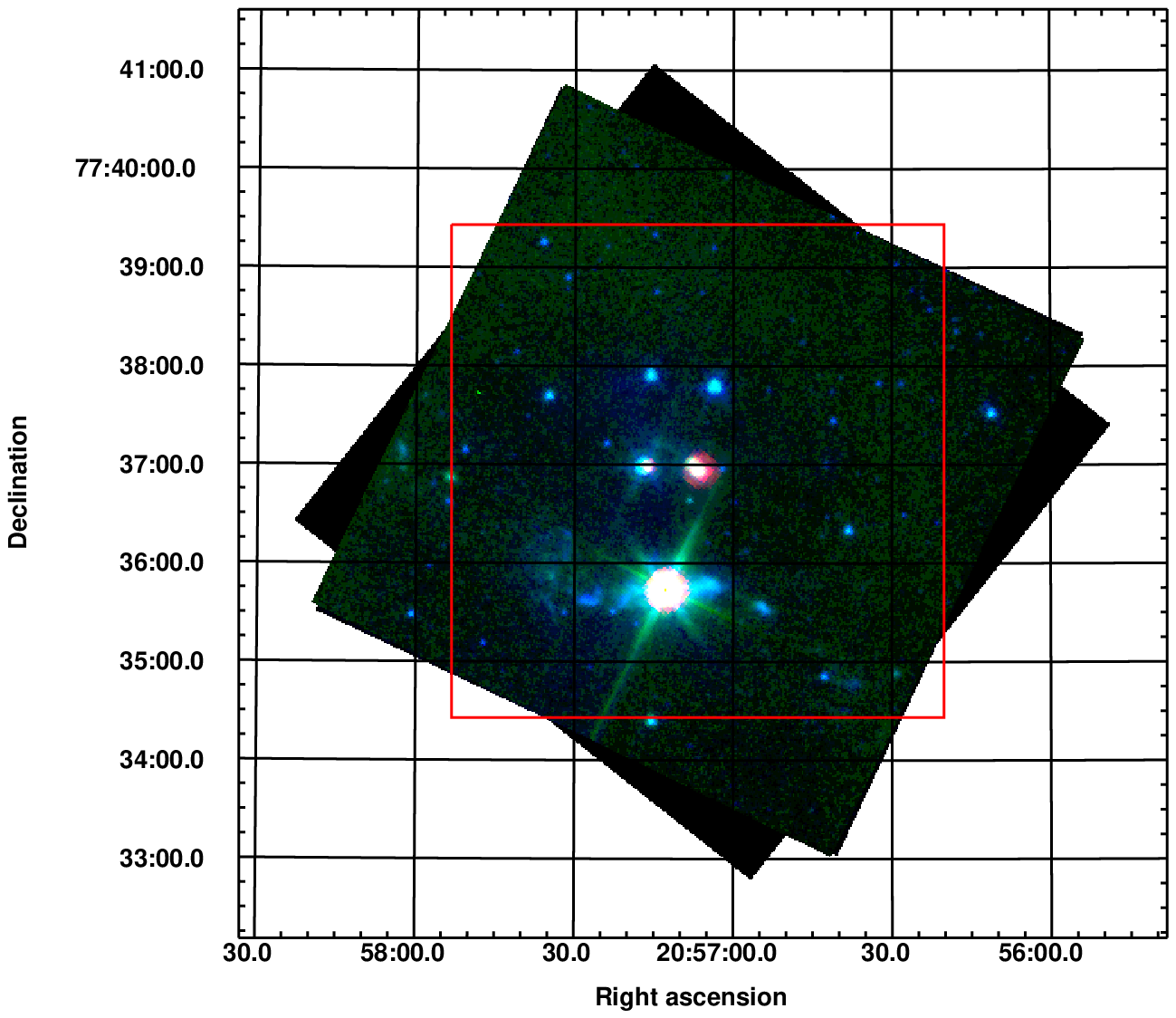}
\figcaption{LEFT: Channel map of outflow emission from L1228 (036). No outflow was detected around this source.
The left column
shows emission from the blue wing with velocity intervals listed 
at the top
of each panel.  The right column shows emission from the red wing.  
Contours
start at 3$\sigma_I$. There is significant confusion in this region and no distinctive outflow signature
from the candidate protostar is seen. UPPER RIGHT: 
Velocity channels used to plot channel maps shown for the 
central 
spectrum.  LOWER RIGHT:  Three color (4.6, 8.0, 24.0 $\mu$m) \textit{Spitzer} image centered on
L1228 (036).
The solid box delineates the mapped region.  }
\label{L1228maps}
\end{figure}
\clearpage


\begin{figure}
\figurenum{18}
\epsscale{1.0}
\vspace*{16cm}
\includegraphics{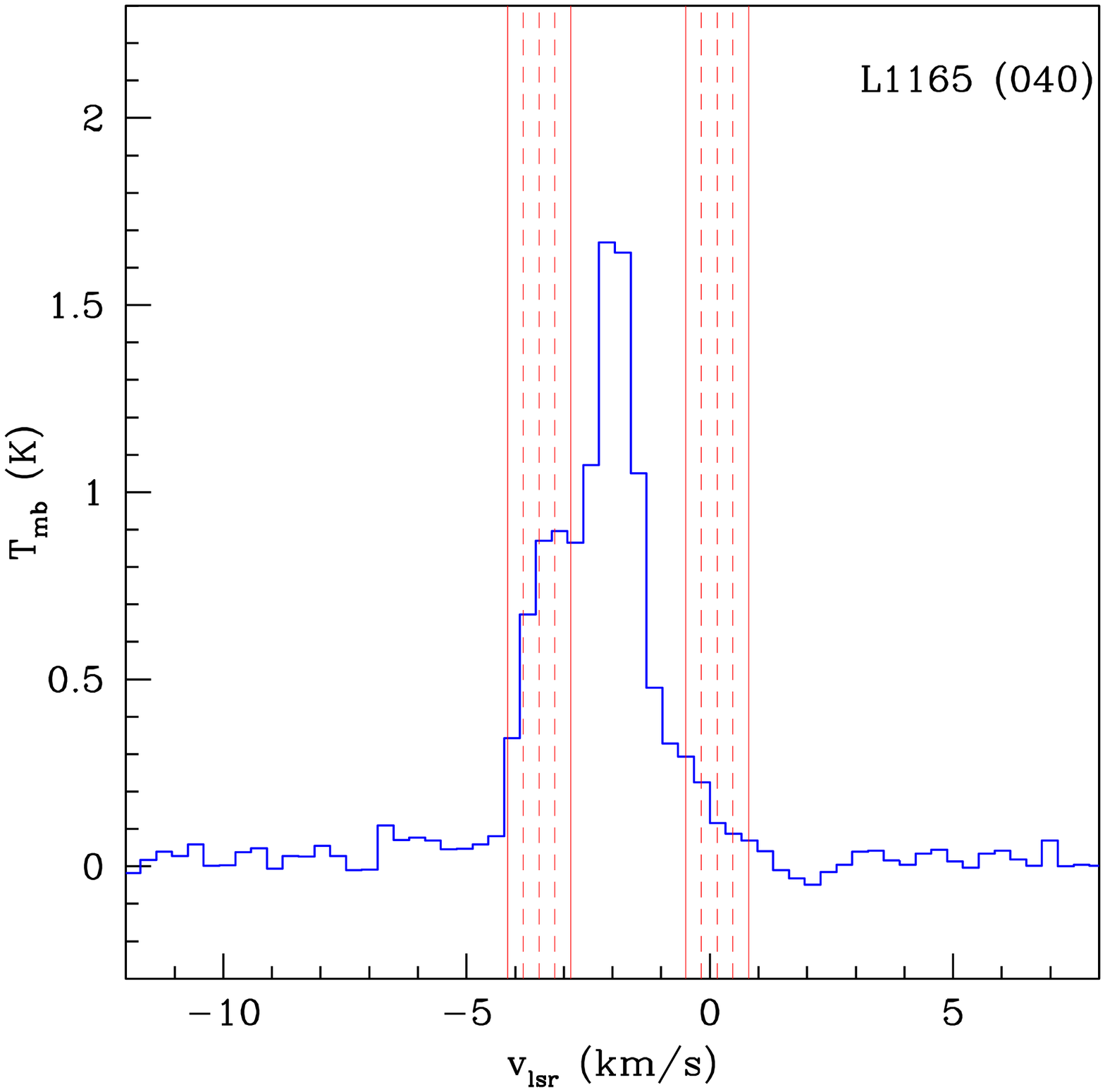}
\includegraphics{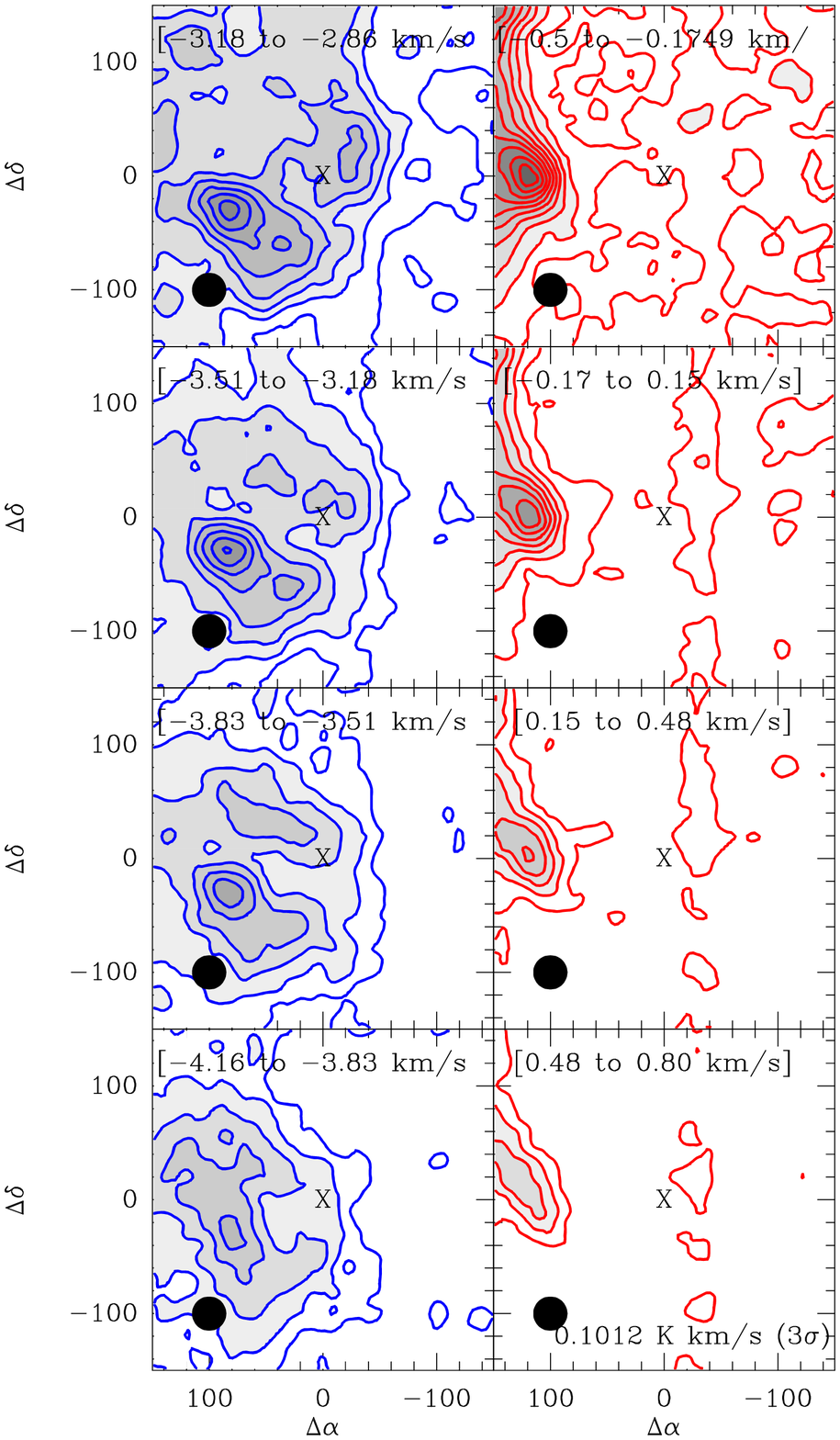}
\includegraphics{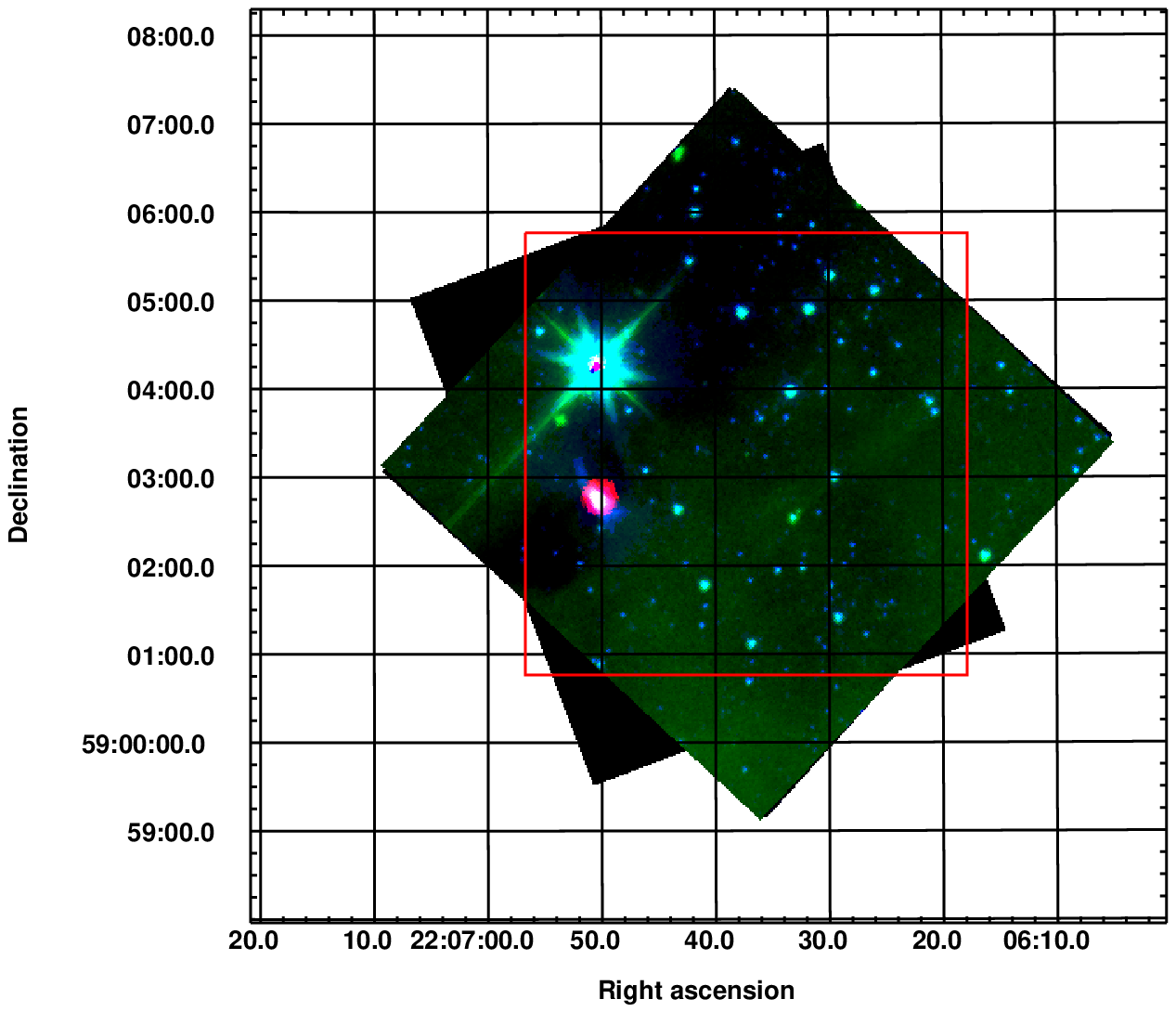}
\figcaption{LEFT: Channel map of outflow emission from L1165 (040). No outflow was detected around this source.
The left column
shows emission from the blue wing with velocity intervals listed 
at the top
of each panel.  The right column shows emission from the red wing.  
Contours
start at 3$\sigma_I$. There is significant confusion in this region and no distinctive outflow signature
from the candidate protostar is seen. UPPER RIGHT: 
Velocity channels used to plot channel maps shown for the 
central 
spectrum.  LOWER RIGHT:  Three color (4.6, 8.0, 24.0 $\mu$m) \textit{Spitzer} image centered on
L1165 (040).
The solid box delineates the mapped region.  }
\label{L1148maps}
\end{figure}
\clearpage


\begin{figure}
\figurenum{19}
\epsscale{.72}
\plotone{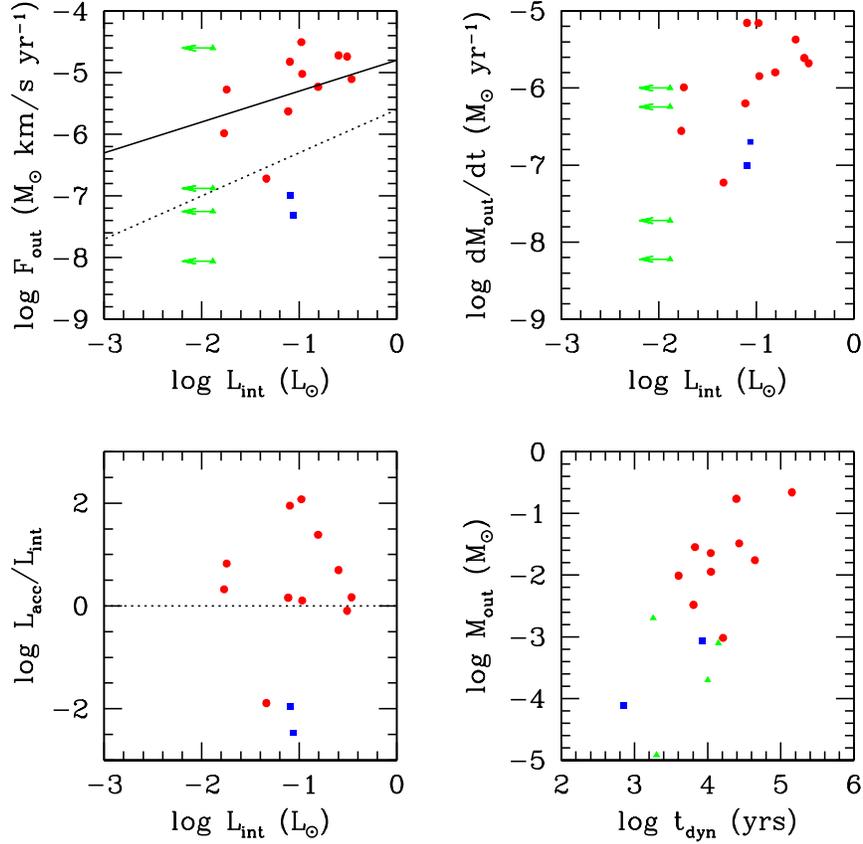}
\caption{The outflow force vs. internal protostellar luminosity is shown in the upper
left panel.  Red circles denote single-dish detected outflow.  Blue squares are
detected with an interferometer only.  Green triangles are first hydrostatic core
candidates with upper limits on their internal luminosity.  Data include the 3 outflows
mapped in this survey plus data culled from the literature for all known VeLLO outflows
and low-luminosity outflows in the Dunham et al. 2008 catalog (IRAM01491 Andr\'e et al. 1999, 
MMS 126 Stanke et al. 2006, L1014 Shirley et al. 2007, L1448-IRS2E1 Chen et al. 2010, 
CB17 MMS Chen et al. 2012, 
Per 071 073 105 106 Curtis et al. 2010, L1251A-IRS3 Lee et al. 2010,
L1148-IRS Kauffmann et al. 2011, Per-Bolo-58 Dunham et al. 2011, 
L1451-MM Pineda et al. 2011, LFAM 26 Nakamura et al. 2011).  
The solid line is the
correlation for Class 0 sources in the Curtis et al. 2010 survey of Perseus while
the dashed line is the correlation for Class I sources.  The outflow mass rate is plotted
versus internal luminosity in the upper right.  The time average accretion luminosity
(assuming $R = 3$\rsun ) vs. internal luminosity is plotted in the lower left.  The dashed
horizontal line denote where the ratio is equal to 1.  The outflow mass is plotted
vs. the dynamical timescale of the outflow in the lower right panel.
}
\label{forcegraph}
\end{figure}
\clearpage




\begin{deluxetable}{lcllcccccccr}
\tabletypesize{\scriptsize}
\rotate
\tablewidth{0pt}
\tablecaption{Candidate Low-Luminosity Embedded Objects in the Cores\tablenotemark{a}}
\label{cores_candidates}
\tablehead{
\colhead{Source} & \colhead{} & \colhead{Date Observed} & \colhead{Spitzer} & \colhead{c2d} & \colhead{Dist.} & \colhead{RA} & \colhead{Dec} & \colhead{$L_{IR}$\tablenotemark{e}} & \colhead{D$_{IRAS}$} & \colhead{Nearest IRAS} & \colhead{HHT} \\
\colhead{Number} & \colhead{Group\tablenotemark{b}} & \colhead{MJD-2450000 } & \colhead{Source Name\tablenotemark{c}} & \colhead{Core Region\tablenotemark{d}} & \colhead{(pc)} & \colhead{(J2000)} & \colhead{(J2000)} & \colhead{(\lsun)} & \colhead{(\as)} & Source & \colhead{Obs.\tablenotemark{f}}
}
\startdata
007 & 6 & 5170 & J054443.94$+$090307.2 & B35A       &    $400\pm  40$ & 05 44 43.94 & $+$09 03 07.2 &  0.006 &  405.4 & 05417+0907 & 5-pt.\\
020 & 5 & 5242 & J163705.11$-$353219.7 & DC 3460+78 &    $150\pm  20$ & 16 37 05.11 & $-$35 32 19.7 &  0.003 &  221.3 & 16335-3528 & 5-pt.\\
022 & 6 & 5242, 5592 & J171111.83$-$272655.0 & B59        &    $125\pm  25$ & 17 11 11.83 & $-$27 26 55.0 &  0.024 &  125.9 & 17082-2724 & OTF \\
023 & 3 & 5242, 5599 & J171122.18$-$272602.0 & B59        &    $125\pm  25$ & 17 11 22.18 & $-$27 26 02.0 &  0.343 &   86.3 & 17081-2721 & OTF\\
026 & 6 & 5171 & J191744.28+191523.7   & L723       &    $200\pm  100$& 19 17 44.28 & $+$19 15 23.7 & 0.048  &  225.4 & 19156+1906 & 5-pt.\\ 
027 & 1 & 5528 & J192025.32$+$112217.4 & L673       &    $300\pm 100$ & 19 20 25.32 & $+$11 22 17.4 &  0.201 &   42.9 & 19180+1116 & 5-pt.\\
028 & 5 & 5529 & J192025.92$+$112221.0 & L673       &    $300\pm 100$ & 19 20 25.92 & $+$11 22 21.0 &  0.168 &   52.5 & 19180+1116 & 5-pt.\\
030 & 3 & 5529 & J192026.54$+$112025.4 & L673       &    $300\pm 100$ & 19 20 26.54 & $+$11 20 25.4 &  0.138 &   35.4 & 19180+1114 & 5-pt.\\
031 & 1 & 5171, 5172, 5179 & J192134.82$+$112123.4 & L673-7     &    $300\pm 100$ & 19 21 34.82 & $+$11 21 23.4 &  0.017 &  427.6 & 19189+1109 & OTF\\
032 & 3 & 5170 & J204056.66$+$672304.9 & L1148      &    $325\pm  25$ & 20 40 56.66 & $+$67 23 04.9 &  0.081 &  222.3 & 20410+6710 & 5-pt.\\
033 & 6 & 5170, 5598 & J204105.95$+$671820.9 & L1148      &    $325\pm  25$ & 20 41 05.95 & $+$67 18 20.9 &  0.003 &  250.9 & 20410+6710 & OTF\\
034 & 6 & 5170 & J204355.51$+$673850.3 & L1155E     &    $325\pm  25$ & 20 43 55.51 & $+$67 38 50.3 &  0.004 &  642.1 & 20423+6736 & 5-pt.\\
035 & 6 & 5170 & J204427.17$+$673835.9 & L1155E     &    $325\pm  25$ & 20 44 27.17 & $+$67 38 35.9 &  0.004 &  768.6 & 20423+6736 & 5-pt.\\
036 & 6 & 5170, 5599 & J205706.72$+$773656.2 & L1228      &    $200\pm  50$ & 20 57 06.72 & $+$77 36 56.2 &  0.199 &   72.7 & 20582+7724 & OTF\\
037 & 6 & 5170 & J205707.85$+$773659.8 & L1228      &    $200\pm  50$ & 20 57 07.85 & $+$77 36 59.8 &  0.038 &   75.4 & 20582+7724 & 5-pt.\\
039 & 6 & 5170 & J220633.22$+$590232.6 & L1165      &    $300\pm  50$ & 22 06 33.22 & $+$59 02 32.6 &  0.030 &  135.6 & 22051+5848 & 5-pt.\\
040 & 6 & 5529, 5592, 5598 & J220637.27$+$590315.8 & L1165      &    $300\pm  50$ & 22 06 37.27 & $+$59 03 15.8 &  0.008 &  107.5 & 22051+5848 & OTF\\
041 & 1 & 5170 & J222807.42$+$690038.9 & L1221      &    $250\pm  50$ & 22 28 07.42 & $+$69 00 38.9 &  0.404 &   41.9 & 22266+6845 & 5-pt.\\
042 & 6 & 5170 & J222933.39$+$751316.0 & L1251      &    $300\pm  50$ & 22 29 33.39 & $+$75 13 16.0 &  0.066 &  119.7 & 22290+7458 & 5-pt.\\
044 & 1 & 5171 & J223031.94$+$751408.9 & L1251      &    $300\pm  50$ & 22 30 31.94 & $+$75 14 08.9 &  0.156 &  120.5 & 22290+7458 & OTF\\
045 & 1 & 5171, 5184, 5599 & J223105.59$+$751337.2 & L1251      &    $300\pm  50$ & 22 31 05.59 & $+$75 13 37.2 &  0.077 &  252.0 & 22290+7458 & OTF \\
046 & 6 & 5171 & J223514.06$+$751502.5 & L1251      &    $300\pm  50$ & 22 35 14.06 & $+$75 15 02.5 &  0.030 &  129.3 & 22343+7501 & 5-pt.\\
047 & 6 & 5171 & J223731.13$+$751041.5 & L1251      &    $300\pm  50$ & 22 37 31.13 & $+$75 10 41.5 &  0.004 &  295.5 & 22376+7455 & 5-pt.\\
178 & 5 & 5592 & J162135.57$-$224351.6 & Ophiuchus   &    $125\pm 25$      & 16 21 35.57 & $-$22 43 51.6 &  0.004 & 1467.3 & 16200-2251 & 5-pt.\\
179 & 5 & 5241 & J162145.12$-$234231.7 & Ophiuchus   &    $125\pm 25$      & 16 21 45.12 & $-$23 42 31.7 &  0.033 &  246.9 & 16187-2339 & 5-pt.\\
181 & 3 & 5241& J162648.48$-$242838.6 & Ophiuchus   &    $125\pm 25$      & 16 26 48.48 & $-$24 28 38.6 &  0.103 &  366.6 & 16235-2416 & 5-pt.\\
183 & 3 & 5241 & J162715.89$-$243843.1 & Ophiuchus   &    $125\pm 25$      & 16 27 15.89 & $-$24 38 43.1 &  0.440 &  163.5 & 16240-2430 & 5-pt.\\
186 & 5 & 5599 & J162843.68$-$231918.1 & Ophiuchus   &    $125\pm 25$      & 16 28 43.68 & $-$23 19 18.1 &  0.000 &  516.2 & 16262-2317 & 5-pt.\\
187 & 5 & 5599 & J163122.87$-$231655.2 & Ophiuchus   &    $125\pm 25$      & 16 31 22.87 & $-$23 16 55.2 &  0.003 & 1354.8 & 16275-2251 & 5-pt.\\
188 & 6 & 5241 & J163131.25$-$242628.0 & Ophiuchus   &    $125\pm 25$      & 16 31 31.25 & $-$24 26 28.0 &  0.013 &   70.9 & 16285-2421 & 5-pt.\\
191 & 5 & 5241 & J163502.45$-$235100.7 & Ophiuchus   &    $125\pm 25$      & 16 35 02.45 & $-$23 51 00.7 &  0.006 & 1074.3 & 16318-2402 & 5-pt.\\
192 & 5 & 5242 & J164001.80$-$244429.0 & Ophiuchus   &    $125\pm 25$      & 16 40 01.80 & $-$24 44 29.0 &  0.004 &  447.9 & 16375-2439 & 5-pt.\\
193 & 5 & 5242 & J164114.35$-$243758.1 & Ophiuchus   &    $125\pm 25$      & 16 41 14.35 & $-$24 37 58.1 &  0.007 &  539.7 & 16377-2426 & 5-pt.\\
194 & 5 & 5242 & J164147.64$-$243030.2 & Ophiuchus   &    $125\pm 25$      & 16 41 47.64 & $-$24 30 30.2 &  0.005 &  591.5 & 16381-2419 & 5-pt.\\
195 & 5 & 5255 & J164206.74$-$233859.3 & Ophiuchus   &    $125\pm 25$      & 16 42 06.74 & $-$23 38 59.3 &  0.005 &  538.5 & 16384-2334 & 5-pt.\\
196 & 5 & 5255 & J164301.87$-$245420.9 & Ophiuchus   &    $125\pm 25$      & 16 43 01.87 & $-$24 54 20.9 &  0.002 &  295.6 & 16403-2447 & 5-pt.\\
197 & 5 & 5255 & J164305.88$-$242357.5 & Ophiuchus   &    $125\pm 25$      & 16 43 05.88 & $-$24 23 57.5 &  0.005 &  345.0 & 16396-2419 & 5-pt.\\
198 & 5 & 5242 & J164433.26$-$233348.2 & Ophiuchus   &    $125\pm 25$      & 16 44 33.26 & $-$23 33 48.2 &  0.005 &  329.6 & 16411-2329 & 5-pt.\\
199 & 5 & 5242 & J164512.50$-$233848.1 & Ophiuchus   &    $125\pm 25$      & 16 45 12.50 & $-$23 38 48.1 &  0.016 &  375.3 & 16420-2327 & 5-pt.\\
\enddata
\tablenotetext{a}{Reproduced from Dunham et al. 2008.}
\tablenotetext{b}{Group indicating likelihood of being an embedded protostar.}
\tablenotetext{c}{All source names are preceded by the prefix ``SSTc2d ''.}
\tablenotetext{d}{Name of the core region from Evans et al. (2007) in which this source is located.}
\tablenotetext{e}{Integrated luminosity using all available detections between 1.25 (2MASS J-band) and 70 \um.  Entries in italics denote that the calculation only extends to 24 \um\ due to no flux information available at 70 \um.}
\tablenotetext{f}{HHT Mapping mode. 5-pt. = 5 point map with 30\as\ spacing.  OTF = On-the-fly map with 10\as\ row spacing.}
\end{deluxetable}

\begin{deluxetable}{lcccc}
\tabletypesize{\scriptsize}
\tablewidth{0pt}
\tablecaption{$\eta_{pol}$ for Observing Shifts}
\label{calibration}
\tablehead{
\colhead{MJD -2450000} & \colhead{V$_{pol}$ LSB} & \colhead{H$_{pol}$ LSB} & 
\colhead{V$_{pol}$ USB} & \colhead{H$_{pol}$ USB} \\
}
\startdata
5170	&	0.89	&	0.92	&	0.87	&	0.89	\\
5172	&	0.97	&	0.94	&	0.94	&	0.90	\\
5179	&	0.91	&	0.95	&	0.90	&	0.93	\\
5184	&	0.88	&	0.92	&	0.87	&	0.89	\\
5527	&	0.80	&	0.53	&	0.79	&	0.53	\\
5242	&	0.99	&	\nodata	&	0.94	&	\nodata	\\
5255	&	1.03	&	\nodata	&	1.00	&	\nodata	\\
5592	&	0.67	&	0.76	&	0.65	&	0.76	\\
5599	&	0.66	&	0.76	&	0.64	&	0.76	\\
\enddata
\end{deluxetable}

\begin{deluxetable}{lcccccc}
\tabletypesize{\scriptsize}
\tablewidth{0pt}
\tablecaption{Spectral properties for the center position of each source.}
\label{specproperties}
\tablehead{
\colhead{Source} & \colhead{\co\ T$_{pk}$ (K)} & \colhead{\co\ I (K km/s)} & \colhead{\coo\ T$_{pk}$} & \colhead{\coo\ I (K km/s)} & \colhead{\coo $\Delta v$ (km/s)} & \colhead{Flags\tablenotemark{a}\tablenotemark{b}} \\
}
\startdata
B35A (007)	&	8.900	(0.034)	&	9.934	(0.047)	& 	1.496	(0.041)	&	1.208	(0.072)	& 	0.892	(0.004)	&	s	\\
DC3460+78 (020)	&	2.717	(0.111)	&	9.934	(0.200)	& 	\nodata	(0.090)	&	1.107	(0.166)	& 	\nodata		&	mc	\\
B59 (022)	&	7.003	(0.089)	&	20.843	(0.160)	& 	4.992	(0.072)	&	7.060	(0.119)	& 	1.098	(0.012)	&	s bw	\\
B59 (023)	&	6.551	(0.092)	&	10.439	(0.148)	& 	4.986	(0.090)	&	9.811	(0.148)	& 	1.549	(0.017)	&	s bw rw	\\
L723 (026)	&	5.626	(0.096)	&	21.485	(0.197)	& 	2.89	(0.097)	&	6.490	(0.179)	& 	1.791	(0.052)	&	mc	\\
L673 (027)	&	3.054	(0.050)	&	22.229	(0.127)	& 	2.412	(0.038)	&	7.509	(0.066)	& 	2.353	(0.018)	&	mc w?	\\
L673 (028)	&	3.197	(0.046)	&	25.798	(0.128)	& 	2.15	(0.048)	&	7.345	(0.093)	& 	2.475	(0.027)	&	mc w?	\\
L673 (030)	&	3.100	(0.073)	&	27.773	(0.186)	& 	2.611	(0.060)	&	9.159	(0.116)	& 	2.298	(0.031)	&	sa rw	\\
L673-7 (031)	&	3.545	(0.073)	&	11.952	(0.150)	& 	2.825	(0.059)	&	3.809	(0.103)	& 	1.504	(0.038)	&	s bw rw	\\
L1148 (032)	&	3.158	(0.040)	&	6.785	(0.072)	& 	1.075	(0.042)	&	1.293	(0.065)	& 	1.944	(0.015)	&	s	\\
L1148 (033)	&	2.295	(0.050)	&	5.561	(0.086)	& 	1.574	(0.039)	&	1.405	(0.060)	& 	1.979	(0.022)	&	s	\\
L1148 (034)	&	3.138	(0.038)	&	8.674	(0.069)	& 	1.64	(0.045)	&	1.536	(0.070)	& 	3.512	(0.024)	&	mc	\\
L1148 (035)	&	3.826	(0.041)	&	9.599	(0.074)	& 	1.357	(0.044)	&	1.619	(0.068)	& 	3.215	(0.020)	&	mc	\\
L1228 (036)	&	6.560	(0.056)	&	11.647	(0.111)	& 	3.373	(0.040)	&	5.262	(0.066)	& 	1.239	(0.084)	&	sa	\\
L1228 (037)	&	6.894	(0.050)	&	12.018	(0.099)	& 	3.546	(0.043)	&	5.543	(0.071)	& 	1.199	(0.072)	&	sa	\\
L1165 (039)	&	0.795	(0.076)	&	1.796	(0.156)	& 	\nodata	(0.043)	&	\nodata	(0.242)	& 	2.394	(0.127)	&	mc sa bw	\\
L1165 (040)	&	1.667	(0.061)	&	5.290	(0.104)	& 	0.7066	(0.038)	&	0.929	(0.063)	& 	0.767	(0.035)	&	s	\\
L1221 (041)	&	5.840	(0.045)	&	28.161	(0.123)	& 	3.435	(0.056)	&	7.756	(0.113)	& 	5.326	(0.033)	&	s	\\
L1251 (042)	&	3.986	(0.072)	&	9.519	(0.116)	& 	1.137	(0.049)	&	1.960	(0.099)	& 	2.229	(0.022)	&	s bw rw	\\
L1251 (044)	&	4.088	(0.048)	&	15.805	(0.102)	& 	1.511	(0.050)	&	2.792	(0.082)	& 	3.006	(0.021)	&	sa	\\
L1251 (045)	&	5.220	(0.057)	&	16.160	(0.122)	& 	1.603	(0.055)	&	2.992	(0.091)	& 	2.612	(0.080)	&	mc	\\
L1251 (046)	&	4.430	(0.061)	&	12.133	(0.120)	& 	2.69	(0.056)	&	4.817	(0.092)	& 	2.627	(0.005)	&	s	\\
L1251 (047)	&	4.184	(0.051)	&	12.862	(0.101)	& 	2.485	(0.046)	&	5.126	(0.076)	& 	2.785	(0.014)	&	s	\\
Oph (178)	&	0.161	(0.049)	&	0.392	(0.068)	& 	\nodata	(0.077)	&	\nodata	(0.432)	& 	\nodata		&	mc	\\
Oph (179)	&	15.910	(0.083)	&	35.408	(0.222)	& 	12.42	(0.083)	&	13.553	(0.153)	& 	0.973	(0.007)	&	mc	\\
Oph (181)	&	16.610	(0.060)	&	59.199	(0.137)	& 	12.51	(0.063)	&	23.853	(0.127)	& 	1.816	(0.011)	&	s	\\
Oph (183)	&	12.040	(0.086)	&	52.213	(0.196)	& 	9.454	(0.059)	&	26.238	(0.109)	& 	2.739	(0.010)	&	s	\\
Oph (186)	&	7.149	(0.057)	&	16.297	(0.086)	& 	0.268	(0.053)	&	0.701	(0.082)	& 	1.685	(0.182)	&	s	\\
Oph (187)	&	0.384	(0.055)	&	0.592	(0.083)	& 	\nodata	(0.051)	&	\nodata	(0.287)	& 	\nodata		&	\nodata	\\
Oph (188)	&	0.505	(0.090)	&	1.086	(0.145)	& 	\nodata	(0.075)	&	\nodata	(0.419)	& 	\nodata		&	\nodata	\\
Oph (191)	&	\nodata	(0.082)	&	\nodata	(0.536)	& 	\nodata	(0.077)	&	\nodata	(0.432)	& 	\nodata		&	\nodata	\\
Oph (192)	&	\nodata	(0.083)	&	\nodata	(0.545)	& 	\nodata	(0.076)	&	\nodata	(0.351)	& 	\nodata		&	\nodata	\\
Oph (193)	&	\nodata	(0.076)	&	\nodata	(0.498)	& 	\nodata	(0.075)	&	\nodata	(0.417)	& 	\nodata		&	\nodata	\\
Oph (194)	&	\nodata	(0.102)	&	\nodata	(0.666)	& 	\nodata	(0.104)	&	\nodata	(0.580)	& 	\nodata		&	\nodata	\\
Oph (195)	&	\nodata	(0.074)	&	\nodata	(0.484)	& 	\nodata	(0.080)	&	\nodata	(0.444)	& 	\nodata		&	\nodata	\\
Oph (196)	&	\nodata	(0.099)	&	\nodata	(0.644)	& 	\nodata	(0.082)	&	\nodata	(0.458)	& 	\nodata		&	\nodata	\\
Oph (197)	&	1.189	(0.081)	&	0.863	(0.131)	& 	\nodata	(0.078)	&	\nodata	(0.438)	& 	\nodata		&	s	\\
Oph (198)	&	\nodata	(0.082)	&	\nodata	(0.539)	& 	\nodata	(0.095)	&	\nodata	(0.533)	& 	\nodata		&	\nodata	\\
Oph (199)	&	\nodata	(0.081)	&	\nodata	(0.531)	& 	\nodata	(0.083)	&	\nodata	(0.463)	& 	\nodata		&	\nodata	\\
\enddata
\tablenotetext{a}{Flags indicating spectral features in \co : rw and bw indicate evidence for red and blue wings, s is for single components, mc for multiple components, and sa for self absorbed peaks. See \S~\ref{detections}.}
\tablenotetext{b}{Ellipses indicate non-detections.}
\end{deluxetable}

\begin{deluxetable}{lcccc}
\tabletypesize{\scriptsize}
\tablewidth{0pt}
\tablecaption{Observed Outflow Properties}
\label{L673properties}
\tablehead{
\colhead{Quantity} & \colhead{Units} & \colhead{B59 (023)} & \colhead{L673-7 (031)} & \colhead{L1251-A (045)}\\
}
\startdata
M$_{out}$ & $\eten{-2} \msun$ & $2.279\pm0.036$ & $1.741\pm0.022$ &  $3.265\pm0.020$ \\
E$_{out}$ & $\eten{42} \ \erg$ $\frac{1}{\cos^{2}i}$   & $3.52\pm0.37$ & $0.99\pm0.17$ &  $1.36\pm0.13$  \\
P$_{out}$ & $\eten{32} \mathrm{g}\ \kms$ $\frac{1}{\cos i}$   & $1.7372\pm0.0016$ & $0.92491\pm.00030$ & $1.266\pm0.062$   \\
t$_{dyn}$ & $\eten{4}\ \yr$ $\frac{\cos i}{\sin i}$   & $1.10\pm0.13$ & $4.43\pm0.23$ & $2.69\pm0.65$ \\
L$_{out}$ & $\eten{-3} \lsun$  $\frac{\sin i}{\cos^{3} i}$ & $2.59\pm0.14$ & $0.182\pm0.033$ & $0.41\pm0.11$  \\
F$_{out}$ & $\eten{-6} \ \msun  \kms\ \yr^{-1}$ $\frac{\sin i}{\cos^{2} i}$  &  $7.86\pm0.25$ & $1.043\pm0.054$ & $2.35\pm0.58$ \\
$\mean{dM_{acc}/dt}$ & $\eten{-7}\ \msun\ \yr^{-1}$ $\frac{\sin i}{\cos^{2} i}$ & $21.0\pm2.5$ & $2.78\pm0.14$ & $6.3\pm1.5$  \\
M$_{proto}$ & $\eten{-2} \ \msun$ $\frac{1}{\cos i}$ & $2.32\pm0.39$ & $1.233\pm0.090$ & $1.69\pm0.58$ \\ 
L$_{acc}$ & $\lsun$ $\frac{\sin i}{\cos^{3} i}$ & $0.51\pm0.10$ & $0.0357\pm0.0032$ & $0.111\pm0.047$ \\
\enddata
\end{deluxetable}

\end{document}